\documentclass[aps,pra,twocolumn,a4paper,groupedaddress,amsmath,amssymb,showpacs,10pt]{revtex4-2}
\usepackage{graphicx}
\usepackage{amsmath}
\usepackage{amssymb}
\usepackage{mathrsfs}
\usepackage{amsfonts}
\usepackage{dsfont}
\usepackage{dcolumn}
\usepackage{bm}
\usepackage{booktabs}
\usepackage{color}
\usepackage{xcolor}
\usepackage{colortbl}
\usepackage{mathtools}
\usepackage{upgreek}

\newcommand{\mE}{\mathscr{E}}

\newcommand{\mC}{\mathscr{C}}
\newcommand{\mD}{\mathscr{D}}
\newcommand{\mM}{\mathscr{M}}

\newcommand{\mV}{\mathscr{V}}
\newcommand{\mDz}{{\mathscr{D}_0}}
\newcommand{\mDI}{{\mathscr{D}_I}}
\newcommand{\mDzN}{{\mathscr{D}_{0N}}}
\newcommand{\mDzM}{{\mathscr{D}_{0M}}}
\newcommand{\hId}{\hat{I}}

\newcommand{\mA}{\mathcal{A}}
\newcommand{\mB}{\mathcal{B}}

\newcommand{\ta}{\widetilde{\alpha}}

\newcommand{\tpsi}{\widetilde{\psi}}
\newcommand{\tvp}{\widetilde{\varphi}}

\newcommand{\hmD}{\hat{\mathcal{D}}}

\newcommand{\hmV}{\hat{\mathcal{V}}}

\newcommand{\hmM}{\hat{\mathcal{M}}}

\newcommand{\hr}{\hat{\mathbf{e}}_r}
\newcommand{\hte}{\hat{\mathbf{e}}_\theta}
\newcommand{\hf}{\hat{\mathbf{e}}_\phi}
\newcommand{\z}{{(0)}}
\newcommand{\un}{{(1)}}
\newcommand{\du}{{(2)}}
\newcommand{\lz}{{l_0}}

\newcommand{\n}{{(n)}}
\providecommand{\abs}[1]{\left|#1\right|}

\providecommand{\ket}[1]{|#1\rangle}
\providecommand{\bra}[1]{\langle#1|}
\providecommand{\brak}[2]{\langle#1|#2\rangle} 
\providecommand{\proj}[2]{|#1\rangle \! \langle#2|} 
\providecommand{\mean}[3]{\langle#1|#2|#3\rangle} 

\newcommand{\ve}{\varepsilon}
\newcommand{\vp}{\varphi}

\everymath{\displaystyle}

\begin{document}

\title{Perturbation theory of nearly spherical dielectric optical resonators}

\author{Julius Gohsrich}
\affiliation{Institute for Theoretical Physics, Department of Physics, University of Erlangen-N\"{u}rnberg, Staudtstrasse 7, 91058 Erlangen, Germany}
\affiliation{Max Planck Institute for the Science of Light, Staudtstrasse 2, 91058
Erlangen, Germany}

\author{Tirth Shah}
\affiliation{Institute for Theoretical Physics, Department of Physics, University of Erlangen-N\"{u}rnberg, Staudtstrasse 7, 91058 Erlangen, Germany}
\affiliation{Max Planck Institute for the Science of Light, Staudtstrasse 2, 91058
Erlangen, Germany}

\author{Andrea Aiello}
\email{andrea.aiello@mpl.mpg.de} 
\affiliation{Max Planck Institute for the Science of Light, Staudtstrasse 2, 91058
Erlangen, Germany}


\date{\today}

\begin{abstract}

Dielectric spheres of various sizes may sustain electromagnetic whispering-gallery modes  resonating at optical frequencies with very narrow linewidths.
Arbitrary small deviations from the spherical shape typically shift and broaden such resonances. Our goal
is to determine these shifted and broadened  resonances.
A boundary-condition perturbation theory for the acoustic vibrations of nearly circular membranes was developed by Rayleigh more than a century ago. We extend this theory to describe the electromagnetic excitations of nearly spherical dielectric cavities.
This approach permits us to avoid  dealing with decaying quasinormal modes.
We explicitly find the frequencies and the linewidths of the optical resonances for arbitrarily deformed nearly spherical dielectric cavities, as power series expansions by a small parameter, up to and including  second-order terms.
We thoroughly discuss the physical conditions for the applicability of perturbation theory.

\end{abstract}

\maketitle

\section{Introduction}\label{Introduction}

In this work we aim at determining  frequencies and linewidths of electromagnetic resonances of nearly spherical dielectric cavities of arbitrary size and (small) deformation. 
In a spherical dielectric cavity, light can excite whispering gallery modes (WGMs) and circulate about any great circle with small attenuation  \cite{Oraevsky}.
In fact, propagation of light along a curved interface between two different dielectric media, is an intrinsically lossy process \cite{SNYDER1974326}. This implies that the resonant optical frequencies associated with the WGMs, have small but finite linewidths. The ratio between the frequency of a mode and its linewidth is proportional to the optical quality factor $Q$ of the mode. This quantifies the number of optical cycles the light in the mode will stay confined within in the cavity. Values of $Q$ around $10^{11}$ have been achieved for silica microspheres  \cite{Collot_1993}. Deviations from the spherical shape change  frequencies and linewidths of the modes, thus modifying their quality factors by an amount depending on the size and the shape of the deformation.
This may be either detrimental or, conversely,  very useful for many applications, ranging from WGMs lasers \cite{LPOR:LPOR200910016}, to dielectric microcavities \cite{RevModPhys.87.61}.
 Therefore, it is highly desirable to have a theory predicting, at least with a certain level of approximation, the frequencies and the linewidths of the electromagnetic resonances of nearly spherical dielectric cavities \cite{Heebner}.

In principle,  determining such resonances is a conceptually simple boundary-value problem: One must solve Maxwell's equations for the fields inside (medium $1$) and outside (medium $2$) the cavity, and match these fields at the interface between the two media. 
However,  satisfying  boundary conditions on interfaces of arbitrarily complicated shape, is typically  a formidable algebraic task. The literature about techniques and methods developed for solving this problem is, without exaggeration,  enormous.
Among the books we found particularly useful the Stratton's and Jackson's classical texts \cite{Stratton,Jackson}, and the perhaps less known but not less valuable works of Grandy \cite{Grandy} and Kristensson \cite{kristensson2016scattering}.
One of the first perturbation approaches to the scattering of electromagnetic waves by dielectric media of arbitrary shape was given by Yeh \cite{PhysRev.135.A1193}. This study was further developed and improved by Erma \cite{PhysRev.179.1238}. Later, an important contribution to the perturbation theory of quasinormal modes in open systems was given by Lai \emph{et al.}, \cite{PhysRevA.41.5187}.
However, a serious problem with  perturbation theory, based upon the analogy between the refractive index in electromagnetism and the potential energy in quantum mechanics, arises from the discontinuity of both the refractive index and  the normal component of the electric field, occurring at the interface between the resonator and the surrounding medium \cite{PhysRevB.24.7112}. Several methods have been proposed to deal with this issue; see, e.g.,  \cite{PhysRevE.65.066611,PhysRevE.77.036611}.

A different approach that avoids this problem, is the so-called boundary-condition perturbation theory.
Recently, Dubertrand \emph{et al.}, presented a boundary-condition perturbation theory for two-dimensional disk resonators \cite{PhysRevA.77.013804}, which may be seen as an extension to electromagnetic waves of the classical work by   Rayleigh for acoustic membranes \cite{strutt_2011}. Such theory was further developed by Wiersig and coworkers, but still limited to   two-dimensional resonators \cite{PhysRevA.94.043850,PhysRevA.99.063825}.

The purpose of our work is to develop a perturbation theory,  for the electromagnetic resonances of three-dimensional nearly spherical dielectric resonators. As we will see, this requires us to fully account for the vector nature of the electromagnetic field in three dimensions, which is a nontrivial technical challenge. However, although we deal with an effectively open system, our method permits us to avoid the use of quasinormal modes \cite{Muljarov10,PhysRevA.90.013834}, and similar techniques \cite{PhysRevLett.125.013901}.  We note the importance of including second-order terms in the theory. In fact, under certain conditions, first-order perturbation theory does not account for the effects  of random deformations, which typically averages to zero. However, such deformations  manifest nonzero correlations, the effects of which are always  disclosed by second-order perturbation theory \cite{PhysRevA.96.063842,PhysRevA.100.023837}. Further details on the theory, omitted here for brevity, can be found in \cite{Gohsrich20}.

The work is organized as follows. In Sec. II we establish the notation and we review the classical Mie solution \cite{Mie} for the scattering of electromagnetic waves by dielectric spheres. This is functional to the perturbation theory to be developed because the Mie solution will be taken as the zeroth-order approximation.
In Sec. III we establish the exact equations for our boundary-condition problem. From Sec. IV to Sec. VIII we thoroughly develop our degenerate  second-order perturbation theory, including the case of highly symmetric problems where the degeneracy is not lifted to first order. In Sec. IX we apply our theory, to the simple case of an oblate spheroid resonator. In Sec. X we summarize our work and draw some conclusions. Three appendixes provide  some detailed calculations.

\section{Notation and scenario}

\begin{figure}[!ht]
\centerline{\includegraphics[scale=3,clip=false,width=.8\columnwidth,trim = 0 0 0 0]{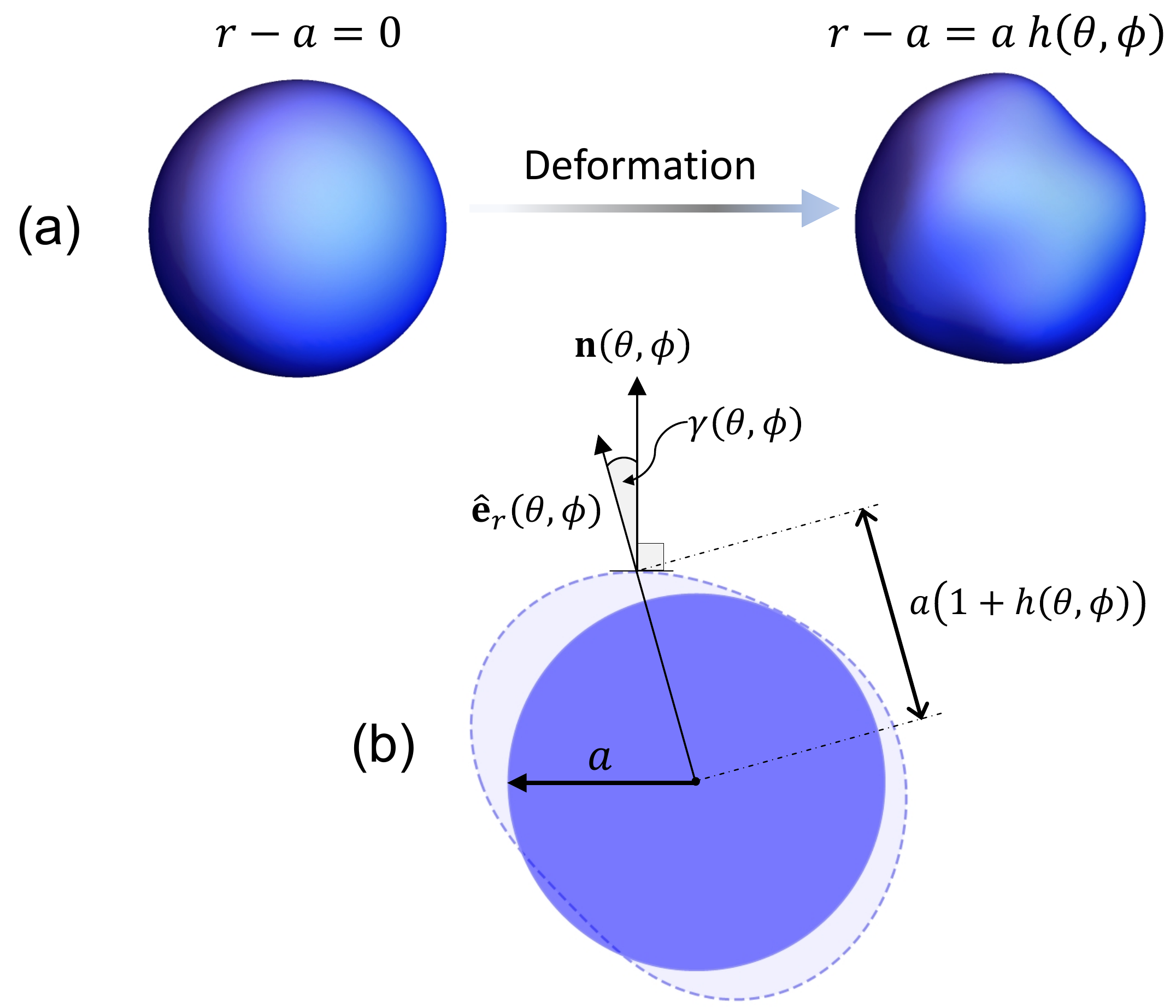}}
\caption{\label{fig0} (a) Illustration of the  dielectric spherical and nearly-spherical resonators, of equations $r-a=0$, and $r-a = a[1+h(\theta,\phi)]$, respectively, where $a$ is the radius of the sphere. (b) Geometry of a section of the spherical (dark blue) and of the nearly spherical (light blue) resonators. Here $\hat{\mathbf{e}}_r$ is the radial unit vector, and $\mathbf{n}(\theta,\phi)$ is the vector normal to the deformed surface \eqref{e400}. From \eqref{e420} it follows that $\cos \gamma = \hat{\mathbf{e}}_r \cdot \mathbf{n}/ \abs{\mathbf{n}}$.}
\end{figure}

In this section we show how to calculate the optical resonances of a dielectric sphere using the method of Debye potentials \cite{doi:10.1119/1.11111}. The sphere has radius ${a}$ and refractive index $n_1$ and it is surrounded by a medium of refractive index $n_2 < n_1$ (typically air or vacuum). Both the sphere and the surrounding medium are nonmagnetic, homogeneous and isotropic.
In the remainder we will benefit from the following definitions:
\begin{itemize}
\setlength\itemsep{-0.1truecm}
  \item $c_0$ is the speed of light in vacuum.
  \item $k_0$  is the (real- or complex-valued) wave number of light in vacuum.
  \item $c_\alpha = c_0 /n_\alpha$ is the speed of light in a medium of refractive index $n_\alpha$, with $\alpha = 1,2$.
  \item $k_\alpha = k_0 n_\alpha $ is the  wave number in a medium of refractive index $n_\alpha$, with $\alpha = 1,2$.
  \item The time-independent Debye potentials $u^E_\alpha = u^E_\alpha(\mathbf{r})$  and $u^M_\alpha = u^M_\alpha(\mathbf{r})$ are scalar fields that, in a medium of refractive index $n_\alpha$, satisfy the Helmholtz equation
\begin{align}\label{e10}
\nabla^2 u^\sigma_\alpha + k^2_\alpha u^\sigma_\alpha =0, \qquad (\sigma=E,M),
\end{align}
with $\alpha = 1,2 $ \cite{Zangwill}.
\item The orbital angular momentum differential operator is defined as
\begin{align}\label{e20}
\mathbf{L} = \frac{1}{i} \, \mathbf{r} \times \bm{\nabla}.
\end{align}

\end{itemize}
Figure \ref{fig0} illustrates our working scenario.
%
%

Without loss of generality, in this work we consider \emph{observable} monochromatic electric and magnetic fields, denoted by $\bm{E}_\alpha(\mathbf{r},t) $ and $\bm{B}_\alpha(\mathbf{r},t)$, respectively,  defined by
\begin{align}
\bm{E}_\alpha(\mathbf{r},t) = & \; \operatorname{Re} \bigl[\mathbf{E}_\alpha(\mathbf{r} ) \exp(- i \omega t) \bigr], \label{e22a} \\[2pt]
\bm{B}_\alpha(\mathbf{r},t) = & \; \operatorname{Re} \bigl[\mathbf{B}_\alpha(\mathbf{r} ) \exp(- i \omega t) \bigr], \label{e22b}
\end{align}
where $\omega = k_0 c_0 = k_1 c_1 = k_2 c_2$.
In a nonmagnetic medium of refractive index $n_\alpha$, the time-independent electric and magnetic fields $\mathbf{E}_\alpha(\mathbf{r}) $ and $\mathbf{B}_\alpha(\mathbf{r})$, can be written in terms of the two Debye potentials $u_\alpha^E(\mathbf{r} )$ and $u_\alpha^M(\mathbf{r} )$,  as \cite{doi:10.1119/1.11111,Zangwill}
\begin{subequations}\label{e30}
\begin{align}
\frac{1}{i} \mathbf{E}_\alpha(\mathbf{r} ) = & \; \left( \mathbf{L} u^E_\alpha \right) + \frac{i}{k_\alpha} \bm{\nabla} \times \left( \mathbf{L} u^M_\alpha \right),\label{e30a} \\[2pt]
\frac{c_\alpha}{i} \mathbf{B}_\alpha (\mathbf{r} ) = & \; \left( \mathbf{L} u^M_\alpha \right) - \frac{i}{k_\alpha}\bm{\nabla} \times  \left( \mathbf{L} u^E_\alpha \right). \label{e30b}
\end{align}
\end{subequations}
In the standard jargon, $u^E_\alpha$ and $u^M_\alpha$ yield T\emph{ransverse} E\emph{lectric} (TE), and T\emph{ransverse} M\emph{agnetic} (TM) waves, respectively \cite{Zangwill}. Note that from \eqref{e30} it follows that
\begin{align}\label{e112}
 c_\alpha \mathbf{B}_\alpha[u^E_\alpha,u^M_\alpha] = \mathbf{E}[u^M_\alpha,-u^E_\alpha],
\end{align}
where the square brackets denote functional dependence.

Using the completeness and orthogonality of the spherical harmonics $Y_{lm}(\theta, \phi) $ \cite{Jackson}, and the spherical coordinates $(r, \theta, \phi)$ with $r \in [0,\infty )$, $\theta \in [0,\pi ]$ and $\phi \in [0, 2 \pi )$, we can write,
\begin{subequations}\label{e50}
\begin{align}
u_\alpha^\sigma(k_\alpha r,\theta,\phi) =  & \; \sum_{l,m} U_{\alpha lm}^\sigma(k_\alpha r) Y_{lm}(\theta, \phi), \label{e50a} \\[2pt]
U_{\alpha lm}^\sigma(k_\alpha r) = & \; \int Y_{lm}^*(\theta, \phi) u_\alpha^\sigma(k_\alpha r,\theta,\phi) \, d \Omega, \label{e50b}
\end{align}
\end{subequations}
with $\sigma = E,M$ and $\alpha =1,2$.   Here and hereafter
\begin{align}\label{w30}
\sum_{l,m} \qquad \text{is shorthand  for} \qquad \sum_{l=0}^\infty\sum_{m=-l}^l,
\end{align}
and for any smooth function $s(\theta,\phi) $,
\begin{align}\label{short1}
\int s(\theta,\phi) \, d \Omega = \int_0^{2\pi} \! \left[ \int_0^\pi  s(\theta,\phi) \sin \theta \, d \theta \right] d \phi .
\end{align}
The radial dependence of the form $k_\alpha r$, is a direct consequence of \eqref{e30}.
Substituting \eqref{e50} into  \eqref{e30}, we obtain
\begin{subequations}\label{e70}
\begin{align}
\mathbf{E}_\alpha = & \; \sum_{l,m} \Biggl\{ U^E_{\alpha l m}(k_\alpha r) \bm{\Phi}_{lm}(\theta, \phi)
\nonumber \\[0pt]
& \phantom{\sum_{l,m} \Biggl\{} - \frac{i}{k_\alpha  r} \biggl[
l(l+1) U^M_{\alpha l m}(k_\alpha r) \mathbf{Y}_{lm}(\theta, \phi)\nonumber \\[0pt]
 & \phantom{\sum_{l,m} \Biggl\{} +
\left[ (k_\alpha  r)  U^M_{\alpha lm}(k_\alpha r) \right]'\bm{\Psi}_{lm}(\theta, \phi)
 \biggr]
 \Biggr\},\label{e70a} \\[2pt]
 c_\alpha \mathbf{B}_\alpha = & \; \sum_{l,m}\Biggl\{
 U^M_{\alpha lm}(k_\alpha r) \bm{\Phi}_{lm}(\theta, \phi) \nonumber \\[0pt]
 & \phantom{\sum_{l,m} \Biggl\{} + \frac{i}{k_\alpha r} \biggl[
 l(l+1) U^E_{\alpha lm}(k_\alpha r) \mathbf{Y}_{lm}(\theta, \phi) \nonumber \\[0pt]
 & \phantom{\sum_{l,m} \Biggl\{}+
\left[ (k_\alpha  r)  U^E_{\alpha lm}(k_\alpha r) \right]' \bm{\Psi}_{lm}(\theta, \phi)
 \biggr]
\Biggr\}, \label{e70b}
\end{align}
\end{subequations}
where the prime denotes the derivative with respect to the argument $k_\alpha r$, and
the three vector spherical harmonics $\mathbf{Y}_{lm}(\theta, \phi)$, $\bm{\Psi}_{lm}(\theta, \phi)  $ and $\bm{\Phi}_{lm}(\theta, \phi) $ are defined as \cite{Carrascal_1991}:
\begin{subequations}\label{e90}
\begin{align}
\mathbf{Y}_{lm}(\theta, \phi) = & \; \hat{\bf{e}}_r \, Y_{lm}(\theta, \phi),\label{e90a} \\[2pt]
\bm{\Psi}_{lm}(\theta, \phi) = & \; r \, \bm{\nabla}  Y_{lm}(\theta, \phi), \label{e90b} \\[2pt]
\bm{\Phi}_{lm}(\theta, \phi) = & \; \hat{\bf{e}}_r \times \bm{\Psi}_{lm}(\hat{\bf{e}}_r). \label{e90c}
\end{align}
\end{subequations}
with  $\hat{\bf{e}}_r = \mathbf{r}/r$.

In  general, the functions $U^E_{\alpha lm}(k_\alpha r)$ and $U^M_{\alpha lm}(k_\alpha r)$ are expressible as linear combinations of spherical Bessel functions $j_l(k_\alpha r)$, $h_l^{(1)}(k_\alpha r)$, and $h_l^{(2)}(k_\alpha r)$, where $j_l(k_\alpha r)$ is finite at $r=0$, and $h_l^{(1)}(k_\alpha r)$ and $h_l^{(2)}(k_\alpha r)$ describe outgoing and ingoing spherical waves, respectively, for  $r \to \infty$ (see, e.g., Appendix A of \cite{GalindoI}).
However, the electric and magnetic fields inside the sphere must be finite everywhere for $0 \leq r \leq {a}$. Moreover, we assume that the field outside the sphere is made of outgoing waves only. This implies that we can write the radial parts of the four Debye potentials $u^E_1(\mathbf{r} ), u^M_1(\mathbf{r} )$ and $u^E_2(\mathbf{r} ), u^M_2(\mathbf{r} )$, in the two media as
\begin{align}\label{e160}
U^\sigma_{\alpha lm}(k_\alpha r) =  a^\sigma_{\alpha lm} R^\sigma_{\alpha l}(k_\alpha r) ,\quad (\sigma = E,M),
\end{align}
where the radial functions
\begin{subequations}\label{eRadial}
\begin{align}
R^E_{\alpha l}(k_\alpha r)  = & \; \frac{b_{\alpha l}(k_\alpha r)}{b_{\alpha l}(k_\alpha a)} ,\label{eRadialA} \\[6pt]
R^M_{\alpha l}(k_\alpha r)  = & \; \frac{b_{\alpha l}(k_\alpha r)}{n_\alpha b_{\alpha l}(k_\alpha a)} , \label{eRadialB}
\end{align}
\end{subequations}
have been defined in terms of the spherical Bessel functions for the fields in media $1$ and $2$,
renamed as
\begin{align}\label{funcB}
b_{1 l} \left( z \right) = j_l(z), \qquad \text{and} \qquad b_{2 l} \left( z \right) = h_l^{(1)}(z).
\end{align}
The  choice of the denominators in \eqref{eRadial}  just fixes an arbitrary normalization which could be absorbed into the definition of the coefficients $a^\sigma_{\alpha lm}$.

Substituting \eqref{e160} into \eqref{e70} we obtain, after a straightforward calculation,
\begin{subequations}\label{p5}
\begin{align}
\mathbf{E}_\alpha(r, \theta, \phi)  = &   \sum_{l,m} \biggl\{
\frac{a^M_{\alpha l m}}{n_\alpha} \Bigl[ F^Y_{\alpha l}(k_\alpha r) \mathbf{Y}_{lm}(\theta,\phi) \nonumber \\[0pt]
& \phantom{  \sum_{l,m} \biggl\{} + F^\Psi_{\alpha l}(k_\alpha r) \mathbf{\Psi}_{lm}(\theta,\phi) \Bigr] \nonumber \\[0pt]
& \phantom{ \sum_{l,m} \biggl\{} + a^E_{\alpha l m} \, F^\Phi_{\alpha l}(k_\alpha r) \mathbf{\Phi}_{lm}(\theta,\phi)
\biggr\} ,\label{p10} \\[0pt]
c_\alpha \mathbf{B}_\alpha(r, \theta, \phi) = &   \sum_{l,m} \biggl\{-a^E_{\alpha l m}
\Bigl[ F^Y_{\alpha l}(k_\alpha r) \mathbf{Y}_{lm}(\theta,\phi) \nonumber \\[0pt]
& \phantom{  \sum_{l,m} \biggl\{} + F^\Psi_{\alpha l}(k_\alpha r) \mathbf{\Psi}_{lm}(\theta,\phi) \Bigr] \nonumber \\[0pt]
& \phantom{ \sum_{l,m} \biggl\{} +\frac{a^M_{\alpha l m}}{n_\alpha} \, F^\Phi_{\alpha l}(k_\alpha r) \mathbf{\Phi}_{lm}(\theta,\phi)
\biggr\} , \label{p20}
\end{align}
\end{subequations}
where we have defined
\begin{subequations}\label{e560}
\begin{align}
F^Y_{\alpha  l}(k_\alpha r) = & \;  \frac{1}{i} \, l(l+1) \frac{1}{(k_\alpha r) } \frac{b_{\alpha l} \bigl( k_\alpha r \bigr)}{b_{\alpha l} ( k_\alpha a )} ,\label{e560a} \\[0pt]
F^\Psi_{\alpha l} (k_\alpha r)  = & \; \frac{1}{i}  \, \frac{\left[ \bigl( k_\alpha r \bigr) b_{\alpha l} \bigl( k_\alpha r \bigr)\right]'}{(k_\alpha r) b_{\alpha l} ( k_\alpha a )}, \label{e560b} \\[0pt]
F^\Phi_{\alpha l}(k_\alpha r)  = & \; \frac{b_{\alpha l} \bigl( k_\alpha r \bigr)}{b_{\alpha l} ( k_\alpha a )}  . \label{e560c}
\end{align}
\end{subequations}

\begin{figure}[!hb]
\centerline{\includegraphics[scale=3,clip=false,width=1\columnwidth,trim = 0 0 0 0]{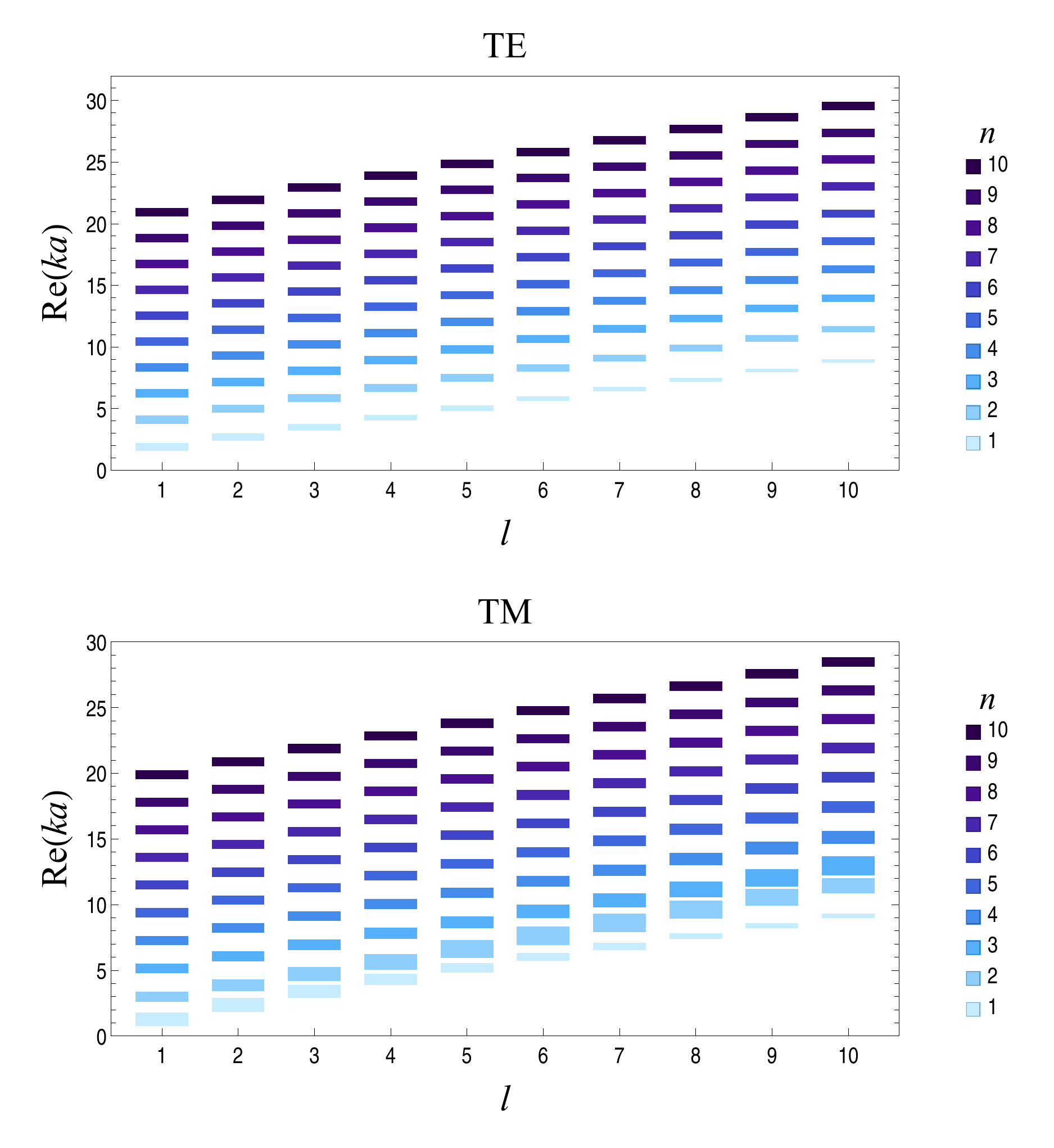}}
\caption{\label{fig1}  Spectrum of the TE and TM resonances of a dielectric sphere of radius $a$ and refractive index $n_1=1.5$, surrounded by vacuum with $n_2=1$. The values of $x_{ln} = k_{ln} a $ for $1 \leq n \leq 10$ and  $1 \leq l \leq 10$ are shown as blue bands. The vertical position of the center of each band  is equal to $\operatorname{Re} (k_{ln} a)$, and
the thickness is equal to $\operatorname{Im} (k_{ln} a)$. For the first radial mode $n=1$ (lighter blue bands) the imaginary part of $k_{nl} a$ quickly decreases as $l$ increases from left to right.  Each resonance characterized by the pair of azimuthal and radial numbers $l$ and $n$ is $2l+1$ times degenerate.}
\end{figure}

The numerical coefficients $a^E_{1lm}, a^E_{2lm}$, and  $a^M_{1lm}, a^M_{2lm}$, are determined by imposing the electromagnetic boundary conditions on the surface of the sphere \cite{Jackson}:
\begin{subequations}\label{e180}
\begin{align}
\hat{\bf{e}}_r \times \left. \left( \mathbf{E}_1 - \mathbf{E}_2 \right) \right|_{r={a}} = & \; 0 ,\label{e180a} \\[2pt]
\hat{\bf{e}}_r \times \left.  \left( \mathbf{B}_1 - \mathbf{B}_2 \right)\right|_{r={a}}  = & \; 0. \label{e180b}
\end{align}
\end{subequations}
It is not difficult to see that using the relations
\begin{align}
\hat{\bf{e}}_r \times \mathbf{Y}_{lm}(\theta, \phi) = & \; 0,\label{e200} \\[2pt]
\hat{\bf{e}}_r \times \bm{\Psi}_{lm}(\theta, \phi) = & \; \bm{\Phi}_{lm}(\theta, \phi), \label{e210} \\[2pt]
\hat{\bf{e}}_r \times \bm{\Phi}_{lm}(\theta, \phi) = & \; -\bm{\Psi}_{lm}(\theta, \phi), \label{e220}
\end{align}
we can rewrite \eqref{e180a} as
\begin{align}\label{e230}
0 = & \; \sum_{l,m}  \left( a^E_{1lm} - a^E_{2lm} \right) \bm{\Psi}_{lm}(\theta, \phi) \nonumber \\[0pt]
& + \frac{i}{k_0 {a}} \sum_{l,m}  \Biggl\{ a^M_{1lm} \frac{\bigl[(k_1 {a}) j_l(k_1{a}) \bigr]'}{n_1^2 j_l(k_1{a})} \nonumber \\[0pt]
& - a^M_{2lm} \frac{\bigl[(k_2{a}) h_l^{(1)}(k_2{a}) \bigr]'}{n_2^2 h_l^{(1)}(k_2{a})} \Biggr\} \bm{\Phi}_{lm}(\theta, \phi),
\end{align}
and \eqref{e180b} as:
\begin{align}\label{e240}
0 = & \; \sum_{l,m}  \left( a^M_{1lm} - a^M_{2lm} \right) \bm{\Psi}_{lm}(\theta, \phi) \nonumber \\[0pt]
& - \frac{i}{k_0 {a}} \sum_{l,m}  \Biggl\{ a^E_{1lm} \frac{\bigl[(k_1 {a}) j_l(k_1{a}) \bigr]'}{ j_l(k_1{a})} \nonumber \\[0pt]
& - a^E_{2lm} \frac{\bigl[(k_2{a}) h_l^{(1)}(k_2{a}) \bigr]'}{ h_l^{(1)}(k_2{a})} \Biggr\} \bm{\Phi}_{lm}(\theta, \phi),
\end{align}
where again the prime  denotes the derivative with respect to the argument. From the orthogonality of the vector spherical harmonics it follows that each term in Eqs. \eqref{e230} and \eqref{e230} that multiplies a vector spherical harmonics must be set equal to zero separately. So the first lines of \eqref{e230} and \eqref{e240} gives
\begin{equation}\label{e250}
a^E_{1lm} = a^E_{2lm} \equiv a^E_{lm},
\end{equation}
and
\begin{equation}\label{e260}
a^M_{1lm} = a^M_{2lm} \equiv a^M_{lm},
\end{equation}
respectively. Substituting \eqref{e250} in \eqref{e240}, we obtain
\begin{align}\label{e270}
 \frac{\bigl[(k_1 {a}) j_l(k_1 {a}) \bigr]'}{ j_l(k_1 {a})} - \frac{\bigl[(k_2 {a}) h_l^{(1)}(k_2 {a}) \bigr]'}{ h_l^{(1)}(k_2 {a})}   = 0,
\end{align}
for TE waves. Similarly, substituting \eqref{e260} in  \eqref{e230}, we find
\begin{align}\label{e280}
 \frac{\bigl[(k_1 {a}) j_l(k_1 {a}) \bigr]'}{n_1^2 \, j_l(k_1 {a})} - \frac{\bigl[(k_2 {a}) h_l^{(1)}(k_2 {a}) \bigr]'}{n_2^2 \, h_l^{(1)}(k_2 {a})}   = 0,
\end{align}
for TM waves. Both Eqs. \eqref{e270} and \eqref{e280} are characterized by the index $l$, so  for each value of $l$ there will be a different set of solutions. To find these solutions,  we write  $k_1 {a} = k_0 {a} n_1 \equiv x n_1$ and $k_2 {a} = k_0 {a} n_2 \equiv x n_2$ in \eqref{e270} and \eqref{e280}, where the dimensionless wave number $x$ is defined as $x \equiv k_0 {a}$. Then, we introduce the compact notation (the irrelevant prefactor $x/i$ is introduced for later notational convenience)
\begin{align}
\frac{x}{i} f_l^{E}(x) \equiv  \frac{\bigl[(n_1 x) j_l(n_1 x) \bigr]'}{ j_l(n_1 x)} - \frac{\bigl[(n_2 x) h_l^{(1)}(n_2 x) \bigr]'}{ h_l^{(1)}(n_2 x)}  \label{e280TE}, \\[0pt]
\frac{x}{i} f_l^{M}(x) \equiv  \frac{\bigl[(n_1 x) j_l(n_1 x) \bigr]'}{n_1^2 \, j_l(n_1 x)} - \frac{\bigl[(n_2 x) h_l^{(1)}(n_2 x) \bigr]'}{n_2^2 \, h_l^{(1)}(n_2 x)}   \label{e280TM},
\end{align}
and we solve (numerically) the two transcendental equations
\begin{align}\label{e282}
f_l^{E}(x) =0, \qquad \text{and} \qquad f_l^{M}(x) = 0,
\end{align}
with respect to $x$ to find the resonant wave numbers for both TE and TM waves. Thus, we obtain two countably infinite sets of solutions denoted by
\begin{align}\label{e284}
\{x_{ln}^{E} \} = \{x_{l1}^{E}, x_{l2}^{E}, \cdots \},
\end{align}
and
\begin{align}\label{e284bis}
\{x_{ln}^{M} \} = \{x_{l1}^{M}, x_{l2}^{M}, \cdots \},
\end{align}
with $x_{ln}^{E},x_{ln}^{M} \in \mathbb{C}$, such that
\begin{align}\label{e282bis}
f_l^{\sigma}(x_{ln}^{\sigma}) =0, \qquad (n = 1,2, \cdots ,) \, ,
\end{align}
with $\sigma = E,M$.
In the remainder, we will refer to \eqref{e284} and \eqref{e284bis}, as the unperturbed spectrum.
A portion of the spectrum of TE and TM resonances of a dielectric sphere with refractive index $n_1=1.5$, surrounded by vacuum with $n_2=1$, is shown in Fig. \ref{fig1}.
%
%

 Equations \eqref{e282} depend on the index $l$ but not on $m$. This implies that for each solution $x_{ln}^{\sigma}$, with $\sigma,l,n$ assigned,  there are $2l+1$ different Debye potentials, denoted by
\begin{align}\label{eDebye0}
\bigl\{ u^\sigma_{\alpha | l m n}(\mathbf{r})  \bigr\} = \bigl\{ u^\sigma_{\alpha |l , -l n}(\mathbf{r}), \cdots, u^\sigma_{\alpha| l, l  n}(\mathbf{r})  \bigr\},
\end{align}
and defined by
\begin{align}\label{eDebye1}
u^\sigma_{\alpha| l m n}(\mathbf{r}) =  R^\sigma_{\alpha l}(n_\alpha x^\sigma_{ln} r/a)  Y_{lm}(\theta,\phi),
\end{align}
($m = -l, -l+1, \ldots, l$),  such that the electric and magnetic fields obtained from \eqref{e30} with $u^E_\alpha= u^E_{\alpha| l m n}(\mathbf{r}) $ and $u^M_\alpha = u^M_{\alpha| l m n}(\mathbf{r})$ will automatically satisfy the boundary conditions \eqref{e180}. The radial functions are still defined by \eqref{eRadial}.

By construction, from the $(2l+1)$-fold degeneracy of $x_{ln}^{\sigma}$ it follows that any linear combination of the form
\begin{multline}\label{eLinComb}
\sum_{m=-l}^l a^\sigma_{lm} u^\sigma_{\alpha| l m n}(\mathbf{r})  \\
 = R^\sigma_{\alpha l}(n_\alpha x^\sigma_{ln} r/a) \sum_{m=-l}^l a^\sigma_{lm} Y_{lm}(\theta,\phi),
\end{multline}
where $a^\sigma_{lm}$ are arbitrary numerical coefficients,
is still an admissible Debye potential associated with the same eigenvalue $x_{ln}^{\sigma}$. Evidently, for a given $x_{ln}^{\sigma}$, it is possible to build $2l+1$ linearly independent of such combinations.  We will make  extensive use of this property when developing a degenerate perturbation theory.

\section{Perturbation of the boundaries}

Now we turn to the more general problem of a dielectric electromagnetic resonator of refractive index $n_1$, surrounded by a medium of refractive index $n_2 < n_1$. Again, both the resonator and the surrounding media are nonmagnetic, homogeneous, and isotropic. The expressions \eqref{e30}-\eqref{e50} and \eqref{e160}-\eqref{e560} for the fields and Debye potentials are perfectly general, so they remain valid also in the present case. What will change are the boundary conditions \eqref{e180} that will be replaced by \eqref{e380}, defined later.

\subsection{Describing the deformation}\label{conditions}

Consider a nearly spherical dielectric resonator, the surface of which can be described by the equation
\begin{align}\label{e290}
F(r,\theta,\phi)=0,
\end{align}
where
\begin{align}\label{e300}
F(r,\theta,\phi)= (r - {a}) - {a} \, h(\theta,\phi),
\end{align}
with $h(\theta,\phi)$  an arbitrary, smooth, single-valued function of $\theta$ and $\phi$ defined on the unit sphere $S$, which describes the deformation of the resonator. The slight departure from the spherical shape is guaranteed by any function $h(\theta,\phi)$, the maximum value of which   is much less than $1$ on $S$:
\begin{align}\label{e310}
\max\{ \abs{h(\theta,\phi)} \}_S \ll 1.
\end{align}
However, as we will see soon, this is not the only condition required for the applicability of the theory. As usual in perturbation theory, it is useful to introduce a formal parameter $0 \leq \varepsilon \ll 1$ defined by
\begin{align}\label{e314}
h(\theta,\phi) = & \; \ve f(\theta,\phi),
\end{align}
where $\abs{f(\theta,\phi)} \leq 1$. This parameter is just a mathematical device used to rewrite $h(\theta,\phi)$ in a more convenient form for later developments. At the end of the calculations we will restore the physical deviation $h(\theta,\phi)$ by replacing everywhere $\ve f(\theta,\phi)$ with  $h(\theta,\phi)$.
With this definition \eqref{e300} becomes
\begin{align}\label{e316}
F(r,\theta,\phi)= (r - {a}) - {a} \, \ve f(\theta,\phi).
\end{align}

The standard electromagnetic boundary conditions on the surface of the dielectric body can now be written as:
\begin{subequations}\label{e380}
\begin{align}
\mathbf{n} \times \left. \left( \mathbf{E}_1 - \mathbf{E}_2 \right) \right|_{r={a}(1+\ve f)} = & \; 0 ,\label{e380a} \\[8pt]
\mathbf{n} \times \left.  \left( \mathbf{B}_1 - \mathbf{B}_2 \right)\right|_{r={a}(1+ \ve f)}  = & \; 0, \label{e390}
\end{align}
\end{subequations}
where $f = f(\theta,\phi)$, and the vector $\mathbf{n} = \mathbf{n}(\theta,\phi) = \bm{\nabla} F(r,\theta, \phi)$ normal to the surface of the dielectric at $r = a \bigl[ 1 + \ve f (\theta,\phi)\bigr]$, is given by
\begin{align}\label{e400}
\mathbf{n}(\theta,\phi) = & \; \hat{\bf{e}}_r - \mathbf{n}_\parallel(\theta,\phi)\nonumber \\[8pt]
= & \; \hat{\bf{e}}_r -  \frac{\ve}{1 + \ve f(\theta,\phi)} \, \mathbf{e}_\parallel (\theta,\phi),
\end{align}
where
\begin{align}\label{e410}
\mathbf{e}_\parallel (\theta,\phi) =  \hat{\bf{e}}_\theta \, \frac{\partial f(\theta,\phi)}{\partial \theta} + \hat{\bf{e}}_\phi \frac{1}{ \sin \theta} \frac{\partial f(\theta,\phi)}{\partial \phi} .
\end{align}

Let $\mathbf{\Delta}=\mathbf{\Delta}(r,\theta,\phi)$ denote either $\mathbf{E}_1 - \mathbf{E}_2$ or  $\mathbf{B}_1 - \mathbf{B}_2$.
Then, we can rewrite the boundary conditions \eqref{e380} in the  suggestive form
\begin{align}\label{e440}
0 = & \; \mathbf{n} \times \bm{\Delta} \left( {a} + {a} \, \ve f,\theta,\phi \right) \nonumber \\[8pt]
 = & \; \hat{\bf{e}}_r  \times \bm{\Delta} \left( {a} ,\theta,\phi \right)   \nonumber \\[8pt]
  & + \Bigl\{\hat{\bf{e}}_r  \times \bigl[
\bm{\Delta} \bigl( {a}  + {a}  \, \ve f,\theta,\phi \bigr) - \bm{\Delta} \left( {a} ,\theta,\phi \right)
\bigr] \nonumber \\[8pt]
  & \phantom{ + \Bigl\{} - \mathbf{n}_\parallel \times \bm{\Delta} \bigl( {a}  + {a}  \, \ve f  ,\theta,\phi \bigr) \Bigr\},
\end{align}
where $f = f(\theta,\phi)$. This expression is \emph{exact};as  no approximations have been done up to now. However, we have written it in such a way to isolate the unperturbed first term $ \hat{\bf{e}}_r  \times \bm{\Delta} \left( {a},\theta,\phi \right)$, coincident with (\ref{e180}), from the perturbed second term delimited by curly brackets.
To develop a meaningful perturbation theory we require this second term to be $ O(\ve) $ with respect to $ \hat{\bf{e}}_r  \times \bm{\Delta} \left( {a},\theta,\phi \right)$. This is certainly true for the tangential part $\hat{\bf{e}}_r  \times \bigl[
\bm{\Delta} \left( {a} + {a} \, \ve f,\theta,\phi \right) - \bm{\Delta} \left( {a},\theta,\phi \right)
\bigr]$ because by definition
\begin{align}\label{e450}
\bm{\Delta}  \left( {a} + {a} \, \ve f,\theta,\phi \right) - \bm{\Delta}  \left( {a},\theta,\phi \right) = O(\ve) .
\end{align}
However, for the radial part we have
\begin{align}\label{e460}
\mathbf{n}_\parallel \times & \bm{\Delta} \left( a + a \, \ve f,\theta,\phi \right) \sim |\mathbf{n}_\parallel| \bigl[ 1 + O(\ve) \bigr] ,
\end{align}
and $ |\mathbf{n}_\parallel|$ is potentially unbounded. To show this, suppose that, for example, in the neighborhood of the direction $(\theta, \phi)$ the deformation of the resonator could be described by
\begin{align}\label{e470}
h(\theta,\phi) = \ve \sin (p \, \theta) \ll 1,
\end{align}
with $p >0$.
This implies that
\begin{align}\label{e490}
|\mathbf{n}_\parallel| = & \;  \frac{\ve \, p \abs{\cos (p \, \theta)}}{1 + \ve \sin (p \, \theta)} \nonumber \\[8pt]
 = & \; \ve \, p \abs{\cos (p \, \theta)} \bigl[ 1 - \ve \sin (p \, \theta)\bigr] + O(\ve^3).
\end{align}
Clearly, $\ve \, p$ can be of the order of unity or bigger if $p \geq 1/\ve$.
In this case the angle $ \gamma(\theta,\phi)$ between the vector $\hat{\bf{e}}_r $ normal  to the unit sphere $S$ along the direction $(\theta,\phi)$, and the vector $\mathbf{n}(\theta,\phi)$ normal to the deformed sphere in the same direction, defined by
\begin{align}\label{e420}
\gamma(\theta,\phi) =   \arctan |\mathbf{n}_\parallel|=   \arctan  \abs{\bm{\nabla} h(\theta,\phi) }_S ,
\end{align}
can be arbitrarily close to $\pi/2$. When this occurs,  we have the so-called \emph{strongly winding} boundaries  \cite{PhysRevA.99.063825}. If this is not the case, then we have \emph{weakly winding} boundaries.   Through this work, we assume that the latter condition is \emph{always} verified.

Thus,  Eqs. \eqref{e310} and  \eqref{e460} imply that the \emph{physical conditions} for the applicability of the perturbation theory, require that the magnitude of the deformation function $h(\theta,\phi)$ and its gradient $\bm{\nabla} h(\theta,\phi)$, both evaluated on the unit sphere $S$,  must be $O(\ve)$, namely
\begin{align}\label{e422}
\max\{ \abs{h(\theta,\phi)} \}_S \approx \max\{\abs{\bm{\nabla} h(\theta,\phi) } \}_S \ll 1.
\end{align}

\subsection{Developments}

To proceed further, we must rewrite \eqref{e380} in a more convenient form. Note, however, that in this section we will not assume the smallness of the deformation as in \eqref{e310}.
Following Sec. II of \cite{PhysRev.179.1238}, we want to show that of the six equations \eqref{e380}, only four are independent. Let us write $ \mathbf{\Delta} = \Delta_r \hr + \Delta_\theta \hte + \Delta_\phi \hf \equiv $ $\Delta_r \hr + \mathbf{\Delta}_\parallel$, and $\mathbf{n} = \hr - n_\theta \hte - n_\phi \hf \equiv$ $ \hr - \mathbf{n}_\parallel$, where, $\mathbf{\Delta}$ denotes either $\mathbf{E}_1 - \mathbf{E}_2$ or  $\mathbf{B}_1 - \mathbf{B}_2$, evaluated at $r = a(1+\ve f)$. Calculating  $\mathbf{n} \times \mathbf{\Delta} = 0$ we obtain the three equations
\begin{subequations}
\begin{align}
n_\theta \Delta_\phi - n_\phi \Delta_\theta   = & \; 0 ,\label{e505a} \\[8pt]
n_\phi \Delta_r + \Delta_\phi   = & \; 0 ,\label{e505b} \\[8pt]
n_\theta \Delta_r + \Delta_\theta    = & \; 0 .\label{e505c}
\end{align}
\end{subequations}
It is easy to see that when \eqref{e505b} and \eqref{e505c} hold true, then \eqref{e505a} is automatically satisfied, because
\begin{align}\label{e506}
n_\theta \Delta_\phi - n_\phi \Delta_\theta = & \; n_\theta \bigl( n_\phi \Delta_r + \Delta_\phi \bigr)  \nonumber \\[6pt]
& -  n_\phi  \bigl( n_\theta \Delta_r + \Delta_\theta  \bigr) .
\end{align}
Therefore, choosing \eqref{e505b} and \eqref{e505c} as independent equations, and multiplying  \eqref{e505b} by $\hf$ and \eqref{e505c} by $\hte$, we can rewrite the four independent boundary conditions \eqref{e505b} and \eqref{e505c} as
\begin{subequations}\label{e510pre}
\begin{align}
 \left( \mathbf{E}_{1 \parallel} - \mathbf{E}_{2 \parallel} \right) + \left(E_{1r} - E_{2r} \right) \mathbf{n}_{\parallel}  = & \; 0 ,\label{e510} \\[8pt]
 \left( \mathbf{B}_{1 \parallel} - \mathbf{B}_{2 \parallel} \right) + \left(B_{1r} - B_{2r} \right) \mathbf{n}_{\parallel} = & \; 0, \label{e520}
\end{align}
\end{subequations}
where all the fields are evaluated  at $r = a \bigl[ 1+\ve f(\theta,\phi) \bigr]$.

We can expand Eqs. \eqref{e510pre}  in terms of $\bm{\Psi}_{l'm'}(\theta, \phi)$ and $\bm{\Phi}_{l'm'}(\theta, \phi)$ solely, the radial components being absent, to obtain
\begin{subequations}\label{w10pre}
\begin{align}
\sum_{l',m'} \Bigl[ \Psi^E_{l' m'} \bm{\Psi}_{l' m'}(\theta, \phi) + \Phi^E_{l' m'} \bm{\Phi}_{l' m'}(\theta, \phi) \Bigr] = & \; 0 ,\label{w10} \\[8pt]
\sum_{l',m'} \Bigl[ \Psi^B_{l' m'} \bm{\Psi}_{l' m'}(\theta, \phi) + \Phi^B_{l' m'} \bm{\Phi}_{l' m'}(\theta, \phi) \Bigr] = & \; 0, \label{w20}
\end{align}
\end{subequations}
respectively, where
\begin{subequations}\label{w40}
\begin{align}
X^E_{l' m'}(x) = & \; \frac{1}{l'(l'+1)} \int \! \mathbf{X}_{l' m'}^* (\theta,\phi) \cdot \Bigl[   \mathbf{E}_{1 \parallel} - \mathbf{E}_{2 \parallel} \nonumber \\[6pt]
&  + \left(E_{1r} - E_{2r} \right) \mathbf{n}_{\parallel} \Bigr] d \Omega , \label{w40a} \\[8pt]
X^B_{l' m'}(x) = & \; \frac{1}{l'(l'+1)}  \int \! \mathbf{X}_{l' m'}^* (\theta,\phi)\cdot \Bigl[  \mathbf{B}_{1 \parallel} - \mathbf{B}_{2 \parallel}\nonumber \\[6pt]
&   + \left( B_{1r} - B_{2r} \right) \mathbf{n}_{\parallel} \Bigr] d \Omega , \label{w40b}
\end{align}
\end{subequations}
with $X = \Psi, \Phi$.
For reasons that will soon  be clear, on the left-hand sides of \eqref{w40}, we have made explicit the dependence on $x = k_0 a$. This arises from the radial dependence of the fields evaluated on the surface of the dielectric resonator:
\begin{align}\label{eKappa}
k_\alpha r \Bigl|_{r = a\left( 1 + \ve f(\theta, \phi) \right)} = & \;  k_\alpha a\bigl[ 1 + \ve f(\theta, \phi) \bigr] \nonumber \\[8pt]
 = & \;  x \, n_\alpha \bigl[ 1 + \ve f(\theta, \phi) \bigr].
\end{align}
 After integration with respect to the angular variables $\theta$ and $\phi$ in \eqref{w40},  we are left with the dependence on $x$ only.

Formally, at this point our problem is perfectly posed: All what we have to do is  determine the values of $x$ (namely, the resonant wave numbers) such that the four equations
\begin{subequations}\label{w60}
\begin{align}
\Psi^E_{l' m'}(x)  =  0 , \qquad \Phi^E_{l' m'}(x)   = 0 ,\label{w60a} \\[6pt]
\Psi^B_{l' m'}(x)  =  0 , \qquad \Phi^B_{l' m'}(x) =  0 ,\label{w60b}
\end{align}
\end{subequations}
possess nontrivial solutions for the coefficients $a^E_{1lm}, a^E_{2lm}$ and $a^M_{1lm}, a^M_{2lm}$. These coefficients enter in \eqref{w40}, via the expressions of the electric and magnetic fields written in terms of the four Debye potentials  $u^E_1(\mathbf{r} ), u^M_1(\mathbf{r} )$ and $u^E_2(\mathbf{r} ), u^M_2(\mathbf{r} )$, defined by \eqref{e50} and \eqref{e160}.

Needless to say, solving the system of nonlinear algebraic equations \eqref{w60},  is a formidable  task.
However, in principle, the way to proceed is direct: Substituting \eqref{p10} and \eqref{p20} into \eqref{w40}, after some long but straightforward calculations we can write explicitly the four equations \eqref{w60} as
\begin{align}
\Psi^E_{l' m'}(x) = & \; \sum_{l,m} \biggl\{
 [A_1^\Phi]_{lm}^{l'm'} a^E_{1lm} -  [A_2^\Phi]_{lm}^{l'm'} a^E_{2lm} \nonumber \\[4pt]
& + \frac{[B_1^\Phi]_{lm}^{l'm'}}{n_1} a^M_{1lm} -  \frac{[B_2^\Phi]_{lm}^{l'm'}}{n_2} a^M_{2lm}  \biggr\}
   = 0, \label{e670A}
\end{align}
\begin{align}
\Psi^B_{l' m'}(x) = & \;   \sum_{l,m} \biggl\{
 -n_1 \, [B_1^\Psi]_{lm}^{l'm'} \, a^E_{1lm} +  n_2 \,[B_2^\Psi]_{lm}^{l'm'}  \, a^E_{2lm} \nonumber \\[4pt]
& + [A_1^\Psi]_{lm}^{l'm'} \, a^M_{1lm} -  [A_2^\Psi]_{lm}^{l'm'}  \, a^M_{2lm} \biggr\}
   = 0, \label{e670B}
\end{align}
\begin{align}
\Phi^E_{l' m'}(x) = & \;   \sum_{l,m} \biggl\{
 [A_1^\Psi]_{lm}^{l'm'} a^E_{1lm} -  [A_2^\Psi]_{lm}^{l'm'} a^E_{2lm} \nonumber \\[4pt]
& + \frac{[B_1^\Psi]_{lm}^{l'm'}}{n_1} a^M_{1lm} -  \frac{[B_2^\Psi]_{lm}^{l'm'}}{n_2} a^M_{2lm}  \biggr\}  = 0,  \label{e670C}
\end{align}
\begin{align}
\Phi^B_{l' m'}(x)  = & \;  \sum_{l,m} \biggl\{
 -n_1 \,[B_1^\Phi]_{lm}^{l'm'} \, a^E_{1lm} +  n_2 \,[B_2^\Phi]_{lm}^{l'm'}  \, a^E_{2lm} \nonumber \\[4pt]
& + [A_1^\Phi]_{lm}^{l'm'} \, a^M_{1lm} -  [A_2^\Phi]_{lm}^{l'm'}  \, a^M_{2lm} \biggr\}
   = 0 , \label{e670D}
\end{align}
where we have defined the matrix elements of types $A$ and  $B$ as, respectively,
\begin{subequations}\label{e640}
\begin{align}
[A_\alpha^X]_{lm}^{l'm'}(x) = &  \; \frac{1}{l'(l'+1)} \int \mathbf{X}_{l'm'}^*(\theta,\phi) \cdot \mathbf{A}_{\alpha lm} \, d \Omega , \label{e640A} \\[8pt]
[B_\alpha^X]_{lm}^{l'm'}(x) = & \; \frac{1}{l'(l'+1)}  \int \mathbf{X}_{l'm'}^*(\theta,\phi) \cdot \mathbf{B}_{\alpha lm} \, d \Omega \label{e640B} ,
\end{align}
\end{subequations}
with $\alpha=1,2$ and $X= \Psi, \Phi$. In \eqref{e640} we have defined
\begin{subequations}\label{e630}
\begin{align}
\mathbf{A}_{\alpha lm} \equiv & \; \mathbf{F}_{\alpha lm}^\Phi(x,\theta,\phi)  , \label{e630A} \\[8pt]
\mathbf{B}_{\alpha lm}\equiv & \; \mathbf{F}_{\alpha lm}^\Psi(x,\theta,\phi) \nonumber \\[8pt]
& + \mathbf{e}_\parallel(\theta,\phi)  Y_{lm}(\theta,\phi) B^Y_{\alpha l}(x,\theta,\phi) \label{e630B} ,
\end{align}
\end{subequations}
where $\mathbf{e}_\parallel(\theta,\phi)$ is defined by \eqref{e410},
\begin{align}\label{e630bis}
\mathbf{F}_{\alpha lm}^X(x,\theta,\phi) \equiv  \mathbf{X}_{lm}(\theta,\phi)F^X_{\alpha l}(x,\theta,\phi), \; (X= \Psi, \Phi),
\end{align}
and
\begin{align}\label{e630quater}
 B^Y_{\alpha l}(x,\theta,\phi) \equiv \frac{\ve}{1 + \ve f(\theta,\phi)} \, F^Y_{\alpha l}(x,\theta,\phi).
\end{align}
Note that in \eqref{e630bis} and \eqref{e630quater} we have used \eqref{e560} to write
\begin{align}\label{notation}
F^W_{\alpha l}(x,\theta,\phi)& = \; F^W_{\alpha l}(k_\alpha r)\Bigl|_{r = a[1+\ve f(\theta, \phi)]} \nonumber \\[6pt]
& = \; F^W_{\alpha l}\bigl(n_\alpha x [1 + \ve f(\theta,\phi)] \bigr),
\end{align}
with $\alpha=1,2$ and $W = \Psi, \Phi, Y$.

It should be noticed that in Eqs. \eqref{e670A} and \eqref{e670D} the pair of indices $l',m'$ comes from \eqref{w60} and the sum with respect to $l,m$, originates from the expressions of the fields \eqref{p10} and \eqref{p20}.
Moreover, for reasons that will soon be clear, it is instructive to rewrite these four equations in the suggestive matrix form
\begin{align}\label{e670}
\sum_{l,m} M^{l'm'}_{lm} \cdot \bm{\psi}_{lm} = 0,
\end{align}
where we have defined the $4 \times 4$ matrix $M^{l'm'}_{lm}$ and the $4 \times 1$ vector $\bm{\psi}_{lm}$ as
\begin{widetext}
\begin{align}\label{e670bis}
M^{l'm'}_{lm} \doteq \begin{bmatrix}
  [A_1^\Phi]_{lm}^{l'm'}(x) & -  [A_2^\Phi]_{lm}^{l'm'}(x) & \frac{[B_1^\Phi]_{lm}^{l'm'}(x)}{n_1} & -  \frac{[B_2^\Phi]_{lm}^{l'm'}(x)}{n_2}
   \\[8pt]
   -n_1 \, [B_1^\Psi]_{lm}^{l'm'}(x)  & n_2 \,[B_2^\Psi]_{lm}^{l'm'}(x)  & [A_1^\Psi]_{lm}^{l'm'} (x) & -[A_2^\Psi]_{lm}^{l'm'}(x)  \\[8pt]
   [A_1^\Psi]_{lm}^{l'm'}(x) & -  [A_2^\Psi]_{lm}^{l'm'}(x) & \frac{[B_1^\Psi]_{lm}^{l'm'}(x)}{n_1} & -  \frac{[B_2^\Psi]_{lm}^{l'm'}(x)}{n_2}
   \\[8pt]
   -n_1 \,[B_1^\Phi]_{lm}^{l'm'} (x) &  n_2 \,[B_2^\Phi]_{lm}^{l'm'} (x)& [A_1^\Phi]_{lm}^{l'm'}(x) & -  [A_2^\Phi]_{lm}^{l'm'} (x) \\
\end{bmatrix} ,
\end{align}
\end{widetext}
and
\begin{align}\label{e670ter}
\bm{\psi}_{lm} \doteq
\begin{bmatrix}
  \psi_{1lm}  \vphantom{a^E_{1lm}} \\[6pt]
  \psi_{2lm}  \vphantom{a^E_{1lm}}  \\[6pt]
  \psi_{3lm}  \vphantom{a^E_{1lm}}  \\[6pt]
  \psi_{4lm}  \vphantom{a^E_{1lm}}  \\
\end{bmatrix}
=
\begin{bmatrix}
  a^E_{1lm} \\[6pt]
  a^E_{2lm} \\[6pt]
  a^M_{1lm} \\[6pt]
  a^M_{2lm} \\
\end{bmatrix},
\end{align}
respectively.
Note that the incognita $x$ in this system is \emph{the same} for all the (infinite) terms of the  sum with respect to $l,m$. Therefore, it is not possible to solve each matrix equation of the sum independently. This is why in the remainder we will develop a perturbation scheme to solve \eqref{e670} with respect to $x$. From a physical point of view, the dependence of $x$ on all indices $(l,m)$ denotes the coupling between all the modes of the resonator, due to the departure from the spherical shape.

We remark that the homogeneous linear system \eqref{e670} is \emph{exact}, and  is valid irrespective of the shape and the magnitude of the deformation, and of the size of the resonator. In principle,  it contains all the information about the resonances of the deformed resonator. Were we able to solve it numerically, we would not need to develop a perturbation theory. However, this is not the case.

\subsection{The unperturbed problem}\label{UnpProb}

As a first step towards a perturbation theory, we must verify that the system \eqref{e670} reduces to Eqs. \eqref{e270} and \eqref{e280} for $\ve \to 0$, that is, when the resonator is perfectly spherical. In this case, from \eqref{e400} it follows that $\mathbf{n}_\parallel = 0$ and Eqs. \eqref{e630} become
\begin{subequations}\label{e632}
\begin{align}
\mathbf{A}_{\alpha lm} \equiv & \; \bm{\Phi}_{lm}(\theta,\phi), \label{e632A} \\[6pt]
\mathbf{B}_{\alpha lm}\equiv & \; \bm{\Psi}_{lm}(\theta,\phi) \, g_{\alpha l}(x)  \label{e632B} ,
\end{align}
\end{subequations}
where Eqs. \eqref{e560} have been used and we have defined
\begin{align}\label{e634}
 g_{\alpha l}(n_\alpha x) \equiv  F^\Psi_{\alpha l} (k_\alpha a) = \frac{1}{i} \, \frac{\left[ \bigl( n_\alpha x \bigr) b_{\alpha l} \bigl( n_\alpha x \bigr)\right]'}{(n_\alpha x) b_{\alpha l} ( n_\alpha x )} ,
\end{align}
with $\alpha = 1,2$. Substituting \eqref{e632} into \eqref{e640} we readily find
\begin{subequations}\label{e636}
\begin{align}
[A_\alpha^\Phi]_{lm}^{l'm'} = & \; \delta_{l l'} \delta_{m m'} , \label{e636a} \\[6pt]
[A_\alpha^\Psi]_{lm}^{l'm'} = & \; 0, \label{e636b} \\[6pt]
[B_\alpha^\Phi]_{lm}^{l'm'} =  & \; 0, \label{e636c} \\[6pt]
[B_\alpha^\Psi]_{lm}^{l'm'} =& \; \delta_{l l'} \delta_{m m'} g_{\alpha l}(n_\alpha x). \label{e636d}
\end{align}
\end{subequations}

Inserting these values into \eqref{e670}, we obtain the algebraic system
\begin{align}\label{system0}
\begin{bmatrix}
  1 & -1 & 0 & 0 \\[6pt]
  -n_1 g_{1l} & n_2 g_{2l} & 0 & 0 \\[6pt]
  0 & 0 & \frac{g_{1l}}{n_1} & -\frac{g_{2l}}{n_2} \\[6pt]
  0 & 0 & 1 & -1 \\
\end{bmatrix} \cdot
\begin{bmatrix}
  a^E_{1lm} \\[6pt]
  a^E_{2lm} \\[6pt]
  a^M_{1lm} \\[6pt]
  a^M_{2lm} \\
\end{bmatrix} = 0,
\end{align}
where $g_{\alpha l} = g_{\alpha l} ( n_\alpha x )$ with $\alpha =1,2$.
The block-diagonal form of this matrix equation reveals that for a spherical dielectric resonator the TE and TM waves are uncoupled. Therefore, the $4 \times 4 $ system \eqref{system0} naturally splits into two independent $2 \times 2$ systems, which are
\begin{align}\label{e700}
\displaystyle{
\begin{bmatrix}
  1 & - 1 \\[6pt]
  -n_1 g_{1l}(n_1 x) & n_2 g_{2l}(n_2 x) \\
\end{bmatrix} \cdot
\begin{bmatrix}
  a^E_{1lm} \\[6pt]
  a^E_{2lm} \\
\end{bmatrix} = 0}
\end{align}
for TE waves and
\begin{align}\label{e720}
\displaystyle{
\begin{bmatrix}
 \frac{g_{1l}(n_1 x)}{n_1} & -\frac{g_{2l}(n_2 x)}{n_2} \\[8pt]
 1 & - 1 \\
\end{bmatrix} \cdot
\begin{bmatrix}
  a^M_{1lm} \\[8pt]
  a^M_{2lm} \\
\end{bmatrix} = 0}
\end{align}
for TM waves.
The first system \eqref{e700},
\begin{subequations}\label{e680}
\begin{align}
a^E_{1lm} - a^E_{2lm} = & \; 0 , \label{e680a} \\[8pt]
-n_1 g_{1l}(n_1 x) \, a^E_{1lm} + n_2 g_{2l}(n_2 x) \, a^E_{2lm} = & \; 0, \label{e680b}
\end{align}
\end{subequations}
possesses the nontrivial solution $a^E_{1lm} = a^E_{2lm}$ if and only if  $-n_1 g_{1l}(n_1 x) + n_2 g_{2l}(n_2 x) =0 $. Using \eqref{e634} it is easy to see that the last condition is equivalent to \eqref{e280TE}. Similarly, the second system  \eqref{e720},
\begin{subequations}\label{e690}
\begin{align}
 \frac{g_{1l}(n_1 x)}{n_1} \, a^M_{1lm} -\frac{g_{2l}(n_2 x)}{n_2} \, a^M_{2lm} = & \; 0, \label{e690a} \\[8pt]
 a^M_{1lm} - a^M_{2lm} = & \; 0, \label{e690b}
\end{align}
\end{subequations}
admits the  solution $a^M_{1lm} = a^M_{2lm}$, provided that $ g_{1l}(n_1 x)/n_1 - g_{2l}(n_2 x)/n_2 =0 $. Again, from \eqref{e634} it follows that this condition is equivalent to \eqref{e280TM}.

We have thus demonstrated that our system of equations \eqref{e670}, correctly reproduces the well-known set of equations for the electromagnetic  resonances of a dielectric sphere.

\subsubsection{A remark}\label{remarks}

The matrix \eqref{e670bis} looks formidable. However, it has in fact a quite simple structure and admits a clear physical picture. To show this, let us introduce the shorthand $M = M^{l'm'}_{lm}$, and omit the indices $l',m'$ and $l,m$ everywhere. Then, it is not difficult to see that we can rewrite \eqref{e670bis} as a block matrix:
\begin{align}\label{rem10}
M = \begin{bmatrix}
      M_0 & \mathcal{T}( V ) \\[6pt]
      V & \mathcal{T}( M_0 )  \\
    \end{bmatrix},
\end{align}
where the matrix-valued function $\mathcal{T}$  is defined by
\begin{align}\label{rem15}
\mathcal{T} \left( \begin{bmatrix}
      a_{11} & a_{12} \\[6pt]
      a_{21} & a_{22}  \\
    \end{bmatrix} \right) =  \begin{bmatrix}
      -a_{21}/n_1^2 & -a_{22}/n_2^2 \\[6pt]
      a_{11} & a_{12}  \\
    \end{bmatrix}
\end{align}
and
\begin{subequations}\label{rem20}
\begin{align}
M_0 = & \; \begin{bmatrix}
      A_1^\Phi & -A_2^\Phi \\[6pt]
    -n_1 B_1^\Psi & n_2 B_2^\Psi  \\
    \end{bmatrix} \sim 1 + O(\ve), \label{rem20a} \\[6pt]
V = & \; \begin{bmatrix}
      A_1^\Psi & -A_2^\Psi \\[6pt]
    -n_1 B_1^\Phi & n_2 B_2^\Phi  \\
    \end{bmatrix} \sim O(\ve). \label{rem20b}
\end{align}
\end{subequations}
Hence, $M$ has only eight different elements of four different types, $\{ A_\alpha^\Psi, A_\alpha^\Phi, B_\alpha^\Psi, B_\alpha^\Phi \}$, four types per each of the two media labeled by $\alpha=1,2$.

From \eqref{system0} it follows that at $\ve = 0$ (spherical resonator),
\begin{align}\label{rem30}
\left. M_0 \right|_{\ve=0} =
\begin{bmatrix}
  1 & - 1 \\[6pt]
  -n_1 g_{1l}(n_1 x) & n_2 g_{2l}(n_2 x) \\
\end{bmatrix} .
\end{align}
This means that at $\ve = 0$,  $M_0$  describes the TE resonances of a perfect sphere. The rest matrix $M_0 - \left. M_0 \right|_{\ve=0} $ gives their corrections  due to self-coupling between TE modes. The same reasoning remains valid if we replace  $M_0$ with $\mathcal{T}( M_0 )$, and TE waves with TM waves.

The off-diagonal  matrices $V$ and $\mathcal{T}( V )$, evidently yield the coupling between TE and TM waves due to the departure from the spherical shape, for they vanishes at $\ve =0$:
\begin{align}\label{rem40}
\left. V \right|_{\ve=0} = 0 = \left. \mathcal{T}( V ) \right|_{\ve=0}.
\end{align}

\section{Quantum-like perturbation theory}

The main goal of this work is to study how the electromagnetic vibrations of a dielectric resonator are affected by a slight departure from the exact spherical form. Such a departure is quantified by the small parameter $ 0 \leq \ve \ll 1 $ defined by \eqref{e314}. Equations \eqref{e282} define the resonances of the unperturbed physical system, which is a dielectric sphere of radius $r=a$. Let us denote by $x^\z$ any solution of either $f_l^\text{TE}(x)=0$ or $f_l^\text{TM}(x)=0$, the type of wave  being irrelevant for the following discussion. We assume the existence of a neighborhood of $\ve = 0$ where the algebraic system of equations \eqref{e670} possesses a nontrivial solution for $x = x(\ve)$, such that
\begin{align}\label{e770}
x(\ve) = x^{(0)} + \ve x^{(1)} + \ve^2 x^{(2)} + \cdots \; .
\end{align}
Following the classical Rayleigh's scheme of perturbation theory \cite{strutt_2011}, we would like to determine $x^{(1)}$ from a set of first-order equations in $\ve$, $x^{(2)}$ from a set of second-order equations in $\ve$, and so on.

To achieve this goal in a systematic and direct manner, we find it convenient at this stage to adopt a quantum-like notation to represent the linear system of (nonlinear) equations \eqref{e670}.  This is possible because we can always associate a linear operator with a matrix  and vice versa. However, we remark that in this work the quantum formalism is just a useful notational tool that permits us to solve an entirely \emph{classical} problem.

\subsection{Linear algebra in quantum-like notation}

 To begin with, let us introduce the fictitious vector states $ \ket{l,m} $ with $l = 0,1, \cdots, \infty$, and $m = -l ,-l+1, \cdots, l$. By hypothesis, they are orthonormal,
\begin{align}\label{e910}
\brak{l,m}{l',m'} = \delta_{ll'} \delta_{mm'},
\end{align}
and form a complete basis in an infinite-dimensional Hilbert space,  denoted  by $\mE_\infty$, that is
\begin{align}\label{e920}
\sum_{l=0}^\infty \sum_{m=-l}^l \proj{l,m}{l,m} = \hId_\infty,
\end{align}
where $\hId_\infty$ is the identity operator in $\mE_\infty$ and here and hereafter the circumflex will mark operators in infinite-dimensional Hilbert spaces. We remark that the vector states $ \ket{l,m} $ are artificial in the sense they do not represent either the scalar spherical harmonics $Y_{lm}(\theta,\phi)$ or the vector spherical harmonics \eqref{e90}. They are a mathematical tool that we use to solve our problem in an efficient way.

Next, we define the four basis vectors $\ket{i}$ with $i=1,2,3,4$. We assume that  they are orthonormal,
\begin{align}\label{e930}
\brak{i}{j} = \delta_{ij},
\end{align}
and span a four-dimensional Hilbert space, denoted by $\mE_4$, where they form a complete basis:
\begin{align}\label{e940}
\sum_{i=1}^4  \proj{i}{i} = I_4,
\end{align}
with $I_4$  the $4 \times 4$ identity matrix in $\mE_4$. Then, the tensor product Hilbert space
\begin{align}\label{e950}
\mE = \mE_\infty \otimes \mE_4
\end{align}
is by construction spanned by the vectors
\begin{align}\label{e960}
\ket{l,m,i} \equiv \ket{l,m} \otimes \ket{i}.
\end{align}
By definition, the completeness relation for $\mE$ reads
\begin{align}\label{e965}
\sum_{l,m,i} \proj{l,m,i}{l,m,i} = \hId_\infty \otimes I_4 \equiv \hat{\mathcal{I}},
\end{align}
where here and hereafter
\begin{align}\label{s74}
\sum_{l,m,i} \qquad \text{stands for} \qquad \sum_{l =0}^\infty \sum_{m=-l}^l \sum_{i=1}^4 \; .
\end{align}

Equipped with this paraphernalia, we can rewrite \eqref{e670} as follows. First, we introduce the vector state $\ket{\psi} \in \mE$, such that
\begin{align}\label{e967}
\ket{\psi} = \sum_{l,m,i} \ket{l,m,i} \brak{l,m,i}{\psi} \equiv \sum_{l,m,i} \psi_{ilm} \ket{l,m,i},
\end{align}
the components of which are
\begin{align}\label{e970}
\brak{l,m,i}{\psi} \equiv \psi_{ilm},
\end{align}
where, using \eqref{e670ter},
\begin{align}\label{e980}
\begin{bmatrix}
  \psi_{1lm}  \vphantom{a^E_{1lm}} \\[6pt]
  \psi_{2lm}  \vphantom{a^E_{1lm}}  \\[6pt]
  \psi_{3lm}  \vphantom{a^E_{1lm}}  \\[6pt]
  \psi_{4lm}  \vphantom{a^E_{1lm}}  \\
\end{bmatrix}
=
\begin{bmatrix}
  a^E_{1lm} \\[6pt]
  a^E_{2lm} \\[6pt]
  a^M_{1lm} \\[6pt]
  a^M_{2lm} \\
\end{bmatrix}.
\end{align}
Second, we define the operator $\hat{\mathcal{M}}=\hat{\mathcal{M}}(x)$ via the matrix elements
\begin{align}\label{e990}
\bra{l',m',i}\hat{\mathcal{M}} \ket{l,m,j} = & \; \bra{i} \Bigl(\bra{l',m'}\hat{\mathcal{M}} \ket{l,m} \Bigr) \ket{j} \nonumber \\[6pt]
= & \; \bra{i} M^{l'm'}_{lm} \ket{j} \nonumber \\[6pt]
\equiv & \; \bigl[ M^{l'm'}_{lm} \bigr]_{ij},
\end{align}
where, according to \eqref{e670bis},
\begin{equation}\label{e1000}
\begin{split}
\bigl[ M^{l'm'}_{lm} \bigr]_{11} = & \; \left[ A_{1}^\Phi \right]_{lm}^{l'm'}(x),  \\[6pt]
\bigl[ M^{l'm'}_{lm} \bigr]_{12} = & \;  -  [A_2^\Phi]_{lm}^{l'm'}(x), \\
\phantom{\bigl[ M^{l'm'}_{lm} \bigr]_{12}}  \vdots & \;  \phantom{-  [A_2^\Phi]_{lm}^{l'm'}(x)}   \\
\bigl[ M^{l'm'}_{lm} \bigr]_{44} = & \;  -  [A_2^\Phi]_{lm}^{l'm'}(x).
\end{split}
\end{equation}

Finally, using this notation we can rewrite \eqref{e670} as
\begin{align}\label{e1010}
\hat{\mathcal{M}} \ket{\psi} = 0.
\end{align}
This can be easily proven by multiplying this equation by $\bra{l',m',i}$ from the left and using the closure relation \eqref{e965},
\begin{align}\label{e1020}
0 = & \; \bra{l',m',i} \hat{\mathcal{M}}\ket{\psi} \nonumber \\[8pt]
= & \; \sum_{l,m,j} \bra{l',m',i} \hat{\mathcal{M}} \proj{l,m,j}{l,m,j}\psi \rangle \nonumber \\[8pt]
= & \; \sum_{l,m,j} \bigl[ M_{lm}^{l'm'} \bigr]_{ij} \psi_{jlm} ,
\end{align}
where \eqref{e970} and \eqref{e990} have been used.

We make an important remark: Unlike the case of quantum mechanics, here we have no guarantee that the operator $\hat{\mathcal{M}}$ is Hermitian. As a matter of fact, in general, it is not. This is why, in the remainder, we will make extensive use of biorthogonal bases generated by the right and left eigenvectors of non-Hermitian operators. The ultimate reason for the presence of non-Hermitian operators in our theory, is that dielectric resonators are intrinsically leaky systems.

\subsection{Formal expansion}\label{formal}

Before diving into the development of a rigorous perturbation theory, in this subsection we provide  a general outline of the theory, irrespective of the precise form of the resonances spectrum.

The  goal is to solve \eqref{e1010}, here rewritten as
\begin{align}\label{s10}
\hat{\mathcal{M}} \left( \ve \right) \ket{\psi(\ve) } = 0,
\end{align}
where $\hat{\mathcal{M}}\left( \ve \right)  $ is defined by \eqref{e990}, and
\begin{align}\label{e770bis}
x(\ve) = x^{(0)} + \ve x^{(1)} + \ve^2 x^{(2)} + \cdots \; ,
\end{align}
has been defined in \eqref{e770}, with $x^{(0)} = x(0)$. Moreover, we assume that also the operator $\hat{\mathcal{M}}\left( \ve \right) $  and the vector $\ket{\psi(\ve)}$ can be expanded as power of $\ve$  as
\begin{align}\label{e1120}
\hat{\mathcal{M}}\left( \ve \right)  = \hmM^{(0)} + \ve \hmM^{(1)} + \ve^2 \hmM^{(2)} + \cdots \; ,
\end{align}
with
\begin{align}\label{e1120b}
\hmM^{(n)} = \frac{1}{n!} \left. \frac{d^n  \hmM(\ve)}{d \, \ve^n} \right|_{\ve=0}, \qquad (n=0,1,\ldots)
\end{align}
and
\begin{align}\label{e1130}
\ket{\psi(\ve)} = \ket{\psi^{(0)}} + \ve \ket{\psi^{(1)}}  + \ve^2 \ket{\psi^{(2)}} + \cdots \; ,
\end{align}
 where, by definition,
\begin{align}\label{eDef}
 \ket{\psi^{(0)}} = \ket{\psi(0)}.
\end{align}
Substituting \eqref{e1120} and \eqref{e1130} into \eqref{s10}, we obtain
\begin{align}\label{e1150}
\hat{\mathcal{M}} & \left( \ve \right) \ket{\psi(\ve)}\nonumber \\[6pt]
&   =   \hmM^{(0)} \ket{\psi^{(0)}}\nonumber \\[6pt]
& \phantom{xx} + \ve \left(  \hmM^{(0)} \ket{\psi^{(1)}}  + \hmM^{(1)} \ket{\psi^{(0)}} \right)\nonumber \\[6pt]
& \phantom{xx}  +  \ve^2 \left(  \hmM^{(0)} \ket{\psi^{(2)}}+ \hmM^{(1)} \ket{\psi^{(1)}}  + \hmM^{(2)} \ket{\psi^{(0)}} \right)  \nonumber \\[6pt]
& \phantom{xx}  + \dots = 0.
\end{align}
All the terms proportional to the same power of $\ve$ must sum to zero. Thus, we obtain the chain of equations,
\begin{subequations}\label{e1160}
\begin{align}
\hmM^{(0)} \ket{\psi^{(0)}} = & \; 0, \label{e1160A} \\[6pt]
\hmM^{(0)} \ket{\psi^{(1)}}  + \hmM^{(1)} \ket{\psi^{(0)}} = & \; 0, \label{e1160B} \\[6pt]
\hmM^{(0)} \ket{\psi^{(2)}}+ \hmM^{(1)} \ket{\psi^{(1)}}  + \hmM^{(2)} \ket{\psi^{(0)}} = & \; 0, \label{e1160C}
\end{align}
\end{subequations}
etc. To solve iteratively these equations, we must first choose the initial state $\ket{\psi^{(0)}}$ (actually, the initial \emph{set of states}) associated with the unperturbed eigenvalue $x^\z$. We will take for $x^\z$  one of the $2l+1$ times degenerate solutions of \eqref{e282bis}, that is $x^\z = x^\sigma_{ln}$. We will see that such a solution is associated with a degenerate subspace of dimension $2l+1$.
However, before starting to solve \eqref{e1160}, it is useful to illustrate some general properties of the operator $\hmM$.

\subsection{General properties of the operator $\hmM$}

The set of operators
\begin{align}\label{ePropOp10}
\bigl\{\hmM^\n \bigr\} = \bigl\{\hmM^\z, \hmM^\un , \hmM^\du, \cdots \bigr\},
\end{align}
possesses some general properties which are key to the development of the perturbation theory. These properties are proven in Appendix \ref{operators}. In this section we will  present the plain results, which are summarized by
\begin{subequations}\label{ePropOp20}
\begin{align}
\hmM^\z = & \; \hmD^\z, \label{ePropOp20a} \\[6pt]
\hmM^\n = & \; \hmV^\n + x^\n \hmD, \label{ePropOp20b}
\end{align}
\end{subequations}
where we have defined
\begin{subequations}\label{ePropOp25}
\begin{align}
\hmV^\n \equiv & \; \left. \hmM^\n \right|_{x^\n \, = \,0}, \label{ePropOp25a} \\[6pt]
\hmD \equiv & \; \frac{d \, \hmM^\n}{d \, x^\n}, \label{ePropOp25b}
\end{align}
\end{subequations}
with $n \geq 1$. We remark again that all the operators in \eqref{ePropOp20} are not necessarily Hermitian.  Note that the operator $\hmD$ is independent of the order index $n$. Both $\hmD^\z$ and $\hmD$ are diagonal with respect to the basis $\ket{l,m}$, that is,
\begin{subequations}\label{s35}
\begin{align}
\mean{l,m,i}{\hmD^\z }{l',m',j} = & \; \delta_{l l'}\delta_{m m'}\mean{i}{ D_{l}^\z}{ j} , \label{s35a}\\[6pt]
\mean{l,m,i}{\hmD }{l',m',j} = & \; \delta_{l l'}\delta_{m m'}\mean{i}{ D_{l}}{ j} , \label{s35b}
\end{align}
\end{subequations}
where the operators $D_{l}^\z$ and $D_{l}$, are represented by a $4 \times 4$ matrix independent of $m$. From \eqref{e990} and \eqref{system0} it follows that $D_{l}^\z$ is equal to
\begin{align}\label{system0bis}
D_{l}^\z \doteq \begin{bmatrix}
  1 & -1 & 0 & 0 \\[6pt]
  -n_1 g_{1l} & n_2 g_{2l} & 0 & 0 \\[6pt]
  0 & 0 & \frac{g_{1l}}{n_1} & -\frac{g_{2l}}{n_2} \\[6pt]
  0 & 0 & 1 & -1 \\
\end{bmatrix} ,
\end{align}
with  $g_{\alpha l} = g_{\alpha l} \bigl( n_\alpha x^\z \bigr)$, $(\alpha =1,2)$,  given by \eqref{e634}. Moreover, Eq. \eqref{f50} gives
\begin{align}\label{f50OLD}
D_l \doteq
\begin{bmatrix}
  0 & 0 & 0 & 0 \\[8pt]
  -n_1^2 \left[ g_{1  l} \right]'  & n_2^2 \left[ g_{2  l} \right]' & 0 & 0 \\[8pt]
  0 & 0 & \left[ g_{1  l} \right]' & - \left[ g_{2  l}\right]' \\[8pt]
  0 & 0 & 0 & 0 \\
\end{bmatrix},
\end{align}
where the prime  denotes the derivative with respect to the argument: $\left[ g_{\alpha  l} \right]'= \left. d g_{\alpha l} (u)/d u \right|_{u = n_\alpha x^\z}$. Note that it is possible to rewrite $D_l$ as
\begin{align}\label{derivative}
D_{l}\bigl( n_\alpha x^\z \bigr) = \frac{d \,D_{l}^\z \bigl( n_\alpha x^\z \bigr)}{d \, x^\z}.
\end{align}
In practice, $\hmV^\n$  may (and, in general, it will) depend on $x^\z, x^\un, \cdots, x^{(n-1)}$, but not on $x^\n$.

\section{Zeroth-order equation}\label{zero}

In this section we will start a systematical analysis of \eqref{e1160}, solving the chained equations order by order.

Using twice the resolution of the identity \eqref{e965} and Eqs. \eqref{s35} and \eqref{system0bis}, we can rewrite   \eqref{e1160A} as
\begin{align}\label{s60}
0 = & \; \hmM^\z \ket{\psi^\z} \nonumber \\[6pt]
 = & \; \hmD^\z \ket{\psi^\z} \nonumber \\[6pt]
= & \; \sum_{l,m,i}  \left(\sum_{j=1}^4 \mean{i}{ D^\z_l }{j} \brak{l,m,j}{\psi^\z} \right)\ket{l,m,i} .
\end{align}
According to \eqref{e970}, in the remainder we will also occasionally use the more compact notation
\begin{align}\label{s67}
\psi^{\n}_{ilm} \equiv \brak{l,m,i}{\psi^{(n)}},
\end{align}
where
\begin{align}\label{e980bic}
\begin{bmatrix}
  \psi_{1lm}^\n  \vphantom{a^E_{1lm}} \\[6pt]
  \psi_{2lm}^\n  \vphantom{a^E_{1lm}}  \\[6pt]
  \psi_{3lm}^\n  \vphantom{a^E_{1lm}}  \\[6pt]
  \psi_{4lm}^\n  \vphantom{a^E_{1lm}}  \\
\end{bmatrix}
=
\begin{bmatrix}
  a^{E (n)}_{1lm} \\[6pt]
  a^{E (n)}_{2lm} \\[6pt]
  a^{M (n)}_{1lm} \\[6pt]
  a^{M (n)}_{2lm} \\
\end{bmatrix}.
\end{align}
Since the vectors $\{ \ket{l,m,i} \}$ form a complete basis in $\mathscr{E}$, then \eqref{s60} is satisfied when all the coefficients of the expansion \eqref{s60} are identically zero, that is when
\begin{align}\label{s70}
 \sum_{j=1}^4  \mean{i}{ D^\z_l }{j} \brak{l,m,j}{\psi^\z} = 0.
\end{align}
In matrix form this equation reads
\begin{align}\label{s73}
\begin{bmatrix}
  1 & -1 & 0 & 0 \vphantom{\psi_{1lm}^\z} \\[6pt]
  -n_1 g_{1l} & n_2 g_{2l} & 0 & 0 \vphantom{\psi_{1lm}^\z}\\[6pt]
  0 & 0 & \frac{g_{1l}}{n_1} & -\frac{g_{2l}}{n_2} \vphantom{\psi_{1lm}^\z}\\[6pt]
  0 & 0 & 1 & -1 \vphantom{\psi_{1lm}^\z}\\
\end{bmatrix} \cdot
\begin{bmatrix}
  \psi_{1lm}^\z   \\[6pt]
  \psi_{2lm}^\z   \\[6pt]
  \psi_{3lm}^\z    \vphantom{\frac{g_{1l}}{n_1} }\\[6pt]
  \psi_{4lm}^\z    \\
\end{bmatrix}  = 0,
\end{align}
where $g_{\alpha l} = g_{\alpha l} \bigl( n_\alpha x^\z \bigr)$, $(\alpha =1,2)$ and \eqref{system0bis} has been used.

We have already solved this system of equations in Sec. \ref{UnpProb}, and we have found two different results for TE and TM waves. Therefore, also now we will consider these two cases separately.

\subsection{TE waves}\label{zeroTE}

Let us choose a pair of values $\bigl( l_0, x^\z_E \bigr)$ such that
\begin{align}\label{zE}
z^E \equiv n_1 g_{1l_0}\bigl( n_1 x^\z_E \bigr) = n_2 g_{2l_0} \bigl(n_2 x^\z_E \bigr) .
\end{align}
In this case \eqref{s73} becomes
\begin{align}\label{u30ter}
 D^\z_\lz\bigl( x^\z_E \bigr) \cdot \bm{\psi}^\z_{{l_0}m} = 0,
\end{align}
where
\begin{align}\label{u30quater}
D^\z_\lz\bigl( x^\z_E \bigr) \equiv
\begin{bmatrix}
  1 & -1 & 0 & 0 \vphantom{\psi_{1lm}^\z} \\[6pt]
  -z^E & z^E & 0 & 0 \vphantom{\psi_{1lm}^\z}\\[6pt]
  0 & 0 & \frac{z^E}{n_1^2} & - \frac{z^E}{n_2^2} \vphantom{\psi_{1lm}^\z}\\[6pt]
  0 & 0 & 1 & -1 \vphantom{\psi_{1lm}^\z}\\
\end{bmatrix}
\end{align}
and
\begin{align}\label{zeroTE5bis}
\bm{\psi}^\z_{{l_0}m} \doteq \begin{bmatrix}
  \psi_{1{l_0}m}^\z   \\[6pt]
  \psi_{2{l_0}m}^\z   \\[6pt]
  \psi_{3{l_0}m}^\z   \\[6pt]
  \psi_{4{l_0}m}^\z    \\
\end{bmatrix} .
\end{align}
Equation \eqref{u30ter} turns into an identity for
\begin{align}\label{zeroTE10}
\bm{\psi}^\z_{{l_0}m} =
\psi_{{l_0}m}^\z \, \begin{bmatrix}
  1 \vphantom{\psi_{1{l_0}m}^\z}    \\[6pt]
  1 \vphantom{\psi_{1{l_0}m}^\z}    \\[6pt]
  0 \vphantom{\psi_{1{l_0}m}^\z}    \\[6pt]
  0 \vphantom{\psi_{1{l_0}m}^\z}    \\
\end{bmatrix}  ,
\end{align}
where $\psi_{l_0 m}^\z$ are, at this stage, arbitrary numbers. By definition, this solution \eqref{zeroTE10} is valid only for  $l=l_0$. However, Eq. \eqref{s70} must be zero for all values of $l$. Therefore, the solutions of \eqref{s70} must be
\begin{align}\label{zeroTE20}
\brak{l,m,j}{\psi^\z} = \left\{
                          \begin{array}{ll}
                            0, & \; l \neq l_0, \\[6pt]
                            \psi_{l_0m}^\z \left(\delta_{j1} + \delta_{j2} \right), & \; l = l_0.
                          \end{array}
                        \right.
\end{align}
Then, we can write $\ket{\psi^\z} $ as
\begin{align}\label{zeroTE30}
\ket{\psi^\z}  =   & \; \sum_{l,m,i}  \ket{l,m ,i} \brak{l,m,i}{\psi^\z}  \nonumber \\[6pt]
=   & \; \sum_{m = -l_0}^{l_0} \sum_{i=1}^2 \psi_{l_0m}^\z  \ket{l_0,m, i}  \nonumber \\[6pt]
\equiv   & \; \ket{\vp^\z} \ket{\alpha_0^E},
\end{align}
where we have defined
\begin{subequations}\label{s180}
\begin{align}
\ket{\vp^\z} \equiv & \;  \sum_{m = -l_0}^{l_0} \psi_{l_0m}^\z  \ket{l_0,m}, \label{s180A} \\[6pt]
\ket{\alpha_0^E} \equiv & \;  \ket{1} + \ket{2}  \doteq
 \begin{bmatrix}
  1 \vphantom{\psi_{1lm}^\z}    \\[0pt]
  1 \vphantom{\psi_{1lm}^\z}    \\[0pt]
  0 \vphantom{\psi_{1lm}^\z}    \\[0pt]
  0 \vphantom{\psi_{1lm}^\z}    \\
\end{bmatrix} . \label{s180B}
\end{align}
\end{subequations}

At this stage, the  $2 \lz +1$ coefficients $\psi_{l_0m}^\z$ in \eqref{s180A} are still undetermined. However, irrespective of their values, we always have
\begin{align}\label{zeroTE50}
\hmD^\z \ket{\psi^\z}  = &  \; \sum_{m = -l_0}^{l_0}\sum_{i=1}^4 \psi_{l_0 m}^\z \ket{l_0,m,i} \mean{i}{D^\z_\lz \bigl( x^\z_E \bigr)}{\alpha_0} \nonumber \\[6pt]
=   & \; 0,
\end{align}
because from \eqref{u30quater} and \eqref{zeroTE10}, it follows that
\begin{align}\label{zeroTE60}
D^\z_\lz \bigl( x^\z_E \bigr)\ket{\alpha_0^E}  = 0.
\end{align}
This equation can be interpreted as an eigenvector equation with eigenvalue equal to $0$. A direct calculation actually shows that
  $\ket{\alpha_0^E} \in \mathscr{E}_4$ belongs to the biorthogonal pair $\bigl\{ \ket{\alpha_0^E}, \ket{\alpha_1^E } \bigr\}$, where
\begin{align}\label{s190}
\ket{\alpha_1^E} \equiv  \frac{1}{z^E + 1} \bigl( \, \ket{1} - z^E \, \ket{2} \, \bigr) \doteq \frac{1}{z^E + 1} \begin{bmatrix}
                                      1 \vphantom{\psi_{1lm}^\z}\\[0pt]
                                      -z^E \vphantom{\psi_{1lm}^\z}\\[0pt]
                                      0 \vphantom{\psi_{1lm}^\z}\\[0pt]
                                      0 \vphantom{\psi_{1lm}^\z}\\
                                    \end{bmatrix} ,
\end{align}
and
\begin{align}\label{s200}
D_\lz^\z\bigl( x^\z_E \bigr) \ket{\alpha_\imath^E}  = & \; \lambda_\imath \ket{\alpha_\imath^E} , \qquad (\imath=0,1),
\end{align}
where $\lambda_0=0$ and $\lambda_1 = z^E + 1$. Note that throughout this paper, we use dotless letters $\,\imath\,$ and $\, \jmath\,$, as indices running from $0$ to $3$, while ordinary letters $\,i\,$ and $\,j\,$ are  indices running from $1$ to $4$. The \emph{left} eigenvectors $\bra{\ta_0^E}$ and $\bra{\ta_2^E}$ of $D^\z_\lz \bigl( x^\z_E \bigr)$ are defined by
\begin{align}\label{s210}
\bra{\ta_\imath^E} D^\z_\lz \bigl( x^\z_E \bigr)  = & \; \lambda_\imath \bra{\ta_\imath^E} , \qquad (\imath=0,1),
\end{align}
where
\begin{subequations}\label{s220}
\begin{align}
 \bra{\ta_0^E} = & \; \frac{1}{1+z^E} \bigl( z^E \bra{1} + \bra{2} \bigr),
 \label{s220A} \\[6pt]
 \bra{\ta_1^E} = & \;  \bra{1} - \bra{2}. \label{s220B}
\end{align}
\end{subequations}
If, additionally,  we define
\begin{equation}\label{s220bis}
\begin{split}
\ket{\alpha_2^E} \equiv & \;  \ket{3}, \\[6pt]
\ket{\alpha_3^E} \equiv & \;   \ket{4},
\end{split}
\end{equation}
and
\begin{equation}\label{s220ter}
\begin{split}
\bra{\ta_2^E} \equiv & \;  \bra{3},  \\[6pt]
\bra{\ta_3^E}  \equiv & \;   \bra{4},
\end{split}
\end{equation}
we can build a complete and biorthogonal set of bases for $\mathscr{E}_4$, denoted by $\bigl\{\ket{\alpha_\imath^E}, \bra{\ta_\imath^E} \bigr\}$ \cite{PhysRevC.6.114}.
It is a simple exercise to verify that these basis vectors satisfy the standard normalization condition
for bi-orthogonal vectors,
\begin{align}\label{s230}
\brak{\ta_\imath^E}{\alpha_\jmath^E} = \delta_{\imath \jmath}, \qquad (\imath,\jmath=0,1,2,3),
\end{align}
and that they form a complete basis for $\mathscr{E}_4$,
\begin{align}\label{s232}
\sum_{\imath=0}^3 \proj{\alpha_\imath^E}{\ta_\imath^E} = I_4,
\end{align}
where $I_4$ is the $4 \times 4$ identity matrix. Such a biorthogonal basis will be very useful for the next steps in perturbation theory.

\subsection{TM waves}\label{zeroTM}

In this case we choose a pair of values $\bigl( l_0, x^\z_M \bigr)$ such that
\begin{align}\label{zM}
z^M  \equiv \frac{1}{n_1} g_{1l_0}\bigl( n_1 x^\z_M \bigr) = \frac{1}{n_2} g_{2l_0}\bigl( n_2 x^\z_M \bigr) .
\end{align}
We proceed as for the TE case and we write again \eqref{s73} as
\begin{align}\label{u30terTM}
 D^\z_\lz\bigl( x^\z_M \bigr) \cdot \bm{\psi}^\z_{{l_0}m} = 0,
\end{align}
with
\begin{align}\label{u30quaterTM}
D^\z_\lz\bigl( x^\z_M \bigr) \equiv
\begin{bmatrix}
  1 & -1 & 0 & 0 \vphantom{\psi_{1lm}^\z} \\[6pt]
  -n_1^2 \, z^M & n_2^2 \, z^M & 0 & 0 \vphantom{\psi_{1lm}^\z}\\[6pt]
  0 & 0 & z^M & - z^M \vphantom{\psi_{1lm}^\z}\\[6pt]
  0 & 0 & 1 & -1 \vphantom{\psi_{1lm}^\z}\\
\end{bmatrix}.
\end{align}
Next, \eqref{u30terTM} turns into an identity for
\begin{align}\label{zeroTM10}
\bm{\psi}^\z_{{l_0}m} =
\psi_{{l_0}m}^\z \, \begin{bmatrix}
  0 \vphantom{\psi_{1{l_0}m}^\z}    \\[6pt]
  0 \vphantom{\psi_{1{l_0}m}^\z}    \\[6pt]
  1 \vphantom{\psi_{1{l_0}m}^\z}    \\[6pt]
  1 \vphantom{\psi_{1{l_0}m}^\z}    \\
\end{bmatrix}  ,
\end{align}
where $\psi_{l_0 m}^\z$ are again arbitrary numbers. Therefore,  Eq. \eqref{s70} becomes an identity for
\begin{align}\label{zeroTM20}
\brak{l,m,j}{\psi^\z} = \left\{
                          \begin{array}{ll}
                            0, & \; l \neq l_0, \\[6pt]
                            \psi_{l_0m}^\z \left(\delta_{j3} + \delta_{j4} \right), & \; l = l_0.
                          \end{array}
                        \right.
\end{align}
Then, we can write $\ket{\psi^\z} $ as
\begin{align}\label{zeroTM30}
\ket{\psi^\z}  =   & \; \sum_{m = -l_0}^{l_0} \sum_{j=3}^4 \psi_{l_0m}^\z  \ket{l_0,m, j}  \nonumber \\[6pt]
\equiv   & \; \ket{\vp^\z} \ket{\alpha_0^M},
\end{align}
where  $\ket{\vp^\z}$ is defined by \eqref{s180A} and
\begin{align}
\ket{\alpha_0^M} \equiv & \; \ket{3} + \ket{4}  \doteq
 \begin{bmatrix}
  0 \vphantom{\psi_{1lm}^\z}    \\[0pt]
  0 \vphantom{\psi_{1lm}^\z}    \\[0pt]
  1 \vphantom{\psi_{1lm}^\z}    \\[0pt]
  1 \vphantom{\psi_{1lm}^\z}    \\
\end{bmatrix} . \label{s180TMB}
\end{align}
By definition, from \eqref{u30quaterTM} and \eqref{s180TMB}, it follows that
\begin{align}\label{zeroTM60}
D^\z_\lz \bigl( x^\z_M \bigr)\ket{\alpha_0^M}  = 0.
\end{align}

Now the complete and biorthogonal set of bases for $\mathscr{E}_4$ is $\bigl\{\ket{\alpha^M_\imath}, \bra{\ta^M_\imath} \bigr\}$ and it is defined by
\begin{subequations}\label{beta10}
\begin{align}
\ket{\alpha^M_0} = & \; \ket{3} + \ket{4},  \label{beta10a} \\[6pt]
\ket{\alpha^M_1} = & \; \frac{1}{z^M-1} \bigl(  z^M \ket{3} +  \ket{4}  \bigr), \label{beta10b} \\[6pt]
\ket{\alpha^M_2} = & \; \ket{1}, \label{beta10c} \\[6pt]
\ket{\alpha^M_3} = & \; \ket{2} \label{beta10d}
\end{align}
\end{subequations}
and
\begin{subequations}\label{beta20}
\begin{align}
\bra{\ta^M_0} = & \; \frac{1}{z^M-1} \bigl( - \bra{3} + z^M \bra{4} \, \bigr), \label{beta20a} \\[6pt]
\bra{\ta^M_1} = & \; \bra{3} - \bra{4}, \label{beta20b} \\[6pt]
\bra{\ta^M_2} = & \;  \bra{1}, \label{beta20c} \\[6pt]
\bra{\ta^M_3} = & \;  \bra{2}, \label{beta20d}
\end{align}
\end{subequations}
where
\begin{align}\label{s200TM}
D_\lz^\z\bigl( x^\z_M \bigr) \ket{\alpha^M_\imath}  = & \; \lambda_\imath \ket{\alpha^M_\imath} , \qquad (\imath=0,1),
\end{align}
and
\begin{align}\label{s210bis}
\bra{\ta^M_\imath} D^\z_\lz \bigl( x^\z_M \bigr)  = & \; \lambda_\imath \bra{\ta^M_\imath} , \qquad (\imath=0,1),
\end{align}
with $\lambda_0 = 0$ and $\lambda_1=z^M-1$. By definition,
\begin{align}\label{s230TM}
\brak{\ta^M_\imath}{\alpha^M_\jmath} = \delta_{\imath \jmath}, \qquad (\imath,\jmath=0,1,2,3),
\end{align}
and
\begin{align}\label{s232TM}
\sum_{\imath=0}^3 \proj{\alpha^M_\imath}{\ta^M_\imath} = I_4.
\end{align}

\section{First-order equations}\label{1order}

\subsection{Some preparatory remarks}\label{preparatory}

In this section we will focus on the degenerate subspace of dimension $N_\lz = 2 \lz +1$, denoted by $\mathscr{D}_0 \subseteq \mE$, generated by a solution $x^\z_E$ of  the TE equation
\begin{align}\label{prep10}
n_1  \, g_{1l_0}\bigl( n_1 x^\z_E \bigr) - n_2 \, g_{2l_0}\bigl( n_2 x^\z_E \bigr) =0,
\end{align}
and the  vectors \eqref{zeroTE10}, or by a solution $x^\z_M$ of the TM equation
\begin{align}\label{prep20}
\frac{1}{n_1}  \, g_{1l_0}\bigl( n_1 x^\z_M \bigr) - \frac{1}{n_2}  \, g_{2l_0}\bigl( n_2 x^\z_M \bigr) = 0,
\end{align}
and the  vectors \eqref{zeroTM10}.

 To build up $\mDz$, consider first the subspace $\mD$, which is naturally spanned by the $ 2 \lz +1$ orthogonal vectors
\begin{align}\label{zeroTE40}
\mD \equiv \operatorname{span}  \bigl\{ \ket{\lz,-\lz}, \ket{\lz, -\lz +1}, \cdots, \ket{\lz, \lz} \bigr\}.
\end{align}
As we will see later, it is actually more convenient to choose a different set of $N_\lz$ orthonormal vectors,
\begin{align}\label{u10}
\bigl\{ \ket{\vp^\z_\mu} \bigr\} =   \bigl\{ \ket{\vp^\z_1}, \ket{\vp^\z_2}, \cdots, \ket{\vp^\z_{N_\lz}} \bigr\},
\end{align}
defined by
\begin{align}\label{u20}
\ket{\vp^\z_\mu} \equiv   \sum_{m = -l_0}^{l_0} \vp_{ \mu m}^\z  \ket{l_0,m},
\end{align}
where the coefficients $\vp_{ \mu m}^\z$ are, at this stage, still undetermined. We may think of this basis as a part of the  biorthogonal set $\bigl\{ \ket{\vp^\z_\mu} ,\bra{\tvp^\z_\mu} \bigr\}$ in $\mE_\infty$, where
\begin{align}\label{u20bis}
\bra{\tvp^\z_\mu} \equiv   \sum_{m = -l_0}^{l_0} \tvp_{ \mu m}^\z  \bra{l_0,m},
\end{align}
with, in general, $\tvp_{ \mu m}^\z \neq {\vp_{ \mu m}^\z}^*$ and
\begin{subequations}\label{u30}
\begin{align}
 \brak{\tvp^\z_\mu}{\vp^\z_\nu} = & \; \sum_{m = -l_0}^{l_0}  \tvp_{ \mu m}^\z \, \vp_{ \nu m}^\z= \delta_{\mu \nu},
 \label{u30A} \\[6pt]
 \brak{l, m}{\vp^\z_\mu} = & \;  0 = \brak{\tvp^\z_\mu}{l, m}, \qquad (l \neq l_0). \label{u30B}
\end{align}
\end{subequations}
Next, we introduce the biorthogonal set $\bigl\{  \ket{\psi^\z_{\mu \imath}} ,\bra{\tpsi^\z_{\mu \imath}} \bigr\}$ in $\mE$,  defined by
\begin{subequations}\label{u40}
\begin{align}
\ket{\psi^\z_{\mu \imath}} \equiv & \; \ket{\vp^\z_\mu} \ket{\alpha_\imath},
 \label{u40A} \\[6pt]
\bra{\tpsi^\z_{\mu \imath}} \equiv & \; \bra{\tvp^\z_\mu} \bra{\ta_\imath}, \label{u40B}
\end{align}
\end{subequations}
with $\mu = 1, \cdots, N_\lz$ and $\imath =0, 1,2,3$. Here $\alpha$ denotes either $\alpha^E$ or $\alpha^M$.
By definition, the subset of vectors $\{ \ket{\psi^\z_{\mu 0}} \}$ spans the sought degenerate subspace $\mDz $, of dimension $N_\lz$:
\begin{align}\label{u50}
\mDz \equiv \operatorname{span}  \bigl\{ \ket{\psi^\z_{\mu 0}}; \; \mu = 1, \cdots, N_\lz \bigr\}.
\end{align}
Therefore, we have
\begin{subequations}\label{u55}
\begin{align}
\brak{\tpsi^\z_{\mu 0}}{\psi^\z_{\nu 0}} = & \; \delta_{\mu \nu},
 \label{u55A} \\[6pt]
\hmD^\z \ket{\psi^\z_{\mu 0}} = & \; 0 = \bra{\tpsi^\z_{\mu 0}} \hmD^\z. \label{u55B}
\end{align}
\end{subequations}
Moreover, from \eqref{u20}, \eqref{u20bis} and \eqref{u30A} and using \eqref{s35} it is possible to show that
\begin{align} \label{u57}
\mean{\tpsi^\z_{\mu \imath}}{\hmD}{\psi^\z_{\nu \jmath}} =   \delta_{\mu \nu} \mean{\ta_\imath}{D_\lz}{\alpha_\jmath},
\end{align}
for $\imath, \jmath=1,2,3$.

By construction, the $3 \times N_\lz$ vectors $\ket{\psi^\z_{\mu \imath}}$, with $(\imath=1,2,3)$, span the subspace $\mD_I$, defined by
\begin{align}\label{u70}
\mDI  \equiv \operatorname{span}  \bigl\{ \ket{\psi^\z_{\mu \imath}};  \mu = 1, \cdots, N_\lz,  \imath =1,2,3 \bigr\}.
\end{align}
This directly implies that $\mDz \oplus \mDI = \mD \otimes \mE_4$.

Finally, the total space $\mE$ defined by \eqref{e950}, is now written as the direct sum
\begin{align}\label{u60}
\mE = \mDz \oplus \mDI \oplus  \mC ,
\end{align}
where  the complement subspace $\mC$ is defined by,
\begin{align}\label{u80}
\mC = & \; \operatorname{span}  \bigl\{ \ket{l,m,i}; \; l = 0, \cdots, \infty, \; m = -l , \cdots, l,
\nonumber \\[6pt]
& \phantom{\operatorname{span}  \bigl\{ .} \wedge  l \neq l_0, \; i=1,2,3,4 \bigr\}.
\end{align}

\subsection{Solving the equations}\label{solving}

We consider now the change of the degenerate vectors $\ket{\psi^\z_{\mu 0}}, \, (\mu = 1, \cdots, N_\lz)$, when the sphere is deformed. Proceeding as in Sec. \ref{formal}, we write
\begin{subequations}\label{u90}
\begin{align}
\ket{\psi^\z_{\mu 0}} \to & \; \ket{\psi_{\mu}(\ve)} =  \ket{\psi^\z_{\mu 0}} + \ve  \ket{\psi^\un_{\mu}} + \ve^2  \ket{\psi^\du_{\mu}} + O(\ve^3),
 \label{u90A} \\[6pt]
x^\z \to & \; x_\mu(\ve) = x^\z + \ve x^\un_\mu + \ve^2 x^\du_\mu + O(\ve^3). \label{u90B}
\end{align}
\end{subequations}
This set of equations \eqref{u90} must hold for all  $\mu = 1, \cdots, N_\lz$. Note that for each value of $\mu$ the corrections  $x_\mu^\n(\ve)$  might be different, because the departure from the spherical shape typically removes the degeneracy.

As it is customary in quantum perturbation theory, we normalize the vector $\ket{\psi_{\mu}(\ve)}$ as
\begin{align}\label{u100}
\brak{\tpsi^\z_{\mu 0}}{\psi_{\mu}(\ve)} =1.
\end{align}
As in quantum mechanics, this normalization is particularly convenient for the later developments of the theory, and does not affect any physical quantity.
Equation \eqref{u100} implies that $\ket{\psi^\n_{\mu}}$ for $n \geq 1$, has no component along $\bra{\tpsi^\z_{\mu 0}}$, that is
\begin{align}\label{u110}
\brak{\tpsi^\z_{\mu 0}}{\psi_{\mu}^\n} = 0 , \qquad \text{for} \qquad n \geq 1.
\end{align}
Note, however, that $\ket{\psi_{\mu}^\n}$ may have components along $\ket{\psi_{\nu 0}^\z}$, with $\nu \neq \mu$.

From \eqref{e1010} it follows that the perturbed vector $\ket{\psi_{\mu 0}(\ve)}$  must satisfy
\begin{align}\label{u120}
\hmM(\ve) \ket{\psi_{\mu}(\ve)} =0.
\end{align}
Substituting \eqref{u90} into this equation and proceeding as in Sec. \ref{formal}, we obtain at first order in $\ve$,
\begin{align}\label{u130}
 \hmD^\z \ket{\psi^\un_{\mu}} + \left[ \hmV^\un  + x^\un_\mu \, \hmD  \right] \ket{\psi^\z_{\mu 0}}
 = 0.
\end{align}
where \eqref{ePropOp20} has been used. To solve this equation we must project it on the three orthogonal subspaces $\mDz, \mDI$, and $\mC$.

\subsubsection{Projecting along $\mDz$}

Multiplying \eqref{u130} from the left by $\bra{\tpsi^\z_{\nu 0}}$ and recalling \eqref{s35}, we obtain
\begin{align}\label{u140}
\mean{\tpsi^\z_{\nu 0}}{\hmV^\un}{\psi^\z_{\mu 0}} + x_\mu^\un \, \delta_{\nu \mu} \, \mean{\ta_0}{D_\lz}{\alpha_0} = 0,
\end{align}
where \eqref{u55B} has been used to cancel the leftmost term in \eqref{u130}. This equation implies that
\begin{align}\label{u150}
\frac{\mean{\tpsi^\z_{\nu 0}}{\hmV^\un}{\psi^\z_{\mu 0}}}{\mean{\ta_0}{D_\lz}{\alpha_0}} = -x_\mu^\un \, \delta_{\nu \mu},
\end{align}
which can be suggestively rewritten as
\begin{align}\label{u160}
\mean{\tvp^\z_{\nu}}{ \;  \frac{\mean{\ta_0}{\hmV^\un}{\alpha_0}}{\mean{\ta_0}{D_\lz}{\alpha_0}} \;
}{\vp^\z_{\mu}} = -x_\mu^\un \, \delta_{\nu \mu}.
\end{align}
The denominator $\mean{\ta_0}{D_\lz}{\alpha_0}$ is just a number, as shown in Appendix \ref{operators}. Conversely, the numerator $\mean{\ta_0}{\hmV^\un}{\alpha_0}$ is an operator in $\mE_\infty$. However, as it is sandwiched between $\bra{\tvp^\z_{\nu}}$ and $\ket{\vp^\z_{\mu}}$ which are in $\mD$, we can equivalently rewrite \eqref{u160} as
\begin{align}\label{u170}
\mean{\tvp^\z_{\nu}}{ \;  \frac{\mean{\ta_0}{P_\mD \,\hmV^\un\, P_\mD}{\alpha_0}}{\mean{\ta_0}{D_\lz}{\alpha_0}} \;
}{\vp^\z_{\mu}} = -x_\mu^\un \, \delta_{\nu \mu},
\end{align}
where
\begin{align}\label{u180}
P_\mD = \sum_{m=-\lz}^\lz \proj{\lz,m}{\lz,m},
\end{align}
is the projector onto the subspace $\mD$. By definition,
\begin{align}\label{u185}
P_\mD \ket{\vp^\z_{\mu}} = \ket{\vp^\z_{\mu}}, \quad \text{and} \quad \bra{\tvp^\z_{\mu}} P_\mD = \bra{\tvp^\z_{\mu}}.
\end{align}

Written in this form, Eq. \eqref{u170} tells us that the biorthogonal set  $\bigl\{ \ket{\vp^\z_\mu} ,\bra{\tvp^\z_\mu} \bigr\}$ must be chosen to make the $N_\lz \times N_\lz$ matrix $\mean{\ta_0}{P_\mD \,\hmV^\un\, P_\mD}{\alpha_0}$ diagonal in the subspace $\mD$. From now on, we assume that the set of vectors $\bigl\{ \ket{\vp^\z_\mu} ,\bra{\tvp^\z_\mu} \bigr\}$ has been chosen in this way (see Appendix \ref{MatEl} for more details).

Finally, setting $\nu = \mu$, we get the value of the first-order correction to the resonance:
\begin{align}\label{u190}
x_\mu^\un = - \frac{\mean{\tpsi^\z_{\mu 0}}{\hmV^\un}{\psi^\z_{\mu 0}}}{\mean{\ta_0}{D_\lz}{\alpha_0}},
\end{align}

The next steps will determine the components of $\ket{\psi^\un_\mu}$,  in $\mDI$ and $\mC$. We recall that for $n \geq 1$,
\begin{align}\label{u200}
\ket{\psi^\n_\mu} = \left. \ket{\psi^\n_\mu} \right|_{\mDz} + \left. \ket{\psi^\n_\mu} \right|_{\mDI} + \left. \ket{\psi^\n_\mu} \right|_{\mC},
\end{align}
where
\begin{subequations}\label{u205}
\begin{align}
\left. \ket{\psi^\n_\mu} \right|_{\mathscr{D}_0}  = & \;  \sum_{\nu = 0}^{N_\lz}  \ket{\psi^\z_{\nu 0}}  \brak{\tpsi^\z_{\nu 0}}{\psi^\n_\mu} , \label{u205a} \\[6pt]
\left. \ket{\psi^\n_\mu} \right|_{\mathscr{D}_I}  = & \;  \sum_{\nu = 1}^{N_\lz}  \sum_{\imath=1}^3 \ket{\psi^\z_{\nu \imath}} \brak{\tpsi^\z_{\nu \imath}}{\psi^\n_\mu}, \label{u205b} \\[6pt]
\left. \ket{\psi^\n_\mu} \right|_{\mathscr{C}}  = & \;  {\sum_{l,m,i}}' \ket{l,m,i} \brak{l,m,i}{\psi^\n_\mu}, \label{u205c}
\end{align}
\end{subequations}
where, here and hereafter,
\begin{align}\label{u210}
{\sum_{l,m,i}}' \qquad \text{stands for} \qquad  \sum_{\substack{l = 0 \\[2pt] \, l \neq l_0}}^{\infty}\sum_{m = -l}^{l} \sum_{i=1}^4.
\end{align}

\subsubsection{Projecting along $\mDI$}

Multiplying \eqref{u130} from the left by $\bra{\tpsi^\z_{\nu \imath}}$, with $\imath =1,2,3$, and recalling \eqref{s35}, we obtain
\begin{align}\label{u220}
0 = & \;  \mean{\tpsi^\z_{\nu \imath}}{\hmD^\z }{\psi^\un_{\mu}} + \mean{\tpsi^\z_{\nu \imath}}{\hat{\mathcal{V}}^\un}{\psi^\z_{\mu 0}} \nonumber \\[6pt]
&  + x^\un_\mu \mean{\tpsi^\z_{\nu \imath}}{ \hmD}{\psi^\z_{\mu 0}},
\end{align}
where
\begin{subequations}\label{u230}
\begin{align}
\mean{\tpsi^\z_{\nu \imath}}{\hmD^\z }{\psi^\un_{\mu}} & = \; \sum_{\jmath =1}^3 \mean{\ta_\imath}{D^\z_\lz}{\alpha_{\jmath}} \brak{\tpsi^\z_{\nu \jmath}}{\psi^\un_\mu}, \label{u230A} \\[6pt]
\mean{\tpsi^\z_{\nu \imath}}{ \hmD}{\psi^\z_{\mu 0}} & = \;  \delta_{\nu \mu} \mean{\ta_\imath}{D_\lz}{\alpha_0} . \label{u230C}
\end{align}
\end{subequations}
Substituting \eqref{u230} into \eqref{u220}, we obtain
\begin{align}\label{u240}
0 = & \;  \sum_{\jmath=1}^3 \mean{\ta_\imath}{D^\z_\lz}{\alpha_{\jmath}} \brak{\tpsi^\z_{\nu \jmath}}{\psi^\un_\mu} + \mean{\tpsi^\z_{\nu \imath}}{\hat{\mathcal{V}}^\un}{\psi^\z_{\mu 0}} \nonumber \\[6pt]
&  + x^\un_\mu \delta_{\nu \mu} \mean{\ta_\imath}{D_\lz}{\alpha_0}, \qquad (\imath=1,2,3),
\end{align}
where $x^\un_\mu$ is given by \eqref{u190}.
For given values of $\mu$ and $\nu$, this is a set of three linear equations in the variables $\brak{\tpsi^\z_{\nu 1}}{\psi^\un_\mu},\brak{\tpsi^\z_{\nu 2}}{\psi^\un_\mu}$, and $\brak{\tpsi^\z_{\nu 3}}{\psi^\un_\mu}$.  We write it concisely as
\begin{align}\label{u250}
\textsf{D} \vec{X} - \vec{Y} = 0,
\end{align}
where we have defined
\begin{subequations}\label{u260}
\begin{align}
X_\jmath  = & \;   \brak{\tpsi^\z_{\nu \jmath}}{\psi^\un_\mu}, \label{u260a} \\[6pt]
Y_\imath = & \; -\mean{\tpsi^\z_{\nu \imath}}{\hat{\mathcal{V}}^\un}{\psi^\z_{\mu 0}} - x^\un_\mu \delta_{\nu \mu} \mean{\ta_\imath}{D_\lz}{\alpha_0}, \label{u260b}
\end{align}
\end{subequations}
$(\imath, \jmath =1,2,3)$, and the $3 \times 3$ matrix $\textsf{D}$ with elements
\begin{align}\label{matD}
\textsf{D}_{\imath \jmath} = \mean{\ta_\imath}{D^\z_\lz}{\alpha_{\jmath}}.
\end{align}
A straightforward calculation gives
\begin{align}\label{u270}
\textsf{D} = \textsf{D}_E = \begin{bmatrix}
            z^E + 1 & 0 & 0 \\[6pt]
            0 & \frac{z^E}{n_1^2} & -\frac{z^E}{n_2^2} \\[8pt]
            0 & 1 & -1 \vphantom{\psi_{1lm}^\z} \\
          \end{bmatrix}
\end{align}
for TE waves, and
\begin{align}\label{u280}
\textsf{D} = \textsf{D}_M = \begin{bmatrix}
            z^M - 1 & 0 & 0 \\[6pt]
            0 & 1 & -1 \\[6pt]
            0 & -n_1^2 z^M & n_2^2 z^M \\
          \end{bmatrix}
\end{align}
for TM waves, where $z^E$ and $z^M$ are given by \eqref{zE} and \eqref{zM}, respectively. These matrices are invertible, because
\begin{subequations}\label{u290}
\begin{align}
\det \textsf{D}_E  = & \;   \left(z^E + 1 \right)f_\lz^M \bigl( x^\z_E \bigr) \neq 0, \label{u290a} \\[6pt]
\det \textsf{D}_M  = & \;   \left(z^M - 1 \right)f_\lz^E \bigl( x^\z_M \bigr) \neq 0, \label{u290b}
\end{align}
\end{subequations}
where $f_l^E(x)$ and $f_l^M(x)$ are given by \eqref{e280TE} and \eqref{e280TM}, respectively, and we have denoted by $x^\z_E$ and $x^\z_M$ the solutions of
\begin{align}\label{u300}
f_\lz^E( x^\z_E \bigr) = 0, \qquad \text{and} \qquad f_\lz^M( x^\z_M \bigr) = 0,
\end{align}
respectively.

 Using these equations, we can eventually write the solution of \eqref{u250},
\begin{align}\label{u310}
 \vec{X} = \textsf{D}^{-1} \vec{Y} ,
\end{align}
as
\begin{subequations}\label{u320}
\begin{align}
 \brak{\tpsi^\z_{\nu 1}}{\psi^\un_\mu}  = & \;   \frac{1}{z^E+1} \, Y_1, \label{u320a} \\[6pt]
 \brak{\tpsi^\z_{\nu 2}}{\psi^\un_\mu}  = & \;   \frac{1}{f_\lz^M ( x^\z_E )} \left( -Y_2 + \frac{z^E}{n_2^2} Y_3\right), \label{u320b} \\[6pt]
 \brak{\tpsi^\z_{\nu 3}}{\psi^\un_\mu}  = & \;   \frac{1}{f_\lz^M ( x^\z_E )} \left( -Y_2 + \frac{z^E}{n_1^2} Y_3\right), \label{u320c}
\end{align}
\end{subequations}
for TE waves and
\begin{subequations}\label{u330}
\begin{align}
 \brak{\tpsi^\z_{\nu 1}}{\psi^\un_\mu}  = & \;   \frac{1}{z^M - 1} \, Y_1, \label{u330a} \\[6pt]
 \brak{\tpsi^\z_{\nu 2}}{\psi^\un_\mu}  = & \;   \frac{1}{f_\lz^E ( x^\z_M )} \left( n_2^2 z^M Y_2 +  Y_3 \right), \label{u330b} \\[6pt]
 \brak{\tpsi^\z_{\nu 3}}{\psi^\un_\mu}  = & \;   \frac{1}{f_\lz^E ( x^\z_M )} \left( n_1^2 z^M Y_2 +  Y_3 \right), \label{u330c}
\end{align}
\end{subequations}
for TM waves.

\subsubsection{Projecting along $\mC$}

Multiplying \eqref{u130} from the left by $\bra{l,m,i}$, with $l\neq \lz$ and $i =1,2,3,4$,  we obtain
\begin{align}\label{u340}
0 = & \;  \mean{l,m,i}{\hmD^\z }{\psi^\un_{\mu}} + \mean{l,m,i}{\hat{\mathcal{V}}^\un}{\psi^\z_{\mu 0}} \nonumber \\[6pt]
&  + x^\un_\mu \mean{l,m,i}{ \hmD}{\psi^\z_{\mu 0}}.
\end{align}
From  \eqref{s35} and $l\neq \lz$, it follows that the last term on the right-hand side is identically zero. Substituting \eqref{u205c} into \eqref{u340} we find, after a little calculation,
\begin{align}\label{u350}
0 =  \sum_{j=1}^4\mean{i}{D^\z_l }{j} \brak{l,m,j}{\psi^\un_\mu} + \mean{l,m,i}{\hat{\mathcal{V}}^\un}{\psi^\z_{\mu 0}},
\end{align}
which can be recast into
\begin{align}\label{u360}
0 = D^\z_l \vec{X} - \vec{Y},
\end{align}
where now we have defined $\vec{X}$ and $\vec{Y}$ as
\begin{subequations}\label{u370}
\begin{align}
X_j  = & \;   \brak{l,m,j}{\psi^\un_\mu}, \label{u370a} \\[6pt]
Y_i = & \; -\mean{l,m,i}{\hat{\mathcal{V}}^\un}{\psi^\z_{\mu 0}}, \label{u370b}
\end{align}
\end{subequations}
$(i,j=1,2,3,4)$. The $4 \times 4$ matrix $D^\z_l$ is defined by \eqref{s73}, with $g_{\alpha l} = g_{\alpha l} \bigl( x^\z \bigr)$, where $x^\z = x^\sigma_{\lz n}$, $(\sigma = E,M)$, is a solution of $f^\sigma_\lz\bigl( x^\z \bigr)=0$. By construction, $D^\z_l$ is invertible because for $l \neq \lz$,
\begin{align}\label{u380}
\det D^\z_l = f_l^E \bigl( x^\sigma_{\lz n}\bigr) f_l^M \bigl( x^\sigma_{\lz n} \bigr) \neq 0.
\end{align}
Finally, a straightforward calculation gives, for $l \neq \lz$,
\begin{subequations}\label{u390}
\begin{align}
\brak{l,m,1}{\psi^\un_\mu} = & \,    \frac{1}{f_l^E ( x^\z )} \left[ n_2 g_{2l}( x^\z ) Y_1 +  Y_2 \right], \label{u390a} \\[6pt]
\brak{l,m,2}{\psi^\un_\mu}  = & \,   \frac{1}{f_l^E ( x^\z )} \left[ n_1 g_{1l}( x^\z ) Y_1 +  Y_2 \right], \label{u390b} \\[6pt]
\brak{l,m,3}{\psi^\un_\mu}  = & \,   \frac{1}{f_l^M ( x^\z )} \left[ -Y_3 + \frac{g_{2l}( x^\z )}{n_2} Y_4 \right], \label{u390c}\\[6pt]
\brak{l,m,4}{\psi^\un_\mu}  = & \,   \frac{1}{f_l^M ( x^\z )} \left[ -Y_3 + \frac{g_{1l}( x^\z )}{n_1} Y_4 \right], \label{u390d}
\end{align}
\end{subequations}
where $x^\z = x^\sigma_{\lz n}$, with $\sigma = E,M$.

\subsection{Summary of the first-order perturbation theory}\label{TreComments}

In this section we have determined the first-order corrections $x^\un_\mu$ [Eq. \eqref{u190}] to the resonances, and the components $\brak{\tpsi^\z_{\nu \imath}}{\psi^\un_\mu}$ [Eqs. \eqref{u320} and \eqref{u330}]  and $\brak{l,m,i}{\psi^\un_\mu}$ [Eqs. \eqref{u390}] of the first-order vector $\ket{\psi^\un_\mu}$ [Eq. \eqref{u200}]. However, similarly to what happens in standard quantum perturbation theory, it is not possible to determine the components $\brak{\tpsi^\z_{\nu 0}}{\psi^\un_\mu}$ of $\ket{\psi^\un_\mu}$ along the degenerate subspace $\mDz$. For this, we need to solve the second-order equation \eqref{e1160C}. This will be done in the next section.

\section{Second-order equations I. Non-degenerate case}\label{secondI}

In this section we are going to solve \eqref{e1160C}, when the degeneracy is lifted to first order, that is when
\begin{align}\label{v10}
x^\un_\mu \neq x^\un_\nu, \quad \text{whenever} \quad \mu \neq \nu,
\end{align}
and $\mu,\nu =1, \cdots, N_\lz$. To begin with, we use \eqref{ePropOp20} to rewrite \eqref{e1160C} as
\begin{multline}\label{v20}
 \hmD^\z \ket{\psi^\du_\mu } +  \left( \hat{\mathcal{V}}^\un  + x^\un_\mu \, \hmD  \right) \ket{\psi^\un_\mu } \\[6pt]
 + \left(\hat{\mathcal{V}}^\du  + x^\du_\mu \,\hmD \right) \ket{\psi^\z_{\mu 0} } = 0.
\end{multline}
Next, we proceed as in first-order theory, projecting this equation on the subspaces $\mDz$ to find $x^\du_\mu$ and  $\brak{\tpsi^\z_{\nu 0}}{\psi^\un_\mu}$. We remind that, at second order we are interested only in the resonance corrections, so we do not need to determine the full vector $\ket{\psi^\du_\mu } $.

\subsubsection{Projecting along $\mDz$}

Multiplying \eqref{v20} from the left by $\bra{\tpsi^\z_{\nu 0}}$ and recalling \eqref{s35}, we obtain
\begin{subequations}\label{v30}
\begin{align}
0 = & \;\mean{\tpsi^\z_{\nu 0}}{\hmD^\z }{\psi^\du_\mu } \label{v30a} \\[6pt]
& + \mean{\tpsi^\z_{\nu 0}}{\hat{\mathcal{V}}^\un}{\psi^\un_\mu} \label{v30b} \\[6pt]
& + x^\un_\mu \, \mean{\tpsi^\z_{\nu 0}}{\hmD }{\psi^\un_\mu} \label{v30c} \\[6pt]
& + \mean{\tpsi^\z_{\nu 0}}{\hmV^\du}{\psi^\z_{\mu 0}} \label{v30d} \\[6pt]
& +  x^\du_\mu \,\mean{\tpsi^\z_{\nu 0}}{\hmD}{\psi^\z_{\mu 0}}.  \label{v30e}
\end{align}
\end{subequations}
Now, for clarity, we calculate the five addends on the right-hand side of \eqref{v30} separately.
\vspace{0.2truecm}
\\
\emph{First addend} [from \eqref{v30a}]:
 \begin{align}\label{v40}
\mean{\tpsi^\z_{\nu 0}}{\hmD^\z }{\psi^\du_\mu }  = 0, \qquad \text{from \eqref{u55B}}.
\end{align}
\vspace{0.2truecm}
\\
\emph{Second addend} [from \eqref{v30b}]:
 \begin{align}\label{v50}
\mean{\tpsi^\z_{\nu 0}}{\hat{\mathcal{V}}^\un}{\psi^\un_\mu} = & \;
- x^\un_\nu \mean{\ta_0}{D_\lz}{\alpha_0} \brak{\tpsi^\z_{\nu 0}}{\psi^\un_\mu} \nonumber  \\[6pt]
& + \sum_{\tau=1}^{N_\lz}\sum_{\imath=1}^3 \mean{\tpsi^\z_{\nu 0}}{\hmV^\un}{\psi^\z_{\tau \imath}} \brak{\tpsi^\z_{\tau \imath}}{\psi^\un_\mu} \nonumber \\[6pt]
& +{\sum_{l,m,i}}' \mean{\tpsi^\z_{\nu 0}}{\hmV^\un}{l,m,i} \brak{l,m,i}{\psi^\un_\mu},
\end{align}
where \eqref{u200} and \eqref{u205} and \eqref{u150} have been used.
\vspace{0.2truecm}
\\
\emph{Third addend} [from \eqref{v30c}]:
 \begin{align}\label{v60}
\mean{\tpsi^\z_{\nu 0}}{\hmD }{\psi^\un_\mu} = & \;
 \mean{\ta_0}{D_\lz}{\alpha_0} \brak{\tpsi^\z_{\nu 0}}{\psi^\un_\mu} \nonumber  \\[6pt]
& + \sum_{\imath=1}^3  \mean{\ta_0}{D_\lz}{\alpha_\imath} \brak{\tpsi^\z_{\nu \imath}}{\psi^\un_\mu} ,
\end{align}
where \eqref{u57} has been used.
\vspace{0.2truecm}
\\
\emph{Fourth addend} [from \eqref{v30d}]:
\begin{align}\label{v70}
\mean{\tpsi^\z_{\nu 0}}{\hmV^\du}{\psi^\z_{\mu 0}} = \mean{\tvp^\z_{\nu}, \ta_0}{\hmV^\du}{\vp^\z_{\mu}, \alpha_0}.
\end{align}
\vspace{0.2truecm}
\\
\emph{Fifth addend} [from \eqref{v30e}]:
\begin{align}\label{v80}
\mean{\tpsi^\z_{\nu 0}}{\hmD}{\psi^\z_{\mu 0}} = \delta_{\nu \mu} \mean{\ta_0}{D_\lz}{\alpha_0},
\end{align}
where \eqref{u57} has been used.

Now that we have all the terms, we can use \eqref{v30} evaluated for $\nu = \mu$ to obtain $x^\du_\mu$. Next we set $\nu \neq \mu$ to get $\brak{\tpsi^\z_{\nu 0}}{\psi^\un_\mu}$. In the first case we find,
\begin{align}\label{v90}
x^\du_\mu = & \; \frac{-1}{\mean{\ta_0}{D_\lz}{\alpha_0}} \Biggl\{\mean{\tpsi^\z_{\mu 0}}{\hmV^\du}{\psi^\z_{\mu 0}} \nonumber \\[6pt]
& + {\sum_{l,m,j}}' \mean{\tpsi^\z_{\mu 0}}{\hmV^\un}{l,m,j} \brak{l,m,j}{\psi^\un_\mu}\nonumber \\[6pt]
& + \sum_{\tau=1}^{N_\lz}\sum_{\imath=1}^3 \mean{\tpsi^\z_{\mu 0}}{\hmV^\un}{\psi^\z_{\tau \imath}} \brak{\tpsi^\z_{\tau \imath}}{\psi^\un_\mu} \nonumber \\[6pt]
& + x^\un_\mu \, \sum_{\imath=1}^3 \mean{\ta_0}{D_\lz}{\alpha_\imath} \brak{\tpsi^\z_{\mu \imath}}{\psi^\un_\mu} \Biggr\}.
\end{align}
In the second case we obtain
\begin{align}\label{v100}
\brak{\tpsi^\z_{\nu 0}}{\psi^\un_\mu} = & \; \frac{1}{\mean{\ta_0}{D_\lz}{\alpha_0}} \frac{1}{x_\nu^\un - x_\mu^\un}  \Biggl\{\mean{\tpsi^\z_{\nu 0}}{\hmV^\du}{\psi^\z_{\mu 0}} \nonumber \\[6pt]
& + {\sum_{l,m,i}}' \mean{\tpsi^\z_{\nu 0}}{\hmV^\un}{l,m,i} \brak{l,m,i}{\psi^\un_\mu}\nonumber \\[6pt]
& + \sum_{\tau=1}^{N_\lz}\sum_{\imath=1}^3 \mean{\tpsi^\z_{\nu 0}}{\hmV^\un}{\psi^\z_{\tau \imath}} \brak{\tpsi^\z_{\tau \imath}}{\psi^\un_\mu} \nonumber \\[6pt]
& + x^\un_\mu \, \sum_{\imath=1}^3 \mean{\ta_0}{D_\lz}{\alpha_\imath} \brak{\tpsi^\z_{\mu \imath}}{\psi^\un_\mu} \Biggr\},
\end{align}
with $\nu \neq \mu$.

It is  instructive to rewrite \eqref{v90} in a compact form using the first-order equation \eqref{u130} in the bra form
\begin{align}\label{u130bis}
\bra{\tpsi^\un_{\mu}} \hmD^\z + \bra{\tpsi^\z_{\mu 0}}\bigl( \hat{\mathcal{V}}^\un  + x^\un_\mu \, \hmD  \bigr)
 = 0.
\end{align}
Multiplying this equation from the right by $\ket{\psi^\un_{\mu}}$, we obtain
\begin{align}\label{u130ter}
\mean{\tpsi^\z_{\mu 0}}{ \hat{\mathcal{V}}^\un  + x^\un_\mu \, \hmD }{\psi^\un_{\mu}}
 = -\mean{\tpsi^\un_{\mu}}{ \hmD^\z }{\psi^\un_{\mu}}.
\end{align}
Substituting this result in \eqref{v30} we find, after a simple manipulation,
\begin{align}\label{u130quater}
x^\du_\mu = \frac{\mean{\tpsi^\un_{\mu}}{ \hmD^\z }{\psi^\un_{\mu}} - \mean{\tpsi^\z_{\mu 0}}{\hmV^\du}{\psi^\z_{\mu 0}}}{\mean{\tpsi^\z_{\mu0}}{ \hmD }{\psi^\z_{\mu0}}}.
\end{align}
This expression is only formal in the sense that it contains unknown coefficients. However, it makes clear what the complicated equation \eqref{v90} actually means.

This completes the calculation for the second-order corrections,  when the degeneracy is lifted to first order. If this is not the case, we need a different procedure, which will be developed in the next section.

\section{Second-order equations II. Degenerate case}\label{secondII}

\subsection{Some preparatory remarks}

Now we consider the case when the degeneracy is only partially removed to first order. Without loss of generality, we assume that the first $N$ first-order corrections $x^\un_\mu$ are equal to each other:
\begin{align}\label{t10}
x^\un_1 = x^\un_2 = \cdots = x^\un_N \equiv x^\un,
\end{align}
where $1 < N \leq N_\lz$, and \eqref{u150} is still valid.
Consequently, the initial degenerate subspace $\mDz$ breaks in two parts, denoted by $\mDzN$ and $\mDzM$, with $N+M = N_\lz$, and defined by
\begin{subequations}\label{t20}
\begin{align}
\mDzN = & \; \operatorname{span} \left\{ \ket{\psi^\z_{\mu0}} , \; \mu = 1, \cdots, N \right\},
 \label{t20A} \\[6pt]
\mDzM = & \; \operatorname{span} \left\{ \ket{\psi^\z_{\mu 0}} , \; \mu = N+1, \cdots, N_\lz \right\}. \label{t20B}
\end{align}
\end{subequations}
As before, because of the remaining $N$-fold degeneracy, it is convenient to define a new orthonormal basis in $\mDzN$, denoted by $\ket{\psi^\z_{A0}}$ and defined by
\begin{align} \label{t30}
\ket{\psi^\z_{A0}} \equiv & \; \ket{\vp_A^\z} \ket{\alpha_0} \nonumber \\[6pt]
= & \; \sum_{\mu=1}^N \vp^\z_{A \mu} \ket{\vp^\z_\mu} \ket{\alpha_0} \nonumber \\[6pt]
= & \; \sum_{\mu=1}^N \vp^\z_{A \mu} \ket{\psi^\z_{\mu0}},
\end{align}
where $A = 1, \cdots, N$. The coefficients $\vp^\z_{A \mu}$ are to be determined. As it should be customary now,
we think of this basis as a part of the  biorthogonal set $\bigl\{ \ket{\psi^\z_{A0}} ,\bra{\tpsi^\z_{A0}} \bigr\}$ in $\mE_\infty$, where
\begin{align} \label{t40}
\bra{\tpsi^\z_{A0}} \equiv \sum_{\mu=1}^N \tvp^\z_{A \mu} \bra{\tpsi^\z_{\mu0}},
\end{align}
with
\begin{align} \label{t50}
\brak{\tpsi^\z_{A0}}{\psi^\z_{B0}} = \sum_{\mu = 1}^{N} \tvp^\z_{A \mu} \, \vp^\z_{B \mu} = \delta_{AB}.
\end{align}

We build up the perturbation theory as usual,
\begin{subequations}\label{t60}
\begin{align}
\ket{\psi_{A}(\ve)} = & \; \ket{\psi^\z_{A 0}} + \ve  \ket{\psi^\un_{A}} + \ve^2  \ket{\psi^\du_{A}} + O(\ve^3),
 \label{t60A} \\[6pt]
x_A(\ve) = & \; x^\z + \ve x^\un + \ve^2 x^\du_A + O(\ve^3). \label{t60B}
\end{align}
\end{subequations}
Note that in \eqref{t60B} the first-order correction $x^\un$ has no label, according to \eqref{t10}. Using the fundamental equation \eqref{s10},
\begin{align}\label{s10bis}
\hat{\mathcal{M}} \left( \ve \right) \ket{\psi_{A}(\ve)} = 0,
\end{align}
we can obtain the familiar chain of equations
\begin{subequations}\label{t70}
\begin{align}
\hmD^{(0)} \ket{\psi^\z_{A 0}} = & \; 0, \label{t70A} \\[6pt]
\hmD^{(0)} \ket{\psi^\un_{A}}  = & \;  -\hmM^{(1)} \ket{\psi^\z_{A 0}} , \label{t70B} \\[6pt]
\hmD^{(0)} \ket{\psi^\du_{A}} = & \;  - \hmM^{(1)} \ket{\psi^\un_{A}} - \hmM^{(2)} \ket{\psi^\z_{A 0}} , \label{t70C}
\end{align}
\end{subequations}
etc., where \eqref{ePropOp20a} has been used. Next, we need to adapt to the present case the normalization condition \eqref{u110}. As shown in Appendix \ref{normalization}, the new condition is
\begin{align}\label{t90}
\brak{\tpsi^\z_{A 0}}{\psi_{A}^\n} = 0 , \qquad \text{for} \qquad 1 \leq A \leq N,
\end{align}
and $n \geq 1$.
Therefore, similarly to \eqref{u200} and \eqref{u205}, we can write now
\begin{align}\label{t110}
\ket{\psi^\n_{A}} =  \left. \ket{\psi^\n_{A}} \right|_{\mDzN}  + \left. \ket{\psi^\n_{A}} \right|_{\mDzM}  + \left. \ket{\psi^\n_{A}}   \right|_{\mDI} + \left. \ket{\psi^\n_{A}}  \right|_{\mC},
\end{align}
where
\begin{subequations}\label{t120}
\begin{align}
\left. \ket{\psi^\n_{A}} \right|_{\mDzN}  = & \;  \sum_{\nu = 1}^{N}  \ket{\psi^\z_{\nu 0}}  \brak{\tpsi^\z_{\nu 0}}{\psi^\n_{A}}, \label{t100d} \\[6pt]
\left. \ket{\psi^\n_{A}} \right|_{\mDzM}  = & \;  \sum_{\nu = N+1}^{N_\lz}  \ket{\psi^\z_{\nu 0}}  \brak{\tpsi^\z_{\nu 0}}{\psi^\n_{A}}, \label{t100a} \\[6pt]
\left. \ket{\psi^\n_{A}} \right|_{\mDI}  = & \;  \sum_{\nu = 1}^{N_\lz}  \sum_{\imath=1}^3 \ket{\psi^\z_{\nu \imath}} \brak{\tpsi^\z_{\nu \imath}}{\psi^\n_{A}}, \label{t120b} \\[6pt]
\left. \ket{\psi^\n_{A}} \right|_{\mathscr{C}}  = & \;  {\sum_{l,m,i}}' \ket{l,m,i} \brak{l,m,i}{\psi^\n_{A}}. \label{t120c}
\end{align}
\end{subequations}

The zeroth-order equation \eqref{t70A} is trivially satisfied. The first-order equation \eqref{t70B}, also becomes an identity when projected on $\mDzN$ and $\mDzM$. However, it furnishes the coefficients $\brak{\tpsi^\z_{\nu \imath}}{\psi^\un_{A}}$ and $\brak{l,m,i}{\psi^\un_{A}}$ when projected upon $\mDI$ and $\mC$, respectively. So, let us study it.

\subsection{First order equations}

\subsubsection{Projecting along $\mDI$}

Multiplying \eqref{t70B} from the left by $\bra{\tpsi^\z_{\nu \imath}}$, with $\nu = 1, \cdots, N_\lz$ and  $\imath =1,2,3$,  we obtain
\begin{align}\label{t130}
0 =  & \; \mean{\tpsi^\z_{\nu \imath}}{\hmD^\z}{\psi^\un_A} \nonumber \\[6pt]
& + \mean{\tpsi^\z_{\nu \imath}}{\hmV^\un}{\psi^\z_{A0}} + x^\un \mean{\tpsi^\z_{\nu \imath}}{\hmD}{\psi^\z_{A0}} ,
\end{align}
where \eqref{ePropOp20b} has been used. The first term of this sum is
\begin{align}\label{t140}
\mean{\tpsi^\z_{\nu \imath}}{\hmD^\z}{\psi^\un_A}  =  & \; \left. \mean{\tpsi^\z_{\nu \imath}}{\hmD^\z}{\psi^\un_A} \right|_{\mDI} \nonumber \\[6pt]
=  & \;  \sum_{\jmath=1}^3 \mean{\ta_\imath}{D^\z_\lz}{\alpha_\jmath} \brak{\tpsi^\z_{\nu \jmath}}{\psi^\un_A} ,
\end{align}
because diagonal operators $\hmD$ cannot connect $\mDI$ with $\mC$, and \eqref{u57} has been used. The second term does not require calculations, being simply
\begin{align}\label{t150}
\mean{\tpsi^\z_{\nu \imath}}{\hmV^\un}{\psi^\z_{A0}}  =    \sum_{\mu=1}^N  \vp^\z_{A\mu} \, \mean{\tpsi^\z_{\nu \imath}}{\hmV^\un}{\psi^\z_{\mu 0}} ,
\end{align}
where the coefficients $ \vp^\z_{A\mu}$ are still to be determined and \eqref{t30} has been used. Finally, the third and last term is
\begin{align}\label{t160}
\mean{\tpsi^\z_{\nu \imath}}{\hmD}{\psi^\z_{A0}}  =  & \; \left. \mean{\tpsi^\z_{\nu \imath}}{\hmD}{\psi^\z_{A 0}} \right|_{\mDI} \nonumber \\[6pt]
=  & \;  \sum_{\mu=1}^N  \vp^\z_{A\mu} \, \delta_{\nu \mu} \,\mean{\ta_\imath}{D_\lz}{\alpha_0},
\end{align}
where  \eqref{u57} has been again used.

Substituting \eqref{t140}-\eqref{t160}  into \eqref{t130}, we obtain
\begin{align}\label{t170}
0 =  & \; \sum_{\jmath=1}^3 \mean{\ta_\imath}{D^\z_\lz}{\alpha_\jmath} \brak{\tpsi^\z_{\nu \jmath}}{\psi^\un_A} \nonumber \\[6pt]
& +   \sum_{\mu=1}^N \vp^\z_{A\mu} \Bigl[\mean{\tpsi^\z_{\nu \imath}}{\hmV^\un}{\psi^\z_{\mu 0}} \nonumber \\[6pt]
& + x^\un  \, \delta_{\nu \mu} \,\mean{\ta_\imath}{D_\lz}{\alpha_0} \Bigr],
\end{align}
for $\nu =1,  \dots, N_\lz$ and $\imath =1,2,3$. For each value of $\nu$, Eq. \eqref{t170} can be written as a matrix equation of the form
\begin{align}\label{t180}
0 = \textsf{D} \vec{X} - \vec{Y},
\end{align}
where  the $3 \times 3$ matrix $\textsf{D}$ is defined by \eqref{matD}, that is, $\textsf{D}_{\imath \jmath} = \mean{\ta_\imath}{D^\z_\lz}{\alpha_{\jmath}}$, and now
\begin{subequations}\label{t190}
\begin{align}
X_\jmath  = & \;   \brak{\tpsi^\z_{\nu \jmath }}{\psi^\un_A}, \label{t190a} \\[6pt]
Y_\imath = & \; -\sum_{\mu=1}^N \vp^\z_{A\mu} \Bigl[\mean{\tpsi^\z_{\nu \imath}}{\hmV^\un}{\psi^\z_{\mu 0}} \nonumber \\[6pt]
& \phantom{-\sum_{\mu=1}^N \vp^\z_{A\mu} \Bigl[} - x^\un  \, \delta_{\nu \mu} \,\mean{\ta_\imath}{D_\lz}{\alpha_0} \Bigr]. \label{t190b}
\end{align}
\end{subequations}
Specifically, $\textsf{D}$ is  given by \eqref{u270} for TE waves, and by \eqref{u280} for TM waves.
Therefore, we know that it is invertible and we can formally write
\begin{align}\label{t200}
\brak{\tpsi^\z_{\nu \imath}}{\psi^\un_A} = &   -\sum_{\mu=1}^N \biggl\{ \sum_{\jmath=1}^3  \textsf{D}^{-1}_{\imath \jmath}
\Bigl[\mean{\tpsi^\z_{\nu j}}{\hmV^\un}{\psi^\z_{\mu 0}} \nonumber \\[6pt]
&  + x^\un  \, \delta_{\nu \mu} \,  \mean{\ta_\jmath}{D_\lz}{\alpha_0} \Bigr]\biggr\} \, \vp^\z_{A\mu} ,
\end{align}
where $\textsf{D}^{-1}$ is the matrix inverse of $\textsf{D}$.
Using the standard notation $\textsf{D}^{-1}(\{\imath,\jmath\})$, $(\imath,\jmath=1,2,3)$, to denote the principal submatrix of $\textsf{D}^{-1}$ that lies between row $\imath$ and row $\jmath$, and between column $\imath$ and column $\jmath$ \cite{Horn1}, we can write $\textsf{D}^{-1} = \textsf{D}^{-1}(\{1,1\}) \oplus \textsf{D}^{-1}(\{2,3\})$, where
\begin{subequations}\label{t210}
\begin{align}
\textsf{D}_E^{-1}(\{1,1\}) = & \; \frac{1}{z^E+1}, \label{t210a} \\[6pt]
\textsf{D}_E^{-1}(\{2,3\}) = & \; \frac{1}{f_\lz^M ( x^\z_E )}
          \begin{bmatrix}
            - 1 \; & \phantom{x} \frac{z^E}{n_2^2}  \\[8pt]
             -1 \; &  \phantom{x} \frac{z^E}{n_1^2}   \\
          \end{bmatrix} \label{t210b}
\end{align}
\end{subequations}
for TE waves, and
\begin{subequations}\label{t220}
\begin{align}
\textsf{D}_M^{-1}(\{1,1\}) = & \; \frac{1}{z^M -1}, \label{t220a} \\[6pt]
\textsf{D}_M^{-1}(\{2,3\}) = & \;  \frac{1}{f_\lz^E ( x^\z_M )}
           \begin{bmatrix}
             n_2^2 z^M & 1 \vphantom{\frac{z^E}{n_2^2} } \\[6pt]
             n_1^2 z^M &  1 \vphantom{\frac{z^E}{n_2^2} }  \\
          \end{bmatrix} \label{t220b}
\end{align}
\end{subequations}
for TM waves, where $z^E$ and $z^M$ are given by \eqref{zE} and \eqref{zM}, respectively.

Thus, the expression between curly brackets in \eqref{t200} is completely determined. In the remainder we will indicate it compactly with
\begin{align}\label{t230}
\mM^\un_{\nu \imath| \mu} =  & \;  -\sum_{\jmath=1}^3  \textsf{D}^{-1}_{\imath \jmath}
\Bigl[\mean{\tpsi^\z_{\nu \jmath}}{\hmV^\un}{\psi^\z_{\mu 0}}  \nonumber \\[6pt]
& \phantom{\sum_{j=1}^3  D^{-1}_{\imath \jmath}
\Bigl[} + x^\un  \, \delta_{\nu \mu} \,  \mean{\ta_\jmath}{D_\lz}{\alpha_0} \Bigr]\nonumber \\[6pt]
=  & \;  -\sum_{\jmath=1}^3  \textsf{D}^{-1}_{\imath \jmath}
\mean{\tpsi^\z_{\nu \jmath}}{\hmM^\un}{\psi^\z_{\mu 0}}  ,
\end{align}
to rewrite \eqref{t200} as
\begin{align}\label{t240}
\brak{\tpsi^\z_{\nu \imath}}{\psi^\un_A} =    \sum_{\mu=1}^N \mM^\un_{\nu \imath| \mu} \, \vp^\z_{A\mu} .
\end{align}

\subsubsection{Projecting along $\mC$}

Multiplying \eqref{t70B} from the left by $\bra{l,m,i}$, with $l \neq \lz$ and  $i =1,2,3,4$,  we obtain
\begin{align}\label{t260}
0 =  & \; \mean{l,m,i}{\hmD^\z}{\psi^\un_A} \nonumber \\[6pt]
& + \mean{l,m,i}{\hmV^\un}{\psi^\z_{A0}} + x^\un \mean{l,m,i}{\hmD}{\psi^\z_{A0}} .
\end{align}
The last term proportional to $x^\un$ is equal to $0$ due to the now familiar properties of the diagonal operators $\hmD$. Using \eqref{s35}, we can directly calculate the first term to get
\begin{align}\label{t270}
\mean{l,m,i}{\hmD^\z}{\psi^\un_A} = \sum_{j=1}^4 \mean{i}{D^\z_l}{j} \brak{l,m,j}{\psi^\un_A}.
\end{align}
Finally, the second term is simply given by
\begin{align}\label{t280}
\mean{l,m,i}{\hmV^\un}{\psi^\z_{A0}} = \sum_{\mu=1}^N \mean{l,m,i}{\hmV^\un}{\psi^\z_{\mu 0}}  \, \vp^\z_{\mu A}.
\end{align}
Substituting \eqref{t270} and \eqref{t280} into \eqref{t260}, we obtain
\begin{align}\label{t290}
0 =  & \; \sum_{j=1}^4 \mean{i}{D^\z_l}{j} \brak{l,m,j}{\psi^\un_A} \nonumber \\[6pt]
& + \sum_{\mu=1}^N \mean{l,m,i}{\hmV^\un}{\psi^\z_{\mu 0}}  \, \vp^\z_{\mu A}.
\end{align}
This equation is analogous to \eqref{u350} with the same $4 \times 4$ invertible matrix $D^\z_l$ with elements $\bigl[D^\z_l\bigr]_{ij} = \mean{i}{D^\z_l}{j}$. Therefore, we do not need to make additional calculations and we can write directly
\begin{align}\label{t300}
\brak{l,m,i}{\psi^\un_A} =    \sum_{\mu=1}^N \mV^\un_{l m i| \mu} \, \vp^\z_{A\mu} ,
\end{align}
where we have defined
\begin{align}\label{t310}
\mV^\un_{l m i| \mu} =  - \sum_{j=1}^4 \bigl[ D^\z_l  \bigr]_{ij}^{-1} \mean{l,m,j}{\hmV^\un}{\psi^\z_{\mu 0}} .
\end{align}
In this expression
\begin{align}\label{t315}
\bigl[ D^\z_l \bigr]^{-1} = \bigl[ D^\z_l \bigr]^{-1}(\{1,2\}) \oplus \bigl[ D^\z_l \bigr]^{-1}(\{3,4\}),
\end{align}
%
%
where
\begin{subequations}\label{t320}
\begin{align}
\bigl[ D^\z_l \bigr]^{-1}(\{1,2\}) = & \;  \frac{1}{f_l^E ( x^\z )}
           \begin{bmatrix}
             n_2 \, g_{2l}( n_2 x^\z ) & 1  \vphantom{\frac{g_{2l}( n_2 x^\z )}{n_2}} \\[6pt]
             n_1 \, g_{1l}( n_1 x^\z ) &  1 \vphantom{\frac{g_{2l}( n_2 x^\z )}{n_2}}  \\
          \end{bmatrix}, \label{t320a} \\[8pt]
\bigl[ D^\z_l \bigr]^{-1}(\{3,4\}) = & \;  \frac{1}{f_l^M ( x^\z )}
           \begin{bmatrix}
             -1 &  \frac{g_{2l}( n_2 x^\z )}{n_2} \vphantom{ i} \\[8pt]
             -1 &  \frac{g_{1l}( n_1 x^\z )}{n_1} \vphantom{ }  \\
          \end{bmatrix}, \label{t320b}
\end{align}
\end{subequations}
where $x^\z = x^\sigma_{\lz n}$, with $\sigma = E,M$.

\subsection{Second order equations}

\subsubsection{Projecting along $\mDzN$}

Let us set $\nu \leq N$. Multiplying \eqref{t70C} from the left by $\bra{\tpsi^\z_{\nu 0}}$ and using \eqref{ePropOp20b}, we obtain
\begin{subequations}\label{x10}
\begin{align}
\mean{\tpsi^\z_{\nu 0}}{\hmD^\z }{\psi^\du_\mu } = &  -\mean{\tpsi^\z_{\nu 0}}{\hat{\mathcal{V}}^\un}{\psi^\un_A} \label{x10a} \\[6pt]
& - x^\un  \mean{\tpsi^\z_{\nu 0}}{\hmD }{\psi^\un_A} \label{x10b} \\[6pt]
& - \mean{\tpsi^\z_{\nu 0}}{\hmV^\du}{\psi^\z_{A 0}} \label{x10c} \\[6pt]
& -  x^\du_A \,\mean{\tpsi^\z_{\nu 0}}{\hmD}{\psi^\z_{A 0}}.  \label{x10d}
\end{align}
\end{subequations}
The left-hand side of this equation vanishes because of \eqref{u55B}. The four addends on the right-hand side of \eqref{x10} are calculated as follows.
\vspace{0.2truecm}
\\
\emph{First addend} [from \eqref{x10a}]:
 \begin{align}\label{x20}
\mean{\tpsi^\z_{\nu 0}}{\hat{\mathcal{V}}^\un}{\psi^\un_A}  = &  - x^\un \mean{\ta_0}{D_\lz}{\alpha_0} \brak{\tpsi^\z_{\nu 0}}{\psi^\un_A} \nonumber \\[6pt]
 & + x^\un  \sum_{\mu=1}^{N_\lz} \sum_{\imath =1}^3 \mean{\tpsi^\z_{\nu 0}}{\hat{\mathcal{V}}^\un}{\psi^\z_{\mu \imath }} \brak{\tpsi^\z_{\mu \imath }}{\psi^\un_A} \nonumber \\[6pt]
 & + {\sum_{l,m,i}}' \mean{\tpsi^\z_{\nu 0}}{\hmV^\un}{l,m,i} \brak{l,m,i}{\psi^\un_A},
\end{align}
where \eqref{u140} has been used. Note that in the first line of \eqref{x20}, the coefficients $\brak{\tpsi^\z_{\nu 0}}{\psi^\un_A}$ are unknown. However, we will see soon that such term is canceled by an analogous one in the second addend \eqref{x10b}.
\vspace{0.2truecm}
\\
\emph{Second addend} [from \eqref{x10b}]:
 \begin{align}\label{x30}
x^\un \, \mean{\tpsi^\z_{\nu 0}}{\hmD }{\psi^\un_A}  = & \;
 x^\un \mean{\ta_0}{D_\lz}{\alpha_0} \brak{\tpsi^\z_{\nu 0}}{\psi^\un_A} \nonumber  \\[6pt]
& +  x^\un  \sum_{\imath =1}^3 \mean{\ta_0}{D_\lz}{\alpha_\imath } \brak{\tpsi^\z_{\nu \imath }}{\psi^\un_A}.
\end{align}
As anticipated, the term $ x^\un \mean{\ta_0}{D_\lz}{\alpha_0} \brak{\tpsi^\z_{\nu 0}}{\psi^\un_A}$ in this expression, cancels with the same term in \eqref{x20}.
\vspace{0.2truecm}
\\
\emph{Third addend} [from \eqref{x10c}]:
 \begin{align}\label{x40}
\mean{\tpsi^\z_{\nu 0}}{\hmV^\du}{\psi^\z_{A 0}}  = \sum_{\mu=1}^N \mean{\tpsi^\z_{\nu 0}}{\hmV^\du}{\psi^\z_{\mu 0} }\vp^\z_{A \mu}.
\end{align}
\vspace{0.2truecm}
\\
\emph{Fourth addend} [from \eqref{x10d}]:
\begin{align}\label{x50}
x^\du_A \mean{\tpsi^\z_{\nu 0}}{\hmD}{\psi^\z_{A 0}} =  x^\du_A  \mean{\ta_0}{D_\lz}{ \alpha_0} \, \vp^\z_{A \nu}.
\end{align}
Summing all these addends, after  straightforward manipulation we eventually obtain
\begin{align}\label{x60}
\sum_{\nu=1}^N \left( M^\du_{\mu \nu} + x^\du_A \, \delta_{\mu \nu} \right) \vp^\z_{A \nu} = 0,
\end{align}
where we have defined the $N \times N$ matrix $M^\du$, by the elements
\begin{align}\label{x70}
M^\du_{\mu \nu} \equiv & \; \frac{1}{\mean{\ta_0}{D_\lz}{ \alpha_0} } \Biggl\{ \sum_{\tau =1}^{N_\lz} \sum_{\imath =1}^3  \Bigl[
\mean{\tpsi^\z_{\mu 0}}{\hmV^\un}{\psi^\z_{\tau \imath }} \nonumber \\[6pt]
& + x^\un \delta_{\tau \mu} \mean{\ta_0}{D_\lz}{\alpha_\imath }
 \Bigr] \mM^\un_{\tau \imath  |\nu} \nonumber \\[6pt]
& + {\sum_{l,m,i}}' \mean{\tpsi^\z_{\mu 0}}{\hmV^\un}{l,m,i}  \mV^\un_{l m i |\nu} \nonumber \\[6pt]
& + \mean{\tpsi^\z_{\mu 0}}{\hmV^\du}{\psi^\z_{\nu 0}} \Biggr\} .
\end{align}
Equation \eqref{x60} is an eigenvalue equation that gives us both the second-order resonance corrections $x^\du_A$, as the eigenvalues of $M^\du$, and the basis vectors $\ket{\psi^\z_{A0}}$, as the associated eigenvectors.  This completes our calculations.

\section{Oblate spheroid}\label{ellipsoid}

In this section we apply our theory to nearly spherical dielectric resonators, which are  rotationally invariant around  the $z$ axis, with $h(\theta,\phi) = h(\theta)$. This permits us to illustrate the use of \emph{degenerate} perturbation theory in the case in which the degeneration is only partially removed to first order. The unperturbed system is, as always in this work, a dielectric sphere of radius $a$ and refractive index $n_1$, surrounded by vacuum or air with $n_2=1$. For practical reasons (the numerical results are more accurate), we will choose $n_1 = 2$.

As a specific example, we consider as nearly spherical resonator, an oblate spheroid with semiaxes $(a(1 + \delta ),a(1 + \delta ),a)$, where $0 < \delta \ll 1$ quantifies the magnitude of the deformation.
The equation of the spheroid in spherical coordinates is
\begin{align}\label{el20}
r = & \; \frac{a}{\displaystyle \sqrt{\cos ^2\theta  + \frac{\sin ^2\theta }{(1 + \delta )^2}}} \nonumber \\[6pt]
=  & \;   a \Bigl[ 1 + \delta \sin^2 \theta  + O \left( \delta^2 \right) \Bigr],
\end{align}
where the Taylor expansion truncated at first order, is a good approximation for $0 < \delta \ll 1$.
Thus, in the remainder we will set
\begin{align}\label{el30}
h(\theta,\phi) = h(\theta) = \delta \sin^2 \theta ,
\end{align}
as deformation function  and
\begin{align}\label{el35}
r = a \left( 1+  \delta \sin^2 \theta \right),
\end{align}
for the equation defining the approximate oblate spheroid, for both the perturbative and the numerical calculations.
 To perform the latter, we used the {\footnotesize COMSOL} Multiphysics$^\circledR$ software (Wave Optics Module) \cite{comsol}.

For illustration purposes, we choose as unperturbed resonances \eqref{e284} and \eqref{e284bis} $x^\z_E =  x_{\lz n_0}^E \simeq 6.826 - i \, 2.535 \times 10^{-3} $, and $x^\z_M =  x_{\lz n_0}^M \simeq 7.248 - i \, 4.325 \times 10^{-3}$, with $\lz = 10$ and $n_0=1$, for TE and TM waves, respectively. We take the magnitude of the deformation to be equal to $\delta = 0.01$ and $\delta=0.05$ for the TE and TM waves, respectively.  The choice of $n_0=1$ (first radial mode), is suggested by the fact that higher-order radial numbers ($n>1$) mark lossy waves not localized near the surface of the resonator, which are of low practical interest \cite{Oraevsky}.
%
\begin{figure}[!h]
\centerline{\includegraphics[scale=3,clip=false,width=.95\columnwidth,trim = 0 0 0 0]{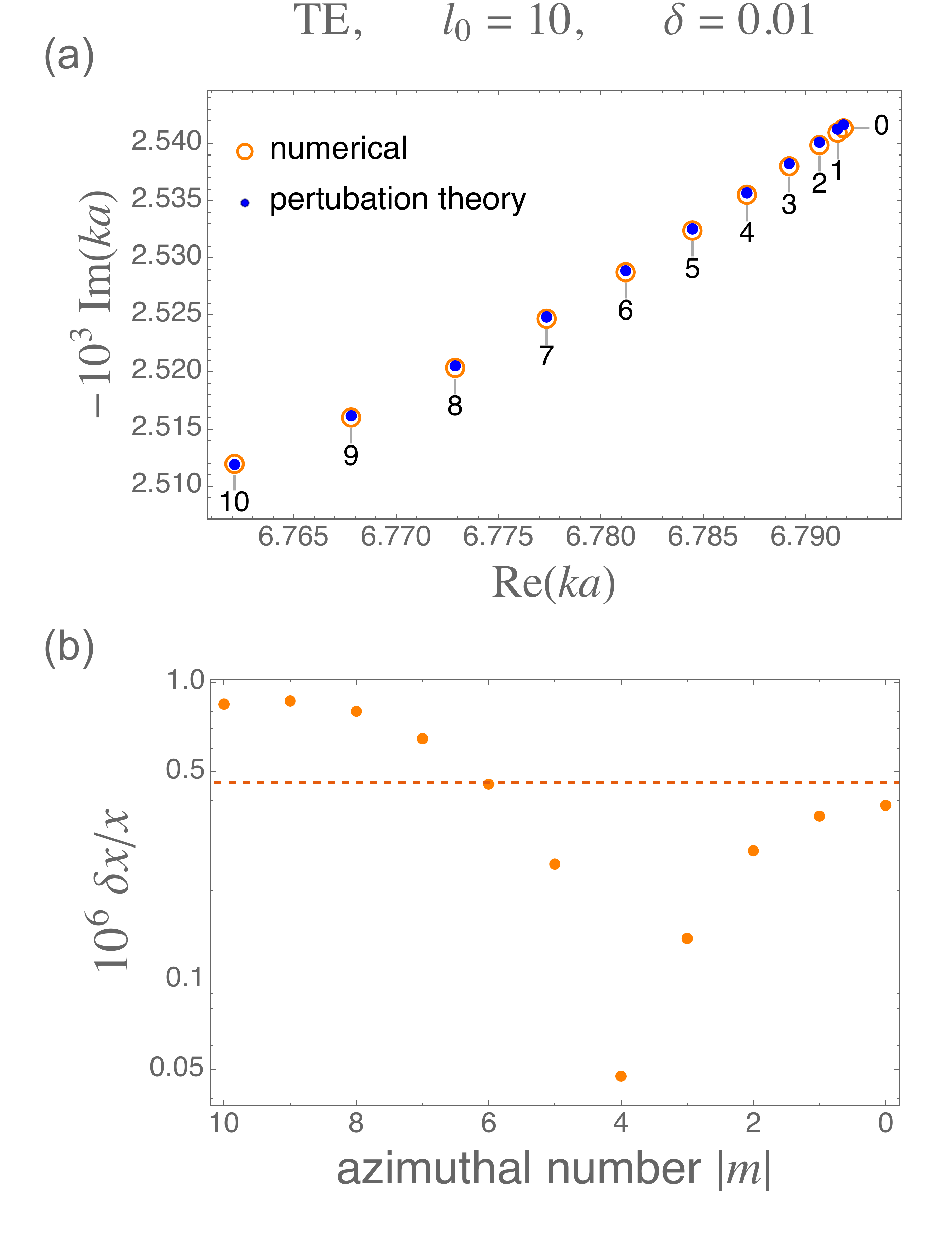}}
\caption{\label{fig3} (a) Comparison between   numerical calculations (orange open circles) and perturbation theory predictions (blue closed circles) for the TE complex-valued resonances $x = k a$ of a dielectric approximate  oblate spheroid  with $\delta = 0.01$ and refractive index $n_1 = 2$. Here $ k =  k_{\lz n_0} $, where $\lz = 10$ and $n_0 = 1$ is the first radial number. The numbers $10, 9, \cdots, 0$ near the resonances mark the values of $0 \leq \abs{m} \leq \lz$. (b) Relative error $\delta x/x$  between numerical and perturbative calculations, calculated according to \eqref{el40}. The dashed orange line gives the average relative error.}
\end{figure}
%

\begin{figure}[!h]
\centerline{\includegraphics[scale=3,clip=false,width=.95\columnwidth,trim = 0 0 0 0]{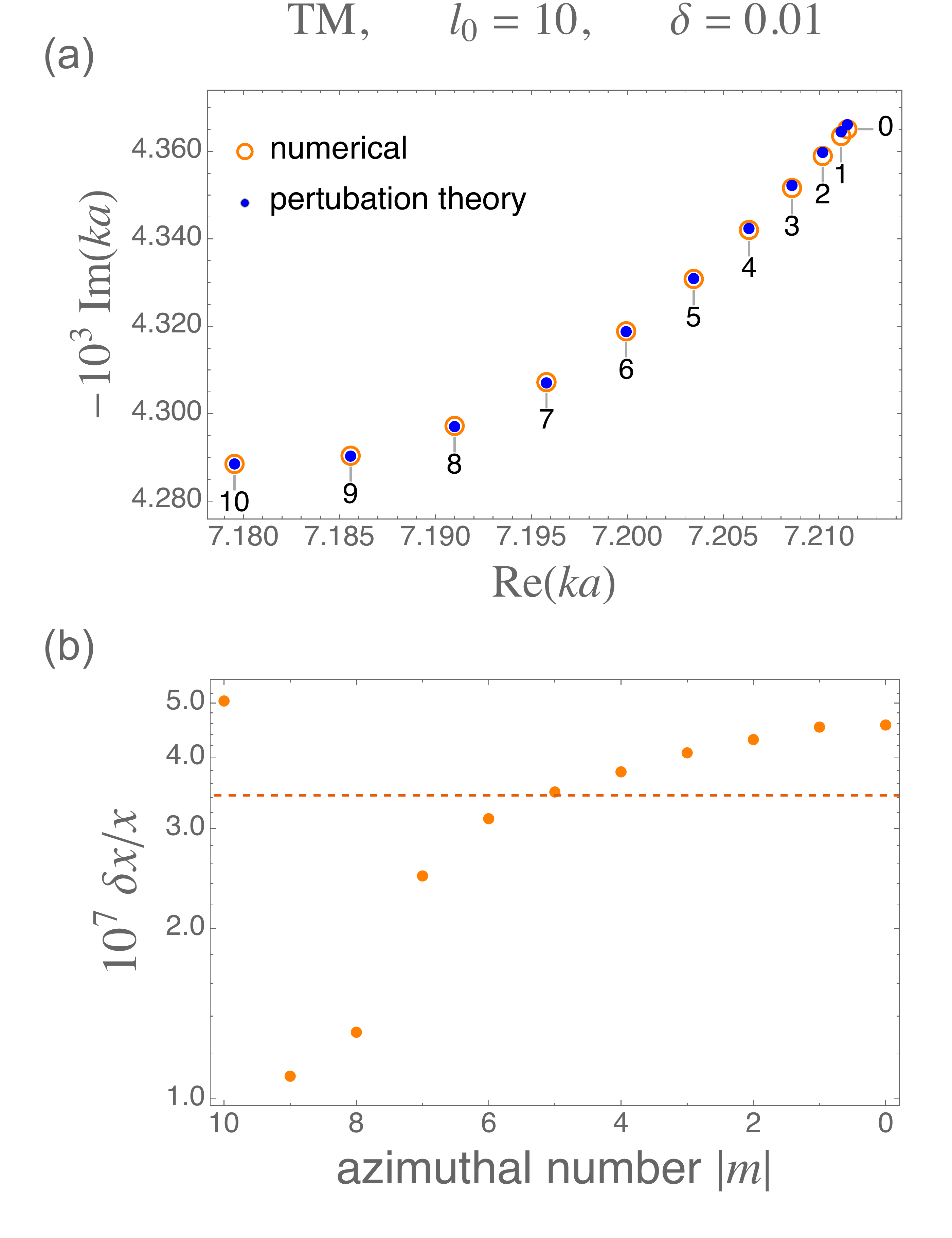}}
\caption{\label{fig4} Same as \eqref{fig3}, but for TM waves. }
\end{figure}

Figures \ref{fig3}(a) and \ref{fig4}(a) show the values of $k a =k_{\lz n_0} a$, 
for TE and TM waves, respectively. The orange open circles are obtained by direct numerical simulations, and the blue closed circles by solving the eigenvalue equation \eqref{x60}.
Deforming the sphere into a spheroid partially lifts the degeneracy, thus yielding $\lz +1 = 11$ distinct resonances, each characterized by a different value of $\abs{m} = 0,1, \cdots, \lz$. The remaining twofold degeneracy is due the rotational invariance of the spheroid with respect to the $z$ axis, which implies that the physics is the same for clockwise ($m>0$) and counterclockwise ($m<0$) waves. Note that waves with $\abs{m} < \lz$  have a polar angle $\theta$ extension, growing with $\lz-\abs{m}$. This implies that they are more sensitive to surface deformations.
%
%

\begin{figure}[!ht]
\centerline{\includegraphics[scale=3,clip=false,width=.9\columnwidth,trim = 0 0 0 0]{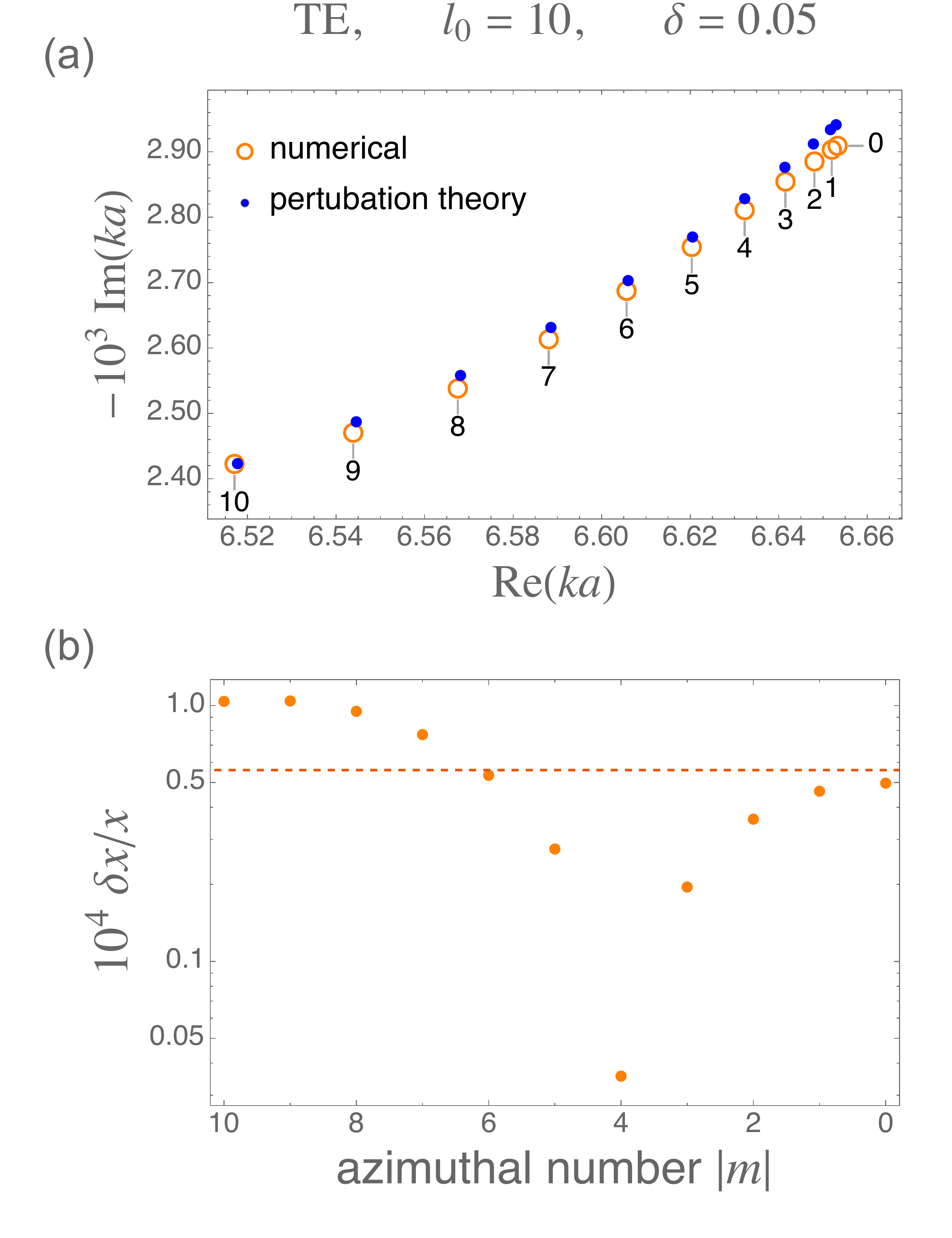}}
\caption{\label{fig5} Same as \eqref{fig3}, but for a bigger value for the magnitude of the deformation $\delta = 0. 05$.}
\end{figure}

\begin{figure}[!ht]
\centerline{\includegraphics[scale=3,clip=false,width=.9\columnwidth,trim = 0 0 0 0]{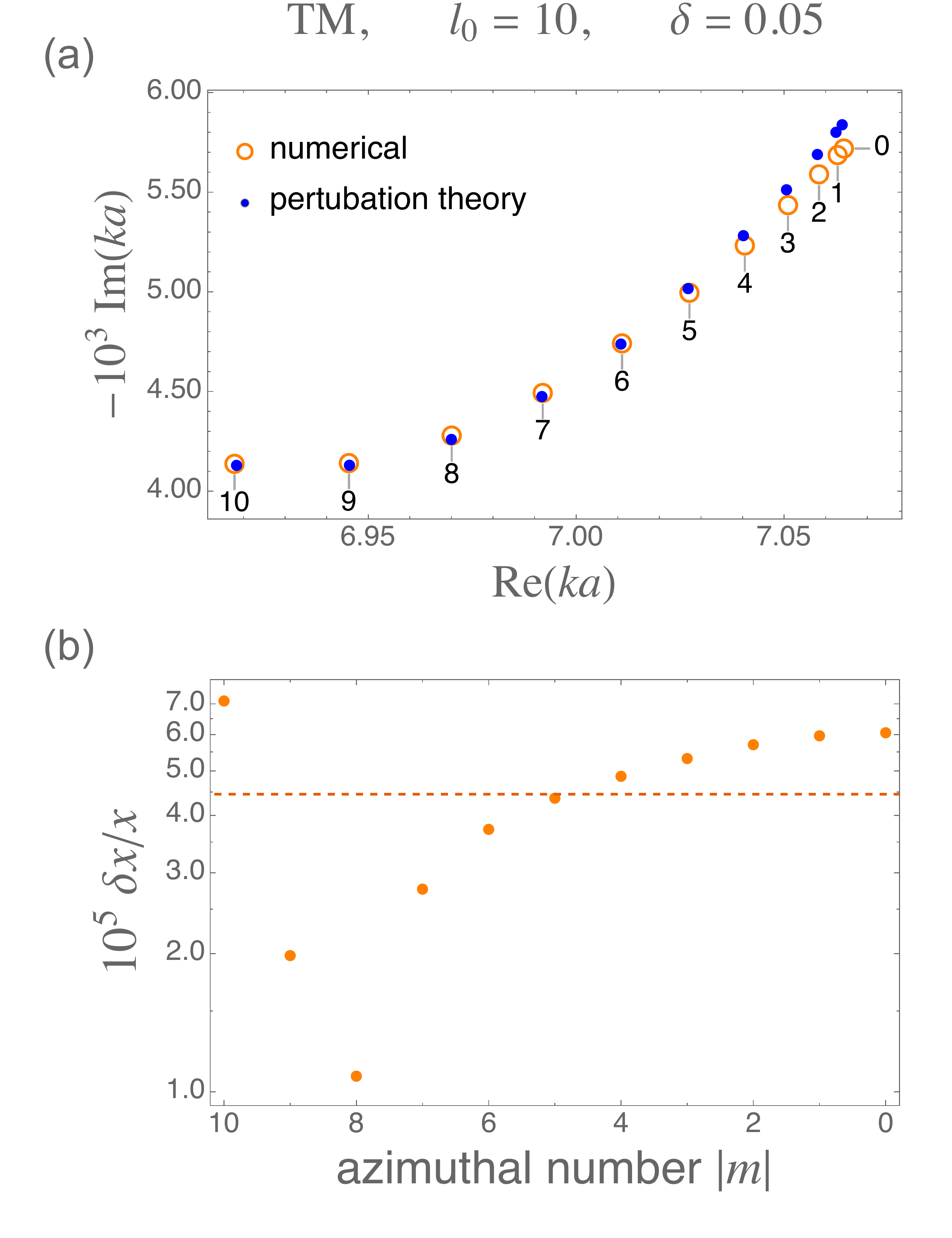}}
\caption{\label{fig6} Same as \eqref{fig5}, but for TM waves.}
\end{figure}

Figures \ref{fig3}(b) and \ref{fig4}(b) display the relative error $\delta x/x$ between numerical ($x_\text{num}$) and perturbative ($x_\text{pert}$) calculations, with $x = k a$, calculated as
\begin{align}\label{el40}
\frac{\delta x}{x} = \left| \frac{x_\text{pert}}{x_\text{num} - \delta x_\text{num}} - 1\right|,
\end{align}
where  $\delta x_\text{num}$ has been estimated as the absolute error between the theoretical (exact, $\abs{m}$-independent) values obtained by solving \eqref{e282} with $l =10, n =1$, and the  ($\abs{m}$-dependent) numerical results for a perfectly spherical cavity.
To give a  quantitative estimate of the error, we have also plotted the average relative error (orange dashed lines).
Figures \ref{fig5} and \ref{fig6} are the same as figs. \ref{fig3} and \ref{fig4}, respectively, but with $\delta = 0.05$.
Overall, all plots exemplify the goodness of the second-order perturbation theory we have developed, even for non-equatorial modes with $\abs{m} < l_0$.
%
%
%
%

To produce the plots above, we greatly benefited from the fact that the infinite sums that appears in \eqref{v90}- \eqref{v100} and \eqref{x70}, actually contain only a finite number of terms according to the rule
\begin{align}\label{u210bis}
{\sum_{l,m,i}}' \quad \to \quad  \sum_{\substack{l = \max(l_0 - L_{\text{max}}, 1) \\[2pt] \, l \neq l_0}}^{l_0+ L_{\text{max}}}\sum_{m = -l}^{l} \sum_{i=1}^4 \; ,
\end{align}
where $L_{\text{max}}$ is determined by the expansion of the deformation function $f(\theta,\phi)$ in terms of the spherical harmonics:
\begin{align}\label{blue10}
f(\theta,\phi) =  \sum_{L=0}^{L_\text{max}} \sum_{M=-L}^L f_{LM} Y_{LM}(\theta,\phi).
\end{align}
For example, from \eqref{el30} it follows that
\begin{align}\label{blue20}
f(\theta,\phi) = &\;  \sin^2 \theta \nonumber \\[4pt]
 = & \;  \frac{4 \sqrt{\pi} }{3} \left[ Y_{00}(\theta,\phi) - \frac{1}{\sqrt{5}}\, Y_{20}(\theta,\phi)\right],
\end{align}
so that $L_\text{max}=2$.

The rule \eqref{u210bis} is empirically determined. However, it could be rigorously proven by writing the product of spherical harmonics in terms of the Wigner $3j$-symbols \cite{Wigner3j-Symbol}. Such products appear in the quantities $ \left\langle \mathbf{X}_{l' m'} , f(\theta,\phi)  \mathbf{\Psi}_{l m} \right\rangle $ and $ \left\langle \mathbf{X}_{l' m'}Y_{lm}(\theta,\phi)  \mathbf{e}_\parallel (\theta,\phi) \right\rangle$, used in \eqref{f10b} and \eqref{f12b}. A practical example of the use of the $3j$-symbols in this kind of calculation, can be found in Sec. VI C of Ref. \cite{PhysRevA.100.023837}, Eqs. (117)-(122).

\section{Summary}\label{Conclusions}

We have developed a boundary conditions perturbation theory to determine the electromagnetic resonances of nearly spherical dielectric resonators. The three-dimensional nature of the resonator and the vector character of the electromagnetic field, dictated the use of vector spherical harmonics  for handling the problem, as opposed to the more familiar scalar spherical harmonics, the latter being typically employed in problems with spherical or nearly-spherical symmetry. By imposing standard electromagnetic boundary conditions at the surface of the resonator separating two different dielectric media, we obtained an exact algebraic homogeneous system of linear equations.
The mathematical correspondence between linear operators and matrices, allowed us to reformulate the problem in the language of quantum mechanics, and to  use the well-known Rayleigh-Schr\"{o}dinger perturbation theory, to build up a perturbation series for the resonances of the electromagnetic field, up to and including second-order terms.
However, as dielectric resonators are de facto  open systems, we had to use the mathematical machinery of non-Hermitian operators and biorthogonal bases.
 We considered both simple and degenerate unperturbed spectra, including the case when degeneracy is not fully removed to first order. For the latter instance, exemplified by the spectrum of an oblate spheroid resonator, we have compared the predictions of our theory with numerical calculations, finding excellent agreement.

The main results are represented by Eqs. \eqref{u190}, \eqref{v90}, \eqref{x60}, and \eqref{x70}. These formulas can be used to calculate the spectrum of the electromagnetic resonances of arbitrarily deformed nearly-spherical dielectric resonators of any size, provided  the conditions \eqref{e422} for the applicability of the perturbation theory are satisfied. Notably, as second-order terms are included, this theory can also be used for the calculation of the spectra of spherical resonators with random surface roughness. This is the case, for example, of helium droplets with thermally excited capillary waves  \cite{PhysRevA.96.063842}.

\begin{acknowledgments}
T.S.  acknowledges support from the European Union's Horizon 2020 research and innovation program under the Marie Sklodowska-Curie Grant Agreement No. 722923 (OMT). The work of A.A. was supported by the European Union’s Horizon 2020 research and innovation programme under grant agreement No. 732894 (FET Proactive HOT).
All the authors thank Florian Marquardt for useful discussions. A.A. also acknowledges financial
support from the Deutsche Forschungsgemeinschaft  Project No. 429529648-TRR 306 QuCoLiMa (``Quantum Cooperativity of Light and Matter'').
\end{acknowledgments}

\appendix

\section{Properties of the matrix elements}\label{operators}

\begin{widetext}

In this appendix we demonstrate some general properties of the operator $\hmM$  defined by \eqref{e990}. In particular, we want to calculate the terms of the expansion
\begin{align}\label{op10}
\mean{l',m', i}{\hmM^\z + \hmM^\un + \hmM^\du + \cdots}{l,m,j}
 = \bigl[ {M_{lm}^{l'm'  \z}} \bigr]_{ij} + \bigl[ {M_{lm}^{l'm'  \un}} \bigr]_{ij} + \bigl[ {M_{lm}^{l'm'  \du}} \bigr]_{ij}+ \cdots \; ,
\end{align}
to demonstrate the validity of \eqref{ePropOp20}-\eqref{s35}.

As $\hmM$ is operatively defined by its matrix elements, defined by \eqref{e990}, we must investigate the properties of $\bigl[ {M_{lm}^{l'm' \, \n}} \bigr]_{ij}$, with $n=0,1,2, \cdots$ and $i,j=1,2,3,4$.
To begin with, we note that from \eqref{e636} it follows  that for $\ve =0 $, the $4 \times 4$ matrix ${M_{lm}^{l'm' \, \z}}$ can be written in terms of
\begin{subequations}\label{e636bis}
\begin{align}
[A_\alpha^{\Phi \z}]_{lm}^{l'm'} = & \; \delta_{l l'} \delta_{m m'} , \label{e636bisa} \\[4pt]
[A_\alpha^{\Psi \z}]_{lm}^{l'm'} = & \; 0, \label{e636bisb} \\[4pt]
[B_\alpha^{\Phi \z}]_{lm}^{l'm'} =  & \; 0, \label{e636bisc} \\[4pt]
[B_\alpha^{\Psi \z}]_{lm}^{l'm'} =& \; \delta_{l l'} \delta_{m m'} g_{\alpha l}(n_\alpha x), \label{e636bisd}
\end{align}
\end{subequations}
where, according to \eqref{e634},
\begin{align}\label{e634bis}
 g_{\alpha l}(n_\alpha x) \equiv  F^\Psi_{\alpha l} (n_\alpha x) =  \frac{1}{i} \, \frac{\left[ \bigl( n_\alpha x \bigr) b_{\alpha l} \bigl( n_\alpha x \bigr)\right]'}{(n_\alpha x) b_{\alpha l} ( n_\alpha x )},
\end{align}
with $\alpha = 1,2$. Using this result and following the discussion in Sec. \ref{remarks}, we can express  the $4 \times 4$ matrix  ${M_{lm}^{l'm'}}$ [Eq. \eqref{e670bis}] in terms of the following elements:
\begin{subequations}\label{e900}
\begin{align}
[A_\alpha^\Phi]_{lm}^{l'm'} (x)= & \; \delta_{l l'} \delta_{m m'}  + \ve \, [A_\alpha^{\Phi(1)}]_{lm}^{l'm'} + \ve^2[A_\alpha^{\Phi(2)}]_{lm}^{l'm'}  + \cdots \quad,\label{e900a} \\[4pt]
[A_\alpha^\Psi]_{lm}^{l'm'}(x) = & \;   \ve \, [A_\alpha^{\Psi(1)}]_{lm}^{l'm'} + \ve^2[A_\alpha^{\Psi(2)}]_{lm}^{l'm'} + \cdots \quad ,\label{e900b} \\[4pt]
[B_\alpha^\Phi]_{lm}^{l'm'}(x) = & \;  \ve \, [B_\alpha^{\Phi(1)}]_{lm}^{l'm'} + \ve^2[B_\alpha^{\Phi(2)}]_{lm}^{l'm'}  + \cdots \quad,\label{e900c} \\[4pt]
[B_\alpha^\Psi]_{lm}^{l'm'}(x) = & \;   \delta_{l l'} \delta_{m m'} g_{ \alpha l}\bigl(n_\alpha x^\z \bigr) + \ve \, [B_\alpha^{\Psi(1)}]_{lm}^{l'm'} + \ve^2[B_\alpha^{\Psi(2)}]_{lm}^{l'm'}  + \cdots  \quad . \label{e900d}
\end{align}
\end{subequations}

Next we  evaluate the terms on the right-hand side of \eqref{e900}. To this end, it is convenient to rewrite the three functions \eqref{e560} isolating their common denominator $b_{\alpha l} ( k_\alpha a )$, that is
\begin{subequations}\label{f5}
\begin{align}
F^Y_{\alpha  l}(k_\alpha r)  =& \;   \frac{1}{i} \, l(l+1) \frac{b_{\alpha l} \bigl( k_\alpha r \bigr)}{k_\alpha r } \,  \frac{1}{b_{\alpha l} ( k_\alpha a )} \equiv \frac{c^Y_{\alpha  l}(k_\alpha r)}{b_{\alpha l} ( k_\alpha a )},\label{f5a} \\[4pt]
F^\Psi_{\alpha l} (k_\alpha r)  = &  \frac{1}{i}  \, \frac{\left[ \bigl( k_\alpha r \bigr) b_{\alpha l} \bigl( k_\alpha r \bigr)\right]'}{k_\alpha r }  \, \frac{1}{b_{\alpha l} ( k_\alpha a )} \equiv  \frac{c^\Psi_{\alpha  l}(k_\alpha r)}{b_{\alpha l} ( k_\alpha a )}, \label{f5b} \\[4pt]
F^\Phi_{\alpha l}(k_\alpha r)  = & \; b_{\alpha l} \bigl( k_\alpha r \bigr) \, \frac{1}{b_{\alpha l} ( k_\alpha a )} \equiv  \frac{c^\Phi_{\alpha  l}(k_\alpha r)}{b_{\alpha l} ( k_\alpha a )} . \label{f5c}
\end{align}
\end{subequations}
Note that in each of these expressions the dependence on $\ve$ enters  in different ways in the numerator and the denominator, because
\begin{align}\label{f7}
\frac{c^W_{\alpha  l}(k_\alpha r)}{b_{\alpha l} ( k_\alpha a )} \to & \; \left. \frac{c^W_{\alpha  l}\bigl(n_\alpha x(\ve) r/a \bigr)}{b_{\alpha l} \bigl( n_\alpha x(\ve) \bigr)} \right|_{r = a(1 + \ve f(\theta,\phi))} \nonumber  \\[4pt]
 & = \frac{c^W_{\alpha  l}\Bigl(n_\alpha \bigl[x^\z + \ve \, x^\un + \ve^2 x^\du + \cdots \bigr] \bigl[ 1 + \ve f(\theta,\phi) \bigl] \Bigr)}{b_{\alpha l} \bigl( n_\alpha [x^\z + \ve \, x^\un + \ve^2 x^\du + \cdots ] \bigr)},
\end{align}
with $W = \Psi, \Phi, Y$.

Substituting \eqref{f5} into \eqref{e630}, making a Taylor expansion around $\ve =0$ using \eqref{f7}, and evaluating the integrals \eqref{e640}, we eventually obtain, up to and including second-order terms,
\begin{subequations}\label{f10}
\begin{align}
[A_\alpha^{X(0)}]_{lm}^{l'm'} = & \; \underbrace{
\frac{c^\Phi_{\alpha  l}\bigl(n_\alpha x^\z \bigr)}{b_{\alpha l} \bigl(n_\alpha x^\z \bigr)}
}_{= \, 1} \left\langle \mathbf{X}_{l' m'} ,  \mathbf{\Phi}_{l m} \right\rangle ,\label{f10a} \\[4pt]
[A_\alpha^{X(1)}]_{lm}^{l'm'} = & \; x^\un \left(n_\alpha \underbrace{\left[\frac{c^\Phi_{\alpha  l}\bigl(n_\alpha x^\z \bigr)}{b_{\alpha l} \bigl(n_\alpha x^\z \bigr)} \right]' }_{ = \, 0} \left\langle \mathbf{X}_{l' m'} ,   \mathbf{\Phi}_{l m} \right\rangle \right)
+ n_\alpha x^\z \frac{\bigl[ c^\Phi_{\alpha l} \bigl(n_\alpha x^\z \bigr) \bigr]' }{b_{\alpha l} \bigl(n_\alpha x^\z \bigr)} \left\langle \mathbf{X}_{l' m'} , f(\theta,\phi)  \mathbf{\Phi}_{l m} \right\rangle ,\label{f10b} \\[4pt]
[A_\alpha^{X(2)}]_{lm}^{l'm'} = & \;  x^\du \left(n_\alpha \underbrace{\left[\frac{c^\Phi_{\alpha  l}\bigl(n_\alpha x^\z \bigr)}{b_{\alpha l} \bigl(n_\alpha x^\z \bigr)} \right]' }_{ = \, 0} \left\langle \mathbf{X}_{l' m'} ,   \mathbf{\Phi}_{l m} \right\rangle \right)
+ \frac{1}{2}  \bigl( n_\alpha x^\un \bigr)^2 \underbrace{\left[\frac{c^\Phi_{\alpha  l}\bigl(n_\alpha x^\z \bigr)}{b_{\alpha l} \bigl(n_\alpha x^\z \bigr)} \right]''}_{ = \, 0}  \left\langle \mathbf{X}_{l' m'} ,  \mathbf{\Phi}_{l m} \right\rangle \nonumber \\[4pt]
& + n_\alpha x^\un \left( \frac{\bigl[ c^\Phi_{\alpha l} \bigl(n_\alpha x^\z \bigr) \bigr]'}{b_{\alpha l} \bigl(n_\alpha x^\z \bigr)}  + n_\alpha x^\z  \left[ \frac{\bigl[ c^\Phi_{\alpha l} \bigl(n_\alpha x^\z \bigr) \bigr]'}{b_{\alpha l} \bigl(n_\alpha x^\z \bigr)} \right]' \right) \left\langle \mathbf{X}_{l' m'} , f(\theta, \phi) \mathbf{\Phi}_{l m} \right\rangle \nonumber \\[4pt]
& + \frac{1}{2}  \bigl( n_\alpha x^\z \bigr)^2 \frac{\bigl[ c^\Phi_{\alpha l} \bigl(n_\alpha x^\z \bigr) \bigr]''}{b_{\alpha l} \bigl(n_\alpha x^\z \bigr)}  \left\langle \mathbf{X}_{l' m'} , f^2(\theta, \phi) \mathbf{\Phi}_{l m} \right\rangle , \label{f10c}
\end{align}
\end{subequations}
and
\begin{subequations}\label{f12}
\begin{align}
[B_\alpha^{X(0)}]_{lm}^{l'm'} = & \; \underbrace{
\frac{c^\Psi_{\alpha  l}\bigl(n_\alpha x^\z \bigr)}{b_{\alpha l} \bigl(n_\alpha x^\z \bigr)}
}_{= \, g_{\alpha l}(n_\alpha x^\z )} \left\langle \mathbf{X}_{l' m'} ,  \mathbf{\Psi}_{l m} \right\rangle ,\label{f12a} \\[4pt]
[B_\alpha^{X(1)}]_{lm}^{l'm'} = & \; x^\un \underbrace{\left(n_\alpha \left[\frac{c^\Psi_{\alpha  l}\bigl(n_\alpha x^\z \bigr)}{b_{\alpha l} \bigl(n_\alpha x^\z \bigr)} \right]'  \left\langle \mathbf{X}_{l' m'} ,   \mathbf{\Psi}_{l m} \right\rangle \right)}_{ \text{contributes to} \; \mean{l',m'}{\hmD }{l,m}}
+ n_\alpha x^\z \frac{\bigl[ c^\Psi_{\alpha l} \bigl(n_\alpha x^\z \bigr) \bigr]' }{b_{\alpha l} \bigl(n_\alpha x^\z \bigr)} \left\langle \mathbf{X}_{l' m'} , f(\theta,\phi)  \mathbf{\Psi}_{l m} \right\rangle\nonumber \\[4pt]
& + \underbrace{\frac{c^Y_{\alpha  l}\bigl(n_\alpha x^\z \bigr)}{b_{\alpha l} \bigl(n_\alpha x^\z \bigr)}}_{ = \, \frac{1}{i} \frac{l(l+1)}{n_\alpha x^\z}} \left\langle \mathbf{X}_{l' m'} , Y_{lm}(\theta,\phi)  \mathbf{e}_\parallel (\theta,\phi) \right\rangle ,\label{f12b} \\[4pt]
[B_\alpha^{X(2)}]_{lm}^{l'm'} = & \;  x^\du \underbrace{\left(n_\alpha \left[\frac{c^\Psi_{\alpha  l}\bigl(n_\alpha x^\z \bigr)}{b_{\alpha l} \bigl(n_\alpha x^\z \bigr)} \right]' \left\langle \mathbf{X}_{l' m'} ,   \mathbf{\Psi}_{l m} \right\rangle \right)}_{ \text{contributes to} \; \mean{l',m'}{\hmD }{l,m}}
+ \frac{1}{2}  \bigl( n_\alpha x^\un \bigr)^2 \left[\frac{c^\Psi_{\alpha  l}\bigl(n_\alpha x^\z \bigr)}{b_{\alpha l} \bigl(n_\alpha x^\z \bigr)} \right]''  \left\langle \mathbf{X}_{l' m'} ,  \mathbf{\Psi}_{l m} \right\rangle \nonumber \\[4pt]
& + n_\alpha x^\un \left( \frac{\bigl[ c^\Psi_{\alpha l} \bigl(n_\alpha x^\z \bigr) \bigr]'}{b_{\alpha l} \bigl(n_\alpha x^\z \bigr)}  + n_\alpha x^\z  \left[ \frac{\bigl[ c^\Psi_{\alpha l} \bigl(n_\alpha x^\z \bigr) \bigr]'}{b_{\alpha l} \bigl(n_\alpha x^\z \bigr)} \right]' \right) \left\langle \mathbf{X}_{l' m'} , f(\theta, \phi) \mathbf{\Psi}_{l m} \right\rangle \nonumber \\[6pt]
& + \frac{1}{2}  \bigl( n_\alpha x^\z \bigr)^2 \frac{\bigl[ c^\Psi_{\alpha l} \bigl(n_\alpha x^\z \bigr) \bigr]''}{b_{\alpha l} \bigl(n_\alpha x^\z \bigr)}  \left\langle \mathbf{X}_{l' m'} , f^2(\theta, \phi) \mathbf{\Psi}_{l m} \right\rangle + n_\alpha x^\un \left[\frac{c^Y_{\alpha  l}\bigl(n_\alpha x^\z \bigr)}{b_{\alpha l} \bigl(n_\alpha x^\z \bigr)} \right]' \left\langle \mathbf{X}_{l' m'} , Y_{lm}(\theta,\phi)  \mathbf{e}_\parallel (\theta,\phi) \right\rangle \nonumber \\[6pt]
& +  \left( - \frac{c^Y_{\alpha  l}\bigl(n_\alpha x^\z \bigr)}{b_{\alpha l} \bigl(n_\alpha x^\z \bigr)} + n_\alpha x^\z  \frac{\bigl[ c^Y_{\alpha l} \bigl(n_\alpha x^\z \bigr) \bigr]'}{b_{\alpha l} \bigl(n_\alpha x^\z \bigr)} \right)
\left\langle \mathbf{X}_{l' m'} , Y_{lm}(\theta,\phi) f(\theta,\phi)  \mathbf{e}_\parallel (\theta,\phi) \right\rangle  , \label{f12c}
\end{align}
\end{subequations}
where single and double primes denote, respectively, first and second derivatives with respect to the argument, and we have introduced the suggestive notation
\begin{align}\label{f30}
\left\langle \mathbf{X}_{l' m'} , \mathbf{H}(r,\theta,\phi) \right\rangle  =   \frac{1}{l'(l'+1)} \int \mathbf{X}_{l'm'}^*(\theta,\phi) \cdot \mathbf{H}(r,\theta,\phi) \, d \Omega,
\end{align}
with $X = \Psi, \Phi$, and $\mathbf{H}(r,\theta,\phi)$ being an arbitrary three-dimensional vector field.
Calculating explicitly the higher-order terms, it is straightforward to see that
\begin{align}\label{f35}
\frac{d\, [A_\alpha^{X(n)}]_{lm}^{l'm'}}{d \, x^\n} =0, \qquad \text{and}
\qquad \frac{d\, [B_\alpha^{X(n)}]_{lm}^{l'm'}}{d \, x^\n} = n_\alpha \left[ F^\Psi_{\alpha  l}\bigl(n_\alpha x^\z \bigr) \right]' \left\langle \mathbf{X}_{l' m'} ,   \mathbf{\Psi}_{l m} \right\rangle , \qquad (n\geq 1),
\end{align}
where \eqref{f5b} has been used, and
\begin{align}\label{f37}
\left\langle \mathbf{X}_{l' m'} ,   \mathbf{\Psi}_{l m} \right\rangle
= \left\{
         \begin{array}{ll}
         0, & \hbox{if} \quad X = \Phi, \\[6pt]
         \delta_{l l'} \delta_{m m'}, & \hbox{if} \quad X = \Psi.
         \end{array}
  \right.
\end{align}
Similarly,
\begin{align}\label{f38}
\left\langle \mathbf{X}_{l' m'} ,   \mathbf{\Phi}_{l m} \right\rangle
= \left\{
         \begin{array}{ll}
         0, & \hbox{if} \quad X = \Psi, \\[6pt]
         \delta_{l l'} \delta_{m m'}, & \hbox{if} \quad X = \Phi.
         \end{array}
  \right.
\end{align}
Other useful properties of the vector spherical harmonics are
\begin{subequations}\label{e902}
\begin{align}
\mathbf{\Phi}_{l' m'}^* \cdot \mathbf{\Phi}_{l m} = & \;  \mathbf{\Psi}_{l' m'}^* \cdot \mathbf{\Psi}_{l m}  , \label{e902a} \\[6pt]
\mathbf{\Psi}_{l' m'}^* \cdot \mathbf{\Phi}_{l m} = & \;  - \mathbf{\Phi}_{l' m'}^* \cdot \mathbf{\Psi}_{l m} \label{e902b} .
\end{align}
\end{subequations}
To evaluate the integrals containing $\mathbf{e}_\parallel (\theta,\phi)$  in \eqref{f12b} and \eqref{f12c}, we find it useful to recast $\mathbf{e}_\parallel (\theta,\phi)$ into the form
\begin{align}\label{ff10}
\mathbf{e}_\parallel (\theta,\phi) = \sum_{L=0}^\infty \sum_{M=-L}^L f_{LM} \mathbf{\Psi}_{LM}(\theta,\phi),
\end{align}
where, from the definition \eqref{e314},
\begin{align}\label{ff20}
f_{LM} = \int Y^*_{LM} (\theta,\phi) \, f (\theta,\phi) \, d \Omega,
\end{align}
and, by definition,
\begin{align}\label{ff30}
 f (\theta,\phi) =  \sum_{L=0}^\infty \sum_{M=-L}^L f_{LM} Y_{LM} (\theta,\phi).
\end{align}

Gathering all these results, we can eventually write
\begin{subequations}\label{e905}
\begin{align}
[A_\alpha^{\Psi (n)}]_{lm}^{l'm'} = & \;  [\mA_\alpha^{\Psi (n)}]_{lm}^{l'm'} , \label{e905a} \\[6pt]
[A_\alpha^{\Phi (n)}]_{lm}^{l'm'} = & \; [\mA_\alpha^{\Phi (n)}]_{lm}^{l'm'}, \label{e905b} \\[6pt]
 [B_\alpha^{\Psi (n)}]_{lm}^{l'm'}= & \;  [\mB_\alpha^{\Psi (n)}]_{lm}^{l'm'}  + x^{(n)} \, \delta_{l l'} \delta_{m m'} d_{\alpha l} , \label{e905c} \\[6pt]
 [B_\alpha^{\Phi (n)}]_{lm}^{l'm'} = & \;  [\mB_\alpha^{\Phi (n)}]_{lm}^{l'm'} , \label{e905d}
\end{align}
\end{subequations}
where the terms denoted by calligraphic letters are independent of $x^{(n)}$ and contribute to $\hmV^\n$. The diagonal operator $\hmD$ is characterized by
\begin{align}\label{f60}
d_{\alpha l} =  n_\alpha \bigl[ F^\Psi_{\alpha l}\bigl(n_\alpha x^\z \bigr) \bigr]' =  n_\alpha \bigl[ g_{\alpha  l}\bigl(n_\alpha x^\z \bigr) \bigr]' , \qquad (\alpha =1,2)  ,
\end{align}
and represented by
\begin{align}\label{f50}
D_l \doteq
\begin{bmatrix}
  0 & 0 & 0 & 0 \\[8pt]
  -n_1^2 \bigl[ g_{1  l}\bigl(n_1 x^\z \bigr) \bigr]'  & n_2^2 \bigl[ g_{2  l}\bigl(n_2 x^\z \bigr) \bigr]' & 0 & 0 \\[8pt]
  0 & 0 & \bigl[ g_{1  l}\bigl(n_1 x^\z \bigr) \bigr]' & - \bigl[ g_{2  l}\bigl(n_2 x^\z \bigr) \bigr]' \\[8pt]
  0 & 0 & 0 & 0 \\
\end{bmatrix}.
\end{align}
This completes the proof of the validity of \eqref{ePropOp20}-\eqref{s35}.

\end{widetext}

\section{Calculation of $\mean{\ta_0}{P_\mD \,\hmV^\un\, P_\mD}{\alpha_0}$}\label{MatEl}

\begin{widetext}

In this appendix we calculate the  elements of the $ N_\lz \times  N_\lz$ matrix $\mean{\ta_0}{P_\mD \,\hmV^\un\, P_\mD}{\alpha_0}$, with $ N_\lz = 2 \lz+1$. The knowledge of this matrix permits us  to evaluate $x_\mu^\un $  from \eqref{u190}, here rewritten as
\begin{align}\label{z10}
 x_\mu^\un = -\mean{\tvp^\z_{\mu}}{ \;  \frac{\mean{\ta_0}{P_\mD \,\hmV^\un\, P_\mD}{\alpha_0}}{\mean{\ta_0}{D_\lz}{\alpha_0}} \;
}{\vp^\z_{\mu}}, \qquad (\mu = 1, 2, \cdots, N_\lz).
\end{align}
In practice, to solve \eqref{z10} we need to solve the right- and left-eigenvalue equations
\begin{subequations}\label{z20}
\begin{align}
  \frac{\mean{\ta_0}{P_\mD \,\hmV^\un\, P_\mD} {\alpha_0} }{\mean{\ta_0}{D_\lz}{\alpha_0}} \, \ket{\vp^\z_{\mu}}  = & \; - x_\mu^\un \ket{\vp^\z_{\mu}}  , \label{z20a} \\[6pt]
 \bra{\tvp^\z_{\mu 0}} \, \frac{\mean{\ta_0}{P_\mD \,\hmV^\un\, P_\mD} {\alpha_0} }{\mean{\ta_0}{D_\lz}{\alpha_0}} = & \; - x_\mu^\un \bra{\tvp^\z_{\mu}} . \label{z20b}
\end{align}
\end{subequations}
The procedure is straightforward: Multiplying \eqref{z20a} from the left by $\bra{\lz, m'}$  and using \eqref{u20}, we obtain
\begin{align}\label{z30}
\sum_{m = -\lz}^\lz   \frac{V^\un_{m'm}}{\mean{\ta_0}{D_\lz}{\alpha_0}} \, \vp^\z_{\mu m}  =   -x_\mu^\un \, \vp^\z_{\mu m'}   ,
\end{align}
where we have used \eqref{u180} twice, to rewrite
\begin{align}\label{y10}
\mean{\ta_0}{P_\mD \,\hmV^\un\, P_\mD}{\alpha_0} = & \; \sum_{m' = -\lz}^\lz \sum_{m = -\lz}^\lz \proj{\lz, m'}{\lz, m} \mean{\lz, m', \ta_0}{\hmV^\un}{l_0, m, \alpha_0} \nonumber \\[6pt]
\equiv & \; \sum_{m' = -\lz}^\lz \sum_{m = -\lz}^\lz \proj{\lz, m'}{\lz, m} \, V^\un_{m'm},
\end{align}
where $V^\un_{m'm}$, with $m',m = -\lz, \lz+1, \cdots, \lz-1, \lz$, denotes the matrix element of the $N_\lz \times N_\lz$ matrix $V^\un$ to be diagonalized. Using the definitions \eqref{e990} and \eqref{ePropOp25a}, we readily find
\begin{align}\label{y20}
V^\un_{m'm} = & \; \left. \mean{\lz, m', \ta_0}{\hmM^\un}{l_0, m, \alpha_0} \right|_{x^\un = 0} \nonumber \\[6pt]
= & \; \left. \mean{\ta_0}{M^{\lz m' \un}_{\lz m} }{\alpha_0} \right|_{x^\un = 0} .
\end{align}
Substituting \eqref{y20} in \eqref{z30}, we can straightforwardly determine the sought eigenvalues $x_\mu^\un$ and eigenvectors $\ket{\vp_{\mu 0}}$. A similar procedure can be repeated to calculate the left eigenvectors $\bra{\tvp_{\mu 0}}$.

Note that $V^\un_{m'm}$ takes a different value for TE and TM waves. Specifically,  we find, for $n \geq 1$,
\begin{align}\label{y30}
 \mean{\ta_0^E}{M^{\lz m' \n}_{\lz m} }{\alpha_0^E}  = \frac{1}{z^E+1}
\Biggl\{ z^E \left( [A_1^{\Phi \n }]_{\lz m}^{\lz m'} - [A_2^{\Phi \n }]_{\lz m}^{\lz m'} \right) -
\left( n_1 [B_1^{\Psi \n }]_{\lz m}^{\lz m'} - n_2 [B_2^{\Psi \n }]_{\lz m}^{\lz m'} \right)\Biggr\},
\end{align}
for TE waves and
\begin{align}\label{y40}
 \mean{\ta_0^M}{M^{\lz m' \n}_{\lz m} }{\alpha_0^M}  =  \frac{1}{z^M - 1}
\left\{-\left( \frac{[B_1^{\Psi \n }]_{\lz m}^{\lz m'}}{n_1} - \frac{[B_2^{\Psi \n }]_{\lz m}^{\lz m'}}{n_2} \right) + z^M
\Biggl([A_1^{\Phi \n }]_{\lz m}^{\lz m'} -  [A_2^{\Phi \n }]_{\lz m}^{\lz m'} \Biggr) \right\},
\end{align}
for TM waves, with $z^E$ and $z^M$ defined by \eqref{zE} and \eqref{zM}, respectively. From the definitions \eqref{e280TE} and \eqref{f5} it follows that, for $n=1$,
\begin{subequations}\label{y50}
\begin{align}
\left. [A_1^{\Phi \un }]_{l m}^{l m'} - [A_2^{\Phi \un }]_{l m}^{l m'} \right|_{x^\un=0}= & \; \frac{1}{i} \, x^\z f^E_l \bigl( x^\z\bigr) \left\langle \mathbf{\Phi}_{l m'} ,  f(\theta,\phi) \mathbf{\Phi}_{l m} \right\rangle ,\label{y50a} \\[6pt]
\left.  n_1 [B_1^{\Psi \un }]_{l m}^{l m'} - n_2 [B_2^{\Psi \un }]_{l m}^{l m'} \right|_{x^\un=0}  = & \;
\left[i \, x^\z \left( n_1^2  - n_2^2 \right)   + f_l^E \bigl( x^\z \bigr) \right]
\left\langle \mathbf{\Psi}_{l m'} ,  f(\theta,\phi) \mathbf{\Psi}_{l m} \right\rangle , \label{y50b} \\[6pt]
\left.  \frac{1}{n_1} [B_1^{\Psi \un }]_{l m}^{l m'} - \frac{1}{n_2} [B_2^{\Psi \un }]_{l m}^{l m'}  \right|_{x^\un=0} = & \;
\left[\frac{l(l+1)}{i \, x^\z }\left( \frac{1}{n_1^2}  - \frac{1}{n_2^2} \right)   + f_l^M \bigl( x^\z \bigr) \right]
\left\langle \mathbf{\Psi}_{l m'} ,  f(\theta,\phi) \mathbf{\Psi}_{l m} \right\rangle \\[6pt]
& + \frac{l(l+1)}{i \, x^\z }\left( \frac{1}{n_1^2}  - \frac{1}{n_2^2} \right)\left\langle \mathbf{\Psi}_{l m'} ,  Y_{lm}(\theta,\phi) \mathbf{e}_\parallel \right\rangle. \label{y50c}
\end{align}
\end{subequations}
To derive these expressions we find it useful to replace second derivatives of the spherical Bessel functions $b_{\alpha l}(u)$ defined by \eqref{funcB}, according to  Bessel's differential equation
\begin{align}\label{y60}
\frac{d^2}{d u^2} \, b_{\alpha l}(u) =  - \left[ \frac{2}{u} \, \frac{d }{d u} + 1 - \frac{l(l+1)}{u^2} \right] b_{\alpha l}(u) .
\end{align}
We can use \eqref{y50} to simplify \eqref{y30} and \eqref{y40}, because $f^E_\lz \bigl( x^\z_E \bigr) = 0 = f_\lz^M \bigl( x^\z_M \bigr)$. After a straightforward calculation we obtain
\begin{align}\label{y30bis}
\left. \mean{\ta_0^E}{M^{\lz m' \un}_{\lz m} }{\alpha_0^E} \right|_{x^\un = 0} = \frac{1}{i}\frac{ 1}{z^E+1} \,  x^\z
\left( n_1^2  - n_2^2 \right)
\left\langle \mathbf{\Psi}_{\lz m'} ,  f(\theta,\phi) \mathbf{\Psi}_{\lz m} \right\rangle
\end{align}
for TE waves and
\begin{align}\label{y40bis}
\left. \mean{\ta_0^M}{M^{\lz m' \un}_{\lz m} }{\alpha_0^M} \right|_{x^\un = 0} = & \; \frac{1}{i} \, \frac{1}{z^M - 1} \Biggl\{ z^M \, x^\z_M f^E_\lz \bigl( x^\z_M \bigr) \left\langle \mathbf{\Phi}_{\lz m'} ,  f(\theta,\phi) \mathbf{\Phi}_{\lz m} \right\rangle \nonumber \\[6pt]
& -  \frac{l(l+1)}{x^\z_M} \left( \frac{1}{n_1^2} - \frac{1}{n_2^2}\right)\biggl[
\langle \mathbf{\Phi}_{\lz m'} ,  f(\theta,\phi) \mathbf{\Phi}_{\lz m} \rangle + \langle \mathbf{\Psi}_{\lz m'} ,  Y_{\lz m}(\theta,\phi) \mathbf{e}_\parallel \rangle
\biggr] \Biggr\}
\end{align}
for TM waves.

Finally, we evaluate the denominator in \eqref{u190}. A lengthy but straightforward calculation gives
\begin{subequations}\label{y70}
\begin{align}
\mean{\ta_0^E}{D_\lz}{\alpha_0^E} = & \; \frac{1}{z^E + 1} \left( n_2^2 \bigl[ g_{2  \lz}\bigl(n_2 x^\z_E \bigr) \bigr]'  - n_1^2 \bigl[ g_{1  \lz}\bigl(n_1 x^\z_E \bigr) \bigr]' \right) \nonumber \\[8pt]
= & \; \frac{1}{i} \, \frac{1 }{z^E+1} \left( n_1^2 - n_2^2 \right) ,\label{y70a} \\[6pt]
\mean{\ta_0^M}{D_\lz}{\alpha_0^M}  = & \; \frac{1}{z^M -1} \left( \bigl[ g_{2  \lz}\bigl(n_2 x^\z_M \bigr) \bigr]'  - \bigl[ g_{1  \lz}\bigl(n_1 x^\z_M \bigr) \bigr]' \right)\nonumber \\[8pt]
= & \;  \frac{1}{i} \, \frac{1}{ z^M-1} \, \left[ - \frac{(2 \lz +1)( \lz+1)}{\bigl(x^\z_M \bigr)^2} +   \left( \frac{j_{ \lz+1}\bigl(n_1 x^\z_M \bigr)}{j_{ \lz}\bigl(n_1 x^\z_M \bigr)} \right)^2 - \left( \frac{h_{ \lz+1}^\un\bigl(n_2 x^\z_M \bigr)}{h_{ \lz}^\un\bigl(n_2 x^\z_M \bigr)} \right)^2  \right] , \label{y70b}
\end{align}
\end{subequations}
for TE and TM waves, respectively, with
\begin{subequations}\label{y80}
\begin{align}
z^E = & \; \frac{\lz +1}{x^\z_E} - n_1 \, \frac{j_{\lz+1}\bigl(n_1 x^\z_E \bigr)}{j_{\lz}\bigl(n_1 x^\z_E \bigr)} ,\label{y80a} \\[6pt]
z^M  = & \; \frac{1}{n_1^2} \left[\frac{\lz+1}{ x^\z_M} - n_1 \, \frac{j_{\lz+1}\bigl(n_1 x^\z_M \bigr)}{j_{\lz}\bigl(n_1 x^\z_M \bigr)} \right]. \label{y80b}
\end{align}
\end{subequations}

Note the common factors
\begin{align}\label{y90}
\frac{1}{i} \, \frac{1 }{z^E+1}, \qquad \text{and} \qquad  \frac{1}{i} \, \frac{1}{z^M -1},
\end{align}
in front of \eqref{y30bis}-\eqref{y70}. They simplify when tacking the ratios, as required by \eqref{u190}. For example, for TE waves using \eqref{y20}, \eqref{y30bis} and \eqref{y70a},  we obtain a particularly simple result:
\begin{align}\label{y100}
\frac{\displaystyle V^\un_{m'm} }{\mean{\ta_0^E}{D_\lz}{\alpha_0^E}} =  x^\z \langle \mathbf{\Psi}_{\lz m'} ,  f(\theta,\phi) \mathbf{\Psi}_{\lz m} \rangle.
\end{align}
Substituting this result into \eqref{z30}, we obtain
\begin{align}\label{z30bis}
\sum_{m = -\lz}^\lz  \langle \mathbf{\Psi}_{\lz m'} ,  f(\theta,\phi) \mathbf{\Psi}_{\lz m} \rangle \, \vp^\z_{\mu m}  =   -\frac{x_\mu^\un}{x^\z} \, \vp^\z_{\mu m'}  \; , \qquad (\mu = 1, 2, \cdots, N_\lz).
\end{align}
Since $f(\theta,\phi)$ is a real-valued function, Eq. \eqref{z30bis} is a Hermitian eigenvalue equation. This implies that the ratio ${x_\mu^\un}/{x^\z}$ is also real valued, in agreement with previous results \cite{PhysRevA.41.5187,PhysRevA.100.023837}.
\end{widetext}

\section{Proof of $\brak{\tpsi^\z_{A 0}}{\psi_{A}^\n} = 0$ }\label{normalization}

Consider the perturbed vector
\begin{align}\label{appB10}
\ket{\psi_{A}(\ve)} =  \ket{\psi^\z_{A 0}} + \ve  \ket{\psi^\un_{A}} + \ve^2  \ket{\psi^\du_{A}} + O(\ve^3),
\end{align}
where, by hypothesis, the vector corrections $\ket{\psi^\n_{A}} $ do not fulfill \eqref{t90}. However, we can always rewrite each $\ket{\psi^\n_{A}} $ as:
\begin{align}\label{appB20}
\ket{\psi^\n_{A}} = & \; \left( \ket{\psi^\n_{A}}  -  \ket{\psi^\z_{A 0}} \brak{\tpsi^\z_{A 0}}{\psi^\n_{A}} \right)  + \ket{\psi^\z_{A 0}} \brak{\tpsi^\z_{A 0}}{\psi^\n_{A}} \nonumber \\[6pt]
& \nonumber \\[6pt]
\equiv & \; \ket{\psi^\n_{A \perp}} + \ket{\psi^\n_{A \parallel}} ,
\end{align}
where, by construction,
\begin{align}\label{appB30}
\brak{\tpsi^\z_{A 0}}{\psi_{A \perp}^\n} = 0.
\end{align}
Substituting \eqref{appB20} into \eqref{appB10}, we obtain
\begin{align}\label{appB40}
\ket{\psi_{A}(\ve)} = & \; \ket{\psi^\z_{A 0}} + \ve  \left( \ket{\psi^\un_{A \perp}}+ \ket{\psi^\un_{A \parallel}} \right)  \nonumber \\[6pt]
& + \ve^2  \left( \ket{\psi^\du_{A \perp}} + \ket{\psi^\du_{A \parallel}} \right) + \cdots \nonumber \\[6pt]
= & \; \left( \ket{\psi^\z_{A 0}} + \ve \, \ket{\psi^\un_{A \parallel}} + \ve^2 \ket{\psi^\du_{A \parallel}}+ \cdots \right)\nonumber \\[6pt]
& + \ve  \, \ket{\psi^\un_{A \perp}} + \ve^2  \ket{\psi^\du_{A \perp}} + \cdots \; ,
\end{align}
where
\begin{align}\label{appB50}
\ket{\psi^\z_{A 0}} + &  \ve \, \ket{\psi^\un_{A \parallel}} + \ve^2 \ket{\psi^\du_{A \parallel}}+ \cdots \nonumber \\[6pt]
& = \left(1 +  \ve \brak{\tpsi^\z_{A 0}}{\psi^\un_{A}} +  \ve^2 \brak{\tpsi^\z_{A 0}}{\psi^\du_{A}} + \cdots \right) \ket{\psi^\z_{A 0}} \nonumber  \\[6pt]
 & \equiv \frac{1}{Z(\ve)}\ket{\psi^\z_{A 0}},
\end{align}
with $Z(\ve)$ a  normalization factor.
Substituting this result back into \eqref{appB40}, we get
\begin{align}\label{appB60}
\ket{\psi_{A}(\ve)} = \frac{1}{Z(\ve)}\ket{\psi^\z_{A 0}}+ \ve  \, \ket{\psi^\un_{A \perp}} + \ve^2  \ket{\psi^\du_{A \perp}} + \cdots \; .
\end{align}
Since $\ket{\psi_{A}(\ve)}$ satisfies
\begin{align}\label{appB70}
\hmM(\ve) \ket{\psi_{A}(\ve)} =  0,
\end{align}
irrespective of its normalization, we can multiply both sides of \eqref{appB60} by $Z(\ve)$ to obtain
\begin{align}\label{appB80}
\ket{\psi_{A}(\ve)}' = & \; \ket{\psi^\z_{A 0}}+ Z(\ve) \Bigl[\ve  \, \ket{\psi^\un_{A \perp}} + \ve^2  \ket{\psi^\du_{A \perp}} + \cdots \Bigr] \nonumber \\[6pt]
 = & \; \ket{\psi^\z_{A 0}} + \ve   \ket{\psi^\un_{A \perp}} \nonumber \\[6pt]
& + \ve^2 \left( \ket{\psi^\du_{A \perp}} - \brak{\tpsi^\z_{A 0}}{\psi^\un_{A}} \ket{\psi^\un_{A \perp}} \right) + \cdots
\end{align}
where $\ket{\psi_{A}(\ve)}' \equiv Z(\ve) \ket{\psi_{A}(\ve)}$ fulfills
\begin{align}\label{appB90}
\hmM(\ve) \ket{\psi_{A}(\ve)}' =  0.
\end{align}
Equation \eqref{appB80}  shows that now all the corrections to the zeroth-order vector $\ket{\psi^\z_{A 0}}$ are orthogonal to it.


\end{document}